\documentclass[%
aip,
pop,
 amsmath,amssymb,
preprint,%
]{revtex4-1}

\usepackage{caption}
\usepackage{subcaption}

\usepackage{graphicx}
\usepackage{dcolumn}
\usepackage{bm}

\usepackage[utf8]{inputenc}
\usepackage[T1]{fontenc}
\usepackage{mathptmx}
\usepackage{etoolbox}

\usepackage[colorlinks,linkcolor=blue,citecolor=blue,urlcolor=blue]{hyperref}

\makeatletter
\def\@email#1#2{%
 \endgroup
 \patchcmd{\titleblock@produce}
  {\frontmatter@RRAPformat}
  {\frontmatter@RRAPformat{\produce@RRAP{*#1\href{mailto:#2}{#2}}}\frontmatter@RRAPformat}
  {}{}
}%
\makeatother

\newcommand{\myrefeq}[1]{Eq.~(\ref{#1})}
\newcommand{\myreftab}[1]{Table~\ref{#1}}
\newcommand{\myreffig}[1]{Fig.~\ref{#1}} 
\newcommand{\myrefsec}[1]{Sec.~\ref{#1}}

\begin{document}


\title[]{Nonthermal electron velocity distribution functions due to 3D kinetic magnetic reconnection for solar coronal plasma conditions}

\author{Xin Yao}
\affiliation{Max Planck Institute for Solar System Research, 37077 {G\"{o}ttingen}, Germany}
\affiliation{Center for Astronomy and Astrophysics, Technical University of Berlin, 10623 Berlin, Germany}
\email{xin.yao@campus.tu-berlin.de}

\author{Patricio A. Mu\~{n}oz}
\affiliation{Center for Astronomy and Astrophysics, Technical University of Berlin, 10623 Berlin, Germany}

\author{J\"{o}rg B\"{u}chner}
\affiliation{Center for Astronomy and Astrophysics, Technical University of Berlin, 10623 Berlin, Germany}
\affiliation{Max Planck Institute for Solar System Research, 37077 {G\"{o}ttingen}, Germany}

\date{\today}

\begin{abstract}
	Magnetic reconnection can convert magnetic energy into kinetic energy of non-thermal electron beams. 
	Those accelerated electrons can, in turn, cause radio emission in astrophysical plasma environments such as solar flares via micro-instabilities.
	The properties of the electron velocity distribution functions (EVDFs) of those non-thermal beams generated by reconnection are, however, still not well understood. 
	In particular properties that are necessary conditions for some relevant micro-instabilities.
	We aim at characterizing the EVDFs generated in 3D magnetic reconnection by means of fully kinetic particle-in-cell (PIC) code simulations. 
	In particular, our goal is to identify the possible sources of free energy offered by the generated EVDFs and their dependence on the strength of the guide field. 
	By applying a machine learning algorithm on the EVDFs, we find that:
	(1) electron beams with positive gradients in their 1D parallel (to the local magnetic field direction) velocity distribution functions are generated in both diffusion region and separatrices. (2) Electron beams with positive gradients in their perpendicular (to the local magnetic field direction) velocity distribution functions are observed in the diffusion region and outflow region near the reconnection midplane. In particular, perpendicular crescent-shaped EVDFs (in the perpendicular velocity space) are mainly observed in the diffusion region. (3) As the guide field strength increases, the number of locations with EVDFs featuring a perpendicular source of free energy significantly decreases.
	The formation of non-thermal electron beams in the field-aligned direction is mainly due to magnetized and adiabatic electrons, while in the direction perpendicular to the local magnetic field it is attributed to unmagnetized electrons.
\end{abstract}

\maketitle


\section{Introduction}

Magnetic reconnection is a fundamental mechanism of energy conversion from magnetic energy into plasma heating, bulk flow kinetic energy and particle acceleration in astrophysical, space and laboratory plasmas \citep[][]{Birn2007,Mann2009,Treumann2013b}. In particular, magnetic reconnection causes electron acceleration and non-thermal velocity electron distribution functions (EVDFs) in collisionless plasmas, for example, stellar coronae. The non-thermal features of EVDFs are not only caused by local plasma processes but they can also carry signatures of processes such as particle's acceleration and their interaction with turbulence in reconnection. Measurement of non-thermal EVDFs can be used to investigate local processes in magnetic reconnection. 
During most of the space age, ion velocity distribution functions are measured with the limited time resolution by instruments on-board spacecrafts, for example, in the solar wind \citep[][]{Demars1990,Marsch2011,Pierrard2010} and in the Earth's magnetosphere \citep[][]{Vaisberg2004,Perri2020,Runov2021}. Recently the availability of high-resolution instruments allows in-situ observation of EVDFs, for example, in the solar wind by the 3DP instrument onboard the WIND spacecraft \citep[][]{Lin1998,Maksimovic2008b} and in the Earth's magnetosphere by the multi-spacecraft Magnetospheric Multiscale Mission (MMS) allows to obtain highly-resolved EVDFs \citep[][]{Burch2016,Burch2016b,Ergun2018}. However, different from in-situ measured magnetospheric EVDFs, it is still impossible to measure in-situ non-thermal EVDFs at the reconnection sites in the solar corona.

Features and consequences of EVDFs formed by magnetic reconnection have been theoretically and numerically investigated under a wide variety of conditions and approaches since decades ago \citep{Hoshino2001,Drake2003,Che2010,Ng2011,Fujimoto2014a,Bessho2016,Munoz2016,Egedal2016,Zenitani2016}. 
The most commonly found EVDFs of beams moving along the reconnection separatrices are due to acceleration by the reconnection electric field \citep[][]{Oieroset2002,Dahlin2015,Munoz2016}. There are other anisotropic and non-gyrotropic EVDFs including but not limited to triangular-shaped with striations near the X-point \citep[][]{Ng2011}, arcs \citep[][]{Bessho2016}, flat-top \citep[][]{Asano2008,Egedal2015}, crescent and U-shaped distribution functions \citep[][]{Bessho2017}. They are attributed to the meandering motion of electrons near the X-point or other processes like pitch angle scattering or magnetic gradients in the outflow region.
Here we focus only on those features that are a necessary condition for the microscopic plasma instabilities that can cause radio emission, e.g., Type III solar radio bursts (SRBs), and thus allow remote diagnostics of magnetic reconnection.
The Type III SRBs are thought to be caused by electron beams accelerated in flare magnetic reconnection. The energetic electron beams then propagate outwards along the open magnetic field, e.g., into the solar wind, and emit radio emission along their way \citep[][]{Melrose1980,Melrose2017,Ni2020,Aschwanden2002}.
Non-thermal EVDFs in the solar corona are not directly measurable but their consequences in the form of electromagnetic waves are, for example, by means of ground-based radio-telescopes \citep[][]{Yan2015,Chen2018}.

The relevant feature of EVDFs that can generate the micro-instabilities leading to formation of radio emission is related to their positive velocity gradients. In more precise terms, the so-called Penrose criterion \citep[][]{Penrose1960} establishes a necessary and sufficient condition for electrostatic instabilities. It is based on an integral of the derivative of the velocity distribution function in the (parallel) velocity space.
It represents a weighted average of the positive gradients of the distribution functions, which contribute to the instability criterion, compared to the rest of the distribution function.
For an electron beam enough separated from the main electron distribution function, it is equivalent to say that a positive gradient makes the distribution unstable to electrostatic waves, in particular Langmuir waves.
The physics behind this criterion is just attributed to the inverse Landau damping.
There are similar but much less known sufficient and necessary conditions for transverse waves
and instabilities propagating along the magnetic field due to positive velocity gradients in the parallel direction as well \citep[see, e.g.,][]{Sestero1971}.
For purely electromagnetic instabilities due to positive gradients in the perpendicular velocity direction, and in particular for non-gyrotropic distribution functions, there seems to be a lack of a general theorem equivalent to the Penrose criterion to our knowledge.
But for the specific case of the electron cyclotron maser instability (explained in detail below), the positive gradient(s) in the perpendicular velocity direction is a necessary condition for the instability, representing the population inversion of the maser mechanism.
Note that positive velocity gradients in the electron distribution functions will always generate unstable waves because of the inverse Landau damping. But their growth rate may be too small compared to the frequency of the unstable waves if those velocity gradients are relatively weak.
Such unstable waves, and so the instability itself, would be negligible in comparison with the normal plasma modes or the surrounding thermal noise.
So in the remainder of this paper we search for positive velocity gradients in the electron distribution functions as a necessary condition for the instabilities discussed in the following.

In the field-aligned direction, EVDFs with parallel velocity space gradients, i.e., $v_{\parallel}\cdot\partial f/\partial v_{\parallel}>0$, can cause streaming-like instabilities, which in turn usually generate unstable electrostatic Langmuir waves at the local plasma frequency and its harmonics, i.e., $\omega=n\omega_{pe}\ (n=1,2,...)$ (see Refs.~\cite{Nicholson1978,Melrose1985,Melrose1987,Melrose2017,Yao2021} and references therein). 
Therefore the typical frequencies of this emission mechanism depend on the local plasma density.
As proposed by the widely accepted plasma emission mechanism, these beam-generated Langmuir waves can further interact with ion-sound waves to produce electromagnetic emission via a multistage nonlinear wave-wave process.
These electromagnetic waves can eventually escape from the ambient plasma and be remotely observed, provided their frequencies are above the (cutoff) local plasma frequency $\omega_{pe}$ and their phase speed is equal to or greater than the speed of light $\omega/k\geq c$ (see Refs.~\cite{Ginzburg1958,Melrose1970,Melrose1970a,Reid2014, Melrose2017,Henri2019,Yi2007,Rhee2009} and references therein).
The plasma emission mechanism due to EVDFs with parallel velocity gradients has been investigated by kinetic simulations (see Refs.~\cite{Ganse2012,Thurgood2015} and references therein). 
This kind of EVDFs leading to streaming instabilities are often found in simulations of magnetic reconnection, their formation is mainly due to the parallel reconnection electric field \citep{Drake2003,Egedal2016,Dahlin2016}.
However, the properties of the emitted waves due to those distribution functions with positive sources of free energy have been comparatively much less studied, with a strong focus on the streaming instabilities in the separatrix region of reconnection \citep{Goldman2008,Goldman2014, Divin2012,Fujimoto2014a,Hesse2018}.

In contrast, EVDFs with a positive velocity gradient in the direction perpendicular to the local magnetic field, i.e., $\partial f/\partial v_{\perp}>0$, can cause the so-called electron cyclotron maser instabilities (ECMIs, see Refs.~\citep[][]{Chu2004a,Hewitt1982,Winglee1986a,Treumann2017} and references therein). The typical frequency of electromagnetic waves caused by ECMIs lies at the electron cyclotron frequency and its harmonics, i.e., $\omega=n\Omega_{ce}\ (n=1,2,...)$. The conventional electron cyclotron maser emission mechanism operates when the electron cyclotron frequency is larger than the plasma frequency, i.e., $\Omega_{ce}\ge\omega_{pe}$. 
This condition occurs in region with relatively low density and strong magnetic field, for example,  density cavities in magnetic reconnection (see Refs.~\cite{Melrose1984,Treumann2006} and references therein). Those generated electromagnetic waves can then directly emit out as X polarized waves from the ambient plasma.
\citet{Treumann2017} pointed out that electromagnetic waves can escape from plasmas with the opposite frequency condition $\Omega_{ce}<\omega_{pe}$, if the harmonics of $\Omega_{ce}$ are excited strongly enough.
In such a case, the remotely observed radio emission is possibly not at the fundamental frequency $\Omega_{ce}$ but at its harmonics.

Analytical studies and kinetic PIC simulations of the ECMI are often relied on the so-called ring EVDF, which resembles a ring in the 2D velocity space perpendicular to the local magnetic field and satisfies $\partial f/\partial v_{\perp}>0$ \citep{Pritchett1984,Melrose1986,Moseev2019,Lee2011,Zhou2020,Yao2021}. Although a very idealized model of EVDF, ring distributions of electrons have been obtained in kinetic magnetic reconnection simulations \citep{Shuster2014,Egedal2016a}.
However, what it is practically more often observed is partial-ring EVDFs in the 2D velocity space perpendicular to the local magnetic field, they are named as perpendicular crescent-shaped EVDFs or electron crescents. Another type of crescent along the parallel velocity direction is often discussed as well \citep[see a comparison of parallel and perpendicular crescents in][]{Burch2016}.
Perpendicular crescent-shaped EVDFs have been observed by the MMS mission in the Earth's magnetopause where magnetic reconnection takes place in asymmetric configurations \citep{Burch2016,Phan2016,Chen2016c,Genestreti2018,Norgren2016} and in the Earth's magnetotail where magnetic reconnection is symmetric \citep[][]{Yu2019}. Kinetic PIC simulations of asymmetric reconnection with conditions similar to those in magnetopause reconnection have generated the perpendicular crescent-shaped EVDFs \citep[][]{Hesse2014,Bessho2016,Bessho2017,Bessho2019,Shay2016,Chen2016,Egedal2016a,Zenitani2017}.
 
There are different proposed formation mechanisms behind those perpendicular crescent-shaped EVDF in magnetic reconnection.
In general, those EVDFs are associated with finite gyroradius effect, which preferentially occurs in the neighborhood of the X-point of reconnection where the magnetic field strength is weak and particles perform meandering motion, or near the separatrices where there are steep gradients \citep[][]{Norgren2016}. Analytical and numerical analysis show meandering electrons have indeed led to crescent-shaped EVDFs \citep{Hesse2014,Zenitani2017} and even earlier to similar crescent-shaped velocity distribution of ions \citep{Buchner1996,Lee2004,Usami2017}.
In particular, some mechanisms attribute the formation of those EVDF to the $\boldsymbol{E}\times\boldsymbol{B}$ drift caused by the electric field associated to asymmetric reconnection  \citep[][]{Shay2016,Bessho2016}, while others have claimed that crescent-shaped EVDFs do not depend on those asymmetric electric fields but rather only on magnetic field gradients \citep{Lapenta2017d}.

Numerical studies of electron bunches crossing tangential discontinuities have indeed revealed formation of crescent-shaped EVDFs due to gradient-$B$ drift \citep{Voitcu2018a}. It is therefore expected that under strong guide magnetic fields, where magnetic field strength gradients are weaker, crescent-shaped EVDFs would tend to be suppressed.
\citet{Bessho2019} investigated the guide field effects on the EVDFs formed in kinetic PIC simulations of asymmetric reconnection. They identified two effects associated with a weak guide field on the crescent-shaped EVDFs: a widening of the opening angle and a cutoff in a reduced EVDF along a direction that is oblique to the magnetic field.
\citet{Egedal2016a} found that the observed crescent-shaped EVDFs in asymmetric reconnection can be accounted for by an extension of their trapping model \citep{Egedal2008}, basically considering an additional population of electrons coming from the magnetosheath side of the current sheet. The physical mechanism is due to an energy cutoff in the distribution functions that can be traced back to gradients and diamagnetic drifts on scales of the order of an electron gyroradius.

Note that most of the aforementioned studies are based on either a test particle approach or 2.5D PIC simulations of kinetic magnetic reconnection.
But the features of 3D magnetic reconnection can be very different to its 2D counterpart. The additional degree of freedom (let us say, z-direction) could allow for a more efficient release of the free energy. Indeed, it is expected that EVDFs with positive gradients in the out-of-plane direction (i.e., along $v_z$ direction) will release their energy via streaming instabilities via unstable waves with a $k_z$ vector. This process is not possible in a 2D configuration. This is particularly relevant under the influence of a (relatively strong) guide field, because the reconnection electric field will accelerate electrons along the z-direction, which direction will be mostly aligned to the total magnetic field pointing also very near the $z-$direction (at least near the reconnection midplane). So the evolution of those electron distribution functions will be clearly different whether the z-direction is allowed (3D) or not (2D).
3D magnetic reconnection kinetic simulations can become more turbulent as a result of field-aligned streaming instabilities \citep{Che2011,Munoz2018}.
But a systematic characterization of the EVDFs in 3D kinetic magnetic reconnection is still lacking. And more so the identification of the EVDFs features relevant to the instabilities that can cause radio emission: i.e., velocity space gradients in EVDFs along the parallel or perpendicular directions to the local magnetic field.
Only a very limited number of 3D kinetic magnetic reconnection simulations have shown evidence of, e.g, crescent-shaped EVDFs \citep{Price2016,Le2017}.
Moreover guide field effects on all those processes remain also unexplored, a parameter that plays a critical role in solar coronal conditions where those instabilities presumably develop.

In addition to all the previous physical considerations, the systematic analysis of the availability of sources of free energy offered by EVDFs generatd by magnetic reconnection is a challenging problem.
Only very recently the first attempt for such a diagnostic has been carried out \citep{Dupuis2020}.
They implemented an unsupervised machine learning technique to the EVDFs formed by 2D magnetic reconnection simulations in different guide field strengths, non-Maxwellian features of EVDFs can be automatically detected. 
By using this method they identified the reconnection region providing an additional signature beyond more traditional criteria like those based on the non-gyrotropy of the electron pressure tensor \citep{Aunai2013} or the dissipation measure in the electron frame of reference \citep{Zenitani2011}.
What is more interesting of this machine learning technique is its capability to automatically identify non-Maxwellian features such as formation of non-thermal electron beams and their possibility to offer sources of free energy causing micro-instabilities, their temperatures and associated anisotropies.

In order to fill the gaps from previous studies, we thus aim at a systematic characterization of the possible sources of free energy from the EVDFs due to 3D kinetic magnetic reconnection.
We restrict ourselves in particular to positive velocity gradients (along the parallel or parallel or perpendicular direction) in those EVDFs since they are necessary conditions for the micro-instabilities eventually leading to radio emission.
Note that this does not necessarily imply that the reconnection sites are unstable and emit radio waves. Rather that radio emission is due to the electron beams generated by reconnection, as per the standard model of solar flares. The electron beams will eventually emit radiation further away from the reconnection region as they propagate in the solar corona and solar wind. A proper description of this radio emission would require a transport model for the electron propagation at AU scales. The objective of our paper is to characterize just the first step in this chain of events, the formation of electron beams.

For this purpose, we extend and apply a similar unsupervised machine learning algorithm already developed by \citep[][]{Dupuis2020} to determine the those features of the EVDFs in reconnection. We in particular assess the guide field effects on those sources of free energy.

The organization of this paper is as follows: in the \myrefsec{sec:simulation}, simulations setup, parameters and diagnostics are described. In the \myrefsec{sec:results}, the results of formation of non-thermal electron beams, possible sources of free energy and their dependence on the strength of guide field are discussed. In the \myrefsec{sec:conclusion}, we summarized the formation of EVDFs and the expected sources of free energy in reconnection plane in dependence on the guide field strength.

\section{Simulations and diagnostics}\label{sec:simulation}

\subsection{Simulation setup}

In order to analyze the EVDFs formed in 3D kinetic magnetic reconnection, we carried out numerical simulations using the 3D fully kinetic particle-in-cell (PIC) code ACRONYM \citep[][]{Kilian2012} (see also \url{http://plasma.nerd2nerd.org/papers.html}). We initialized our simulations with a Harris current sheet equilibrium \citep[][]{Harris1962} with different guide field strengths. Note that some parameters and initial conditions of the simulations are similar to those in Refs.~\cite{Munoz2016,Munoz2018}.

The simulations were initialized with two current sheets sufficiently separated to avoid their interaction at short timescales. The main reason to apply two current sheets is due to the implement of periodic boundary conditions in all three directions. 
Note that we only concentrate on one of the CSs for all investigations of EVDFs. 

The initial magnetic field of the two current sheets is expressed as follows:
\begin{align}
    \boldsymbol{B}(x)&=B_{\infty}\left[\tanh\left(\frac{x-L_x/4}{L}\right)-\tanh\left(\frac{x-3L_x/4}{L}\right)-1\right]\boldsymbol{e}_y+B_g\boldsymbol{e}_z
    \label{eq:cs_harris_B}
\end{align}
here $L=0.25d_i$ is the current sheet halfwidth with $d_i$ the ion skin depth. $B_{\infty}$ is the asymptotic reconnection magnetic field, $B_g$ is the strength of the guide field in the out-of-the-reconnection-plane direction (or simply the out-of-plane direction, here it is the $z$ direction) and $b_g=B_g/B_{\infty}$ is defined as the relative guide field. 
$L_x$ is the simulation box size along $x$ direction (while $L_y$ and $L_z$ are the sizes along $y$ and $z$ directions respectively).
We chose homogeneously distributed plasma and equal ion and electron temperatures, i.e., $T_i/T_e=1.0$. The ion-to-electron mass ratio is set to be $\mu=m_i/m_e=100$. Provided the Harris sheet equilibrium and the associated equilibrium between magnetic and thermal pressures, the asymptotic magnetic field is constrained by the electron thermal speed  $v_{the}=\sqrt{k_BT_e/m_e}$ (here  $v_{the}=0.1c$) in the following way:
\begin{align}
    B_{\infty} = \frac{m_e}{e}\omega_{pe}v_{the}\sqrt{2\left(1+\frac{T_i}{T_e}\right)}
\end{align}
here $\omega_{pe}=\sqrt{4\pi n_0e^2/m_e}$ is the electron plasma frequency calculated based on $n_0$: the central peak density of each species defined via the following density profile derived from the Harris sheet equilibrium for the two current sheets:
\begin{align}
    n(x)&=n_0\left[\cosh^{-2}\left(\frac{x-L_x/4}{L}\right)+\cosh^{-2}\left(\frac{x-3L_x/4}{L}\right)\right] +n_{bg}
    \label{eq:cs_harris_ne}
\end{align}
here $n_{bg}$ is the background population density added to avoid vacuum out of the current sheets. Note that the expression of particle number density \myrefeq{eq:cs_harris_ne} is valid for both ions and electrons due to the neutrality condition $n_i(x)=n_e(x)=n(x)$.

Another constraint of the Harris sheet equilibrium is a relation between the drift speeds of ions and electrons and the current sheet halfwidth $L$. Assuming equal ion and electron temperatures, this constrain implies that the drift velocities of the ions and electrons are equal in magnitude but of opposite signs in the out-of-plane direction, i.e., $\boldsymbol{U}_e =-\boldsymbol{U}_i=U_z\boldsymbol{e}_z$, with $U_z=-v_{th,e}d_i/(\sqrt{\mu}L)$.
The expression is valid for the left half of the domain in the $x$ direction (i.e., for the left current sheet), while the right half-domain (with the right current sheet) has opposite drift velocities for both species. In our simulation $U_z=0.04c$. This implies that the total out-of-plane current density is oppositely directed for either current sheet in the following way:
\begin{align}
    J_z(x)&=2en_0|\boldsymbol{U}_e|\left[\cosh^{-2}\left(\frac{x-L_x/4}{L}\right) -\cosh^{-2}\left(\frac{x-3L_x/4}{L}\right)\right]
    \label{eq:cs_harris_j}
\end{align}
the electron and ion current densities equally contribute to the total current density in each current sheet.

The systems are initialized with a perturbation in the magnetic field components $B_x$ and $B_y$ to accelerate the reconnection onset by seeding an O and X point at either current sheet. It can be derived from the following out-of-reconnection-plane vector potential:
\begin{align}
    \delta A_z=\delta P\cdot B_{\infty}\cdot\frac{L_y}{2\pi}&\sin^2\left(\frac{2\pi x}{L_x}\right) \sin\left(\frac{2\pi\left(y+L_y/4\right)}{L_y}\right)\cosh^{-2}\left(\frac{z-L_z/2}{\sigma_z}\right)
\end{align}
with an amplitude of $\delta P=0.07$ and $\sigma_z=0.25d_i$. In particular, the localization in the $z$ direction is to trigger reconnection at the middle of the simulation box (in contrast to trigger reconnection at all points along the $z$ direction).

All the simulations are calculated on a 3D mesh with $N_x\times N_y\times N_z=256\times 512\times 1024$ grid points. The 3D simulation box covers a physical domain $L_x\times L_y\times L_z=4d_i\times 8d_i\times 16d_i$. The grid cell size is $\Delta x=1.56\lambda_D$ with the Debye length $\lambda_D=\sqrt{k_BT_e/(4\pi n_0e^2)}$, and the time step is chosen to be $\Delta t=0.56\Delta x/c$ in order to fulfill the Courant-Friedrichs-Lewy (CFL) condition for light wave propagation. The number of particles per cell for the current carrying population of electrons and ions is 80 per species at the center of the current sheet (peak value of the Harris current sheet density) and 13 for the background species.

We carried out three simulations of magnetic reconnection with varying external magnetic fields (see ~\myreftab{tab:reconnections_bg}). The initial relative guide field $b_g$ is determined by the critical magnetic field $b_c$ \citep[][]{Daughton2005}. This quantity, which plays an important role on the linear tearing mode instability, represents the relative guide field that is necessary to fully magnetize the electrons with thermal speed $v_{the}$ in the central layer of the current sheet and is equal to $b_c=1/\sqrt{2}\cdot\left(\sqrt{2}r_i/L\right)^{1/2}\cdot\left(T_em_e/(T_im_i)\right)^{1/4}$, where $r_i=v_{thi}/\Omega_{ci}$ is the thermal ion gyroradius derived from $B_{\infty}$. In this study, the critical magnetic field value is $b_c=0.376$. 
The regime $b_g<b_c$ is denoted as the weak guide field limit in which electrons trajectories are dominantly affected by the Harris magnetic field $\boldsymbol{B}_{\infty}(x)$, while the opposite limit $b_g\gg b_c$ refers to the strong guide field regime, in which electrons are fully magnetized by the guide field. The transitional regime between the unmagnetized and fully magnetized electron trajectories is denoted in the following as the intermediate range of guide fields \citep[][]{Daughton2005}.

\begin{table}
    \caption{\label{tab:reconnections_bg}Parameters of 3D PIC code simulations of magnetic reconnection. Here $b_c\approx 0.376$ is the critical magnetic field, and $\kappa_{min}$ is the minimum of curvature parameter calculated by \myrefeq{eq:kappa_parameter} on the reconnection plane $z=8d_i$ for each simulation.}
    \begin{ruledtabular}
        \begin{tabular}{cccl}
            Run  &  $b_g$  & guide field  &  $\kappa_{min}$\\
            \hline
            1    &   $0$   & none         & $\sim 0.1$\\
            2   &    $b_c\approx 0.376$   & weak  & $\sim 1.0$\\
            3   & $2b_c\approx 0.752$   & intermediate & $\sim 3.0$\\
        \end{tabular}
    \end{ruledtabular}
\end{table}

\subsection{Diagnostics}

In this study, resulting EVDFs are evaluated in the following local velocity reference frame at each location in the reconnection plane \citep[][]{Goldman2016}:
\begin{equation}
    \left\{
    \begin{aligned}
        \boldsymbol{e}_{\parallel}&=\boldsymbol{b}\\
        \boldsymbol{e}_{\perp 1}&=\left(\boldsymbol{b}\times\boldsymbol{u}/|\boldsymbol{u}|\right)\times\boldsymbol{b}\\
        \boldsymbol{e}_{\perp 2}&=\boldsymbol{b}\times\boldsymbol{u}/|\boldsymbol{u}|
    \end{aligned}
    \right.
    \label{eq:local_frame}
\end{equation}
here $\boldsymbol{b}=\boldsymbol{B}/|\boldsymbol{B}|$ is the unit vector of the local magnetic field $\boldsymbol{B}$ and $\boldsymbol{u}$ is the local electron bulk flow velocity. These three basis unit vectors determine a local right-handed orthogonal velocity coordinate system. $\boldsymbol{e}_{\parallel}$ is the unit vector along (or parallel) to the local magnetic field, while $\boldsymbol{e}_{\perp 1}$ and $\boldsymbol{e}_{\perp 2}$ are other two unit vectors perpendicular to the local magnetic filed.

The EVDFs to be investigated are in essence the probability density functions, namely, they are estimated as the frequency of particle number per bin in the 2D or 1D velocity space and normalized by the total number of particles. As a result, the integral of the EVDF over the whole space where it is estimated will yield 1. For example, for the 2D EVDF calculated in $v_{\parallel}-v_{\perp}$ plane, we have $\displaystyle\iint f(v_{\parallel},v_{\perp}) dv_{\parallel}dv_{\perp}=1$, while for the 1D EVDF of $v_{\parallel}$, we get $\displaystyle\int f(v_{\parallel}) dv_{\parallel}=1$. Here $v_{\parallel}$ and $v_{\perp}=\sqrt{v_{\perp1}^2+v_{\perp2}^2}$ are the local parallel and perpendicular electron velocities in the local reference frame determined by \myrefeq{eq:local_frame}.

As discussed above, a possible source of free energy and necessary condition for micro-instabilities eventually leading to radio emission is the presence of positives gradients in the 1D EVDFs $f(v_{\parallel})$ and $f(v_{\perp})$ separately. However, different from investigating positive gradients in the field-aligned EVDF $f(v_{\parallel})$, we look for positive gradients in the perpendicular EVDF $f(v_{\perp})/2\pi v_{\perp}$ rather than in $f(v_{\perp})$. The reason is because the thermal electron population always contributes an additional positive gradient in $f(v_{\perp})$ besides the possible gradient(s) caused by non-thermal electrons. Positive gradients in  $f(v_{\perp})/2\pi v_{\perp}$, on the other hand, only are due to non-thermal electrons.

In this study, an unsupervised machine learning algorithm, namely, the Bayesian Gaussian mixture model (BGMM) \citep[][]{Bishop2006}, is extended to fit the 1D EVDFs and to identify the possible sources of free energy generated in simulations of magnetic reconnection.

We now briefly describe why we used this approach. The appropriate fitting distributions to EVDFs should take both physical and mathematical considerations into account. In low-energy regions, like those away from the current sheet, the plasma is Maxwellian (or bi-Maxwellian) distributed.
Some common deviations from this distribution, prone to happen in collisionless magnetic reconnection, are electron beams with eventually anisotropic temperatures, which are usually associated to more high-energy regions.
Their total EVDFs can then be represented as a superposition of Maxwellian EVDFs, at least to a first order approximation, although more complex features can also exist.
This simplified assumption could be justified based on the fact that large deviations from Maxwellian EVDFs in the form of distribution functions with positive gradients are not sustainable for a long time in collisionless plasmas.
This is because those gradients tend to be reduced due to the inverse Landau damping effect and consequently the deviations from a Maxwellian are reduced as well.

A distribution that combines both low-energy Maxwellian and high-energy non-Maxwellian features is the Kappa distribution, which has been extensively applied to collisionless space and astrophysical plasmas \citep[][]{Livadiotis2017,Livadiotis2018}.
Mathematically speaking, a single Kappa distribution population can be approximately fitted by the sum of a central Gaussian distribution function to fit the core and a sum of Gaussian distributions function with large widths to fit its wide tail. In this way, many EVDFs in space and astrophysical plasma can be fitted by a sum of Gaussian distributions \citep[][]{Dupuis2020}.

Based on this assumption, we firstly fit the velocity distribution function of electrons by a sum of Gaussians by the so-called Gaussian mixture model (GMM) \citep[][]{Bishop2006} as follows:
\begin{align}
    f(\boldsymbol{v}|\boldsymbol{\Phi})&=\sum_{k=1}^K A_k\mathcal{N}(\boldsymbol{v}|\boldsymbol{\mu}_k,\boldsymbol{\Sigma}_k)
    \label{eq:PDFGMM}
\end{align}
here $A_k$ corresponds to the weight of the $k$-th component of the Gaussian distribution, $\mathcal{N}$ indicates the multivariate Gaussian distribution parameterized by the mean vector $\boldsymbol{\mu}_k$ and the covariance matrix $\boldsymbol{\Sigma}_k$, the mixture parameter $\boldsymbol{\Phi}=\{A_1,A_2,...A_K,\boldsymbol{\mu}_1,\boldsymbol{\mu}_2,...,\boldsymbol{\mu}_K,\boldsymbol{\Sigma}_1,\boldsymbol{\Sigma}_2,...,\boldsymbol{\Sigma}_K\}$ includes all the parameters of the GMM. The quantity $|A_i|^2$ represents the intensity of the $k$-th component of the Gaussian mixture, and $\boldsymbol{\mu}_k$ and $\boldsymbol{\Sigma}_k$ indicate the bulk flow velocity and thermal speed, respectively.
The GMM model is ideal for our purposes because it allows to fit the most common distribution functions expected in reconnection and easily determine their properties such as mean bulk flow speed or temperature.

In order to compensate for the overfitting of GMM models, the second step is to assess the appropriate number of sub-populations based on the model selection technique named Bayesian information criterion (BIC). By introducing a penalty term associated with the number of parameters in the model, this model selection method can automatically determine the appropriate number of sub-populations \citep[][]{Bishop2006}.
See a detailed discussion about the BGMM algorithm applied to the  EVDFs in Appendix \ref{app:BGMM}.

The formation of EVDFs for different guide field regimes critically depend on the electron dynamics. During the non-linear process of magnetic reconnection, the electrons crossing the magnetic field reversal can be characterized by a curvature parameter in the following form \citep[][]{Buchner1989,Buchner1991a}:
\begin{align}
    \kappa=\min\sqrt{\frac{R_B}{\rho_{eff}}}
    \label{eq:kappa_parameter}
\end{align}
here $R_B=|\boldsymbol{b}\cdot\nabla \boldsymbol{b}|^{-1}$ is the curvature radius of a local magnetic field line, $\boldsymbol{b}=\boldsymbol{B}/|\boldsymbol{B}|$ is the unit vector of local magnetic field, $\rho_{eff}=\sqrt{Tr\left(\boldsymbol{P_e}\right)/3m_en_e}/\Omega_{ce}$ is the effective electron Larmor radius in the local magnetic field $\boldsymbol{B}$, $\Omega_{ce}$ is the local electron cyclotron frequency and $\boldsymbol{P_e}$ is electron pressure tensor. The $\kappa$ parameter dynamically changes during the course of evolving magnetic reconnection and it depends on the magnetic field geometry \citep[][]{Buchner1989,Buchner1991a,Munoz2016}: for $\kappa\le 1$, electrons are mostly non-adiabatic and form non-gyrotropic distributions like the meandering (or Speiser) motion across the central reconnection plane. 
For $\kappa\ge 2.5$, electrons are fully magnetized or adiabatic, and their distributions are mostly gyrotropic.
The transition regime between those two limits is in between  $1\le\kappa\le 2.5$.

\section{Results}\label{sec:results}

\subsection{Global evolution and selection of time snapshots for analysis}

\begin{figure*}
    \centering
    \begin{subfigure}{0.46\textwidth}
        \centering
       \includegraphics[height=0.8\textwidth,width=\linewidth]{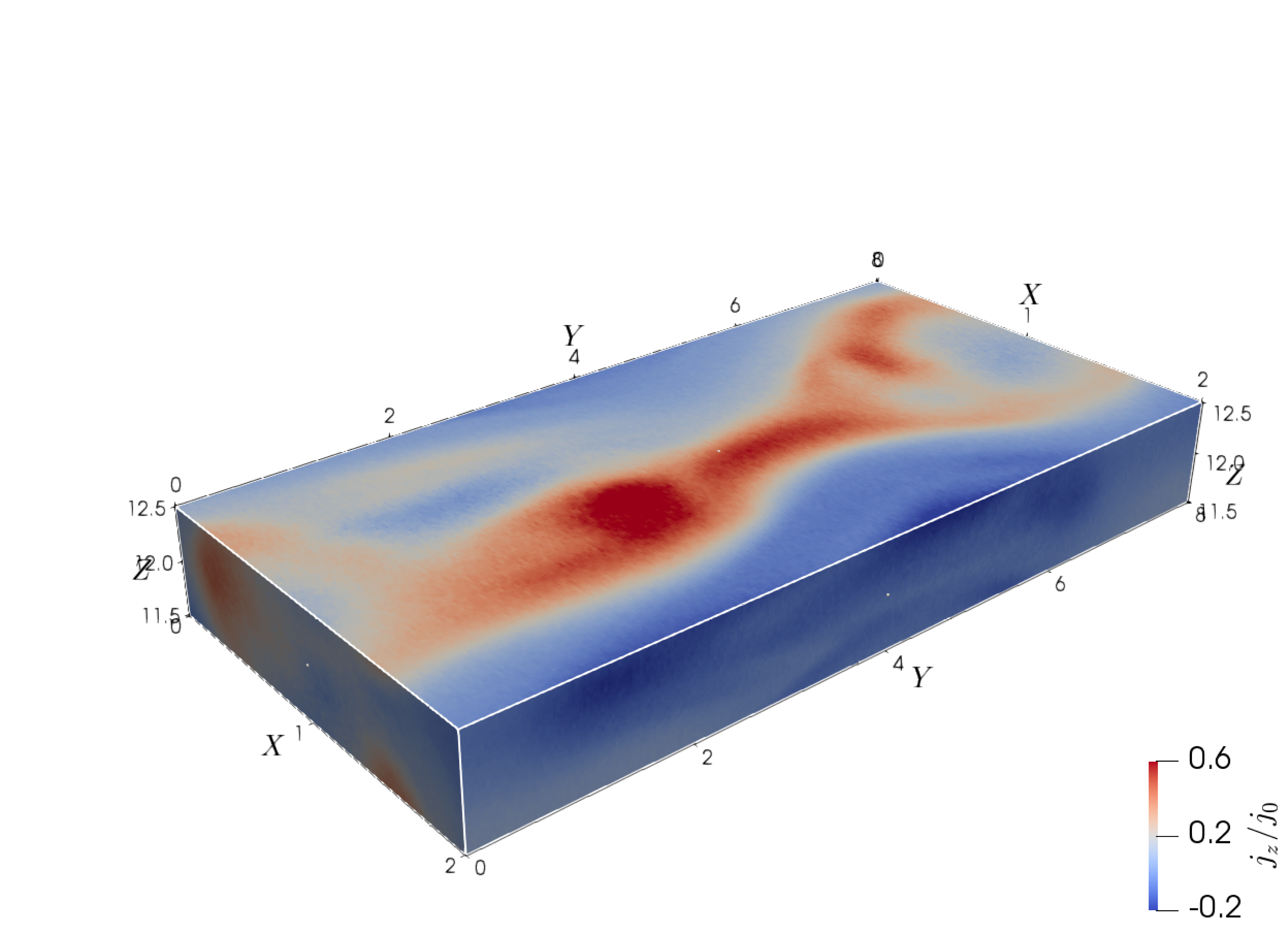}
        \caption[]{$j_z$ at $z=12\pm 0.5d_i$.}
    \end{subfigure}
    \centering
    \begin{subfigure}{0.46\textwidth}
        \centering
       \includegraphics[height=0.8\textwidth,width=\linewidth]{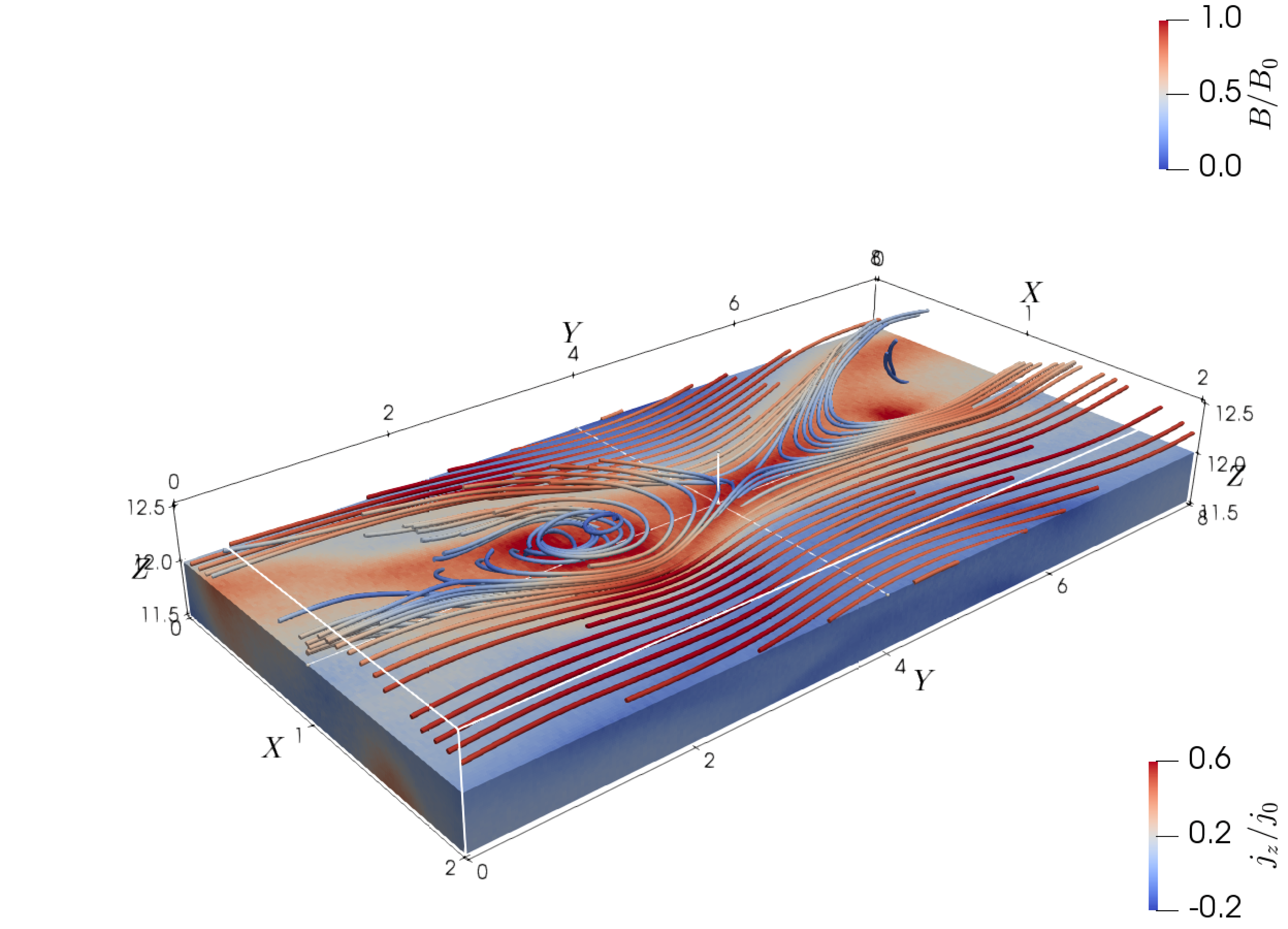}
        \caption[]{Magnetic field lines within $z=12\pm 0.5d_i$.}
    \end{subfigure}
    \centering
    \begin{subfigure}{0.46\textwidth}
        \centering
       \includegraphics[height=0.8\textwidth,width=\linewidth]{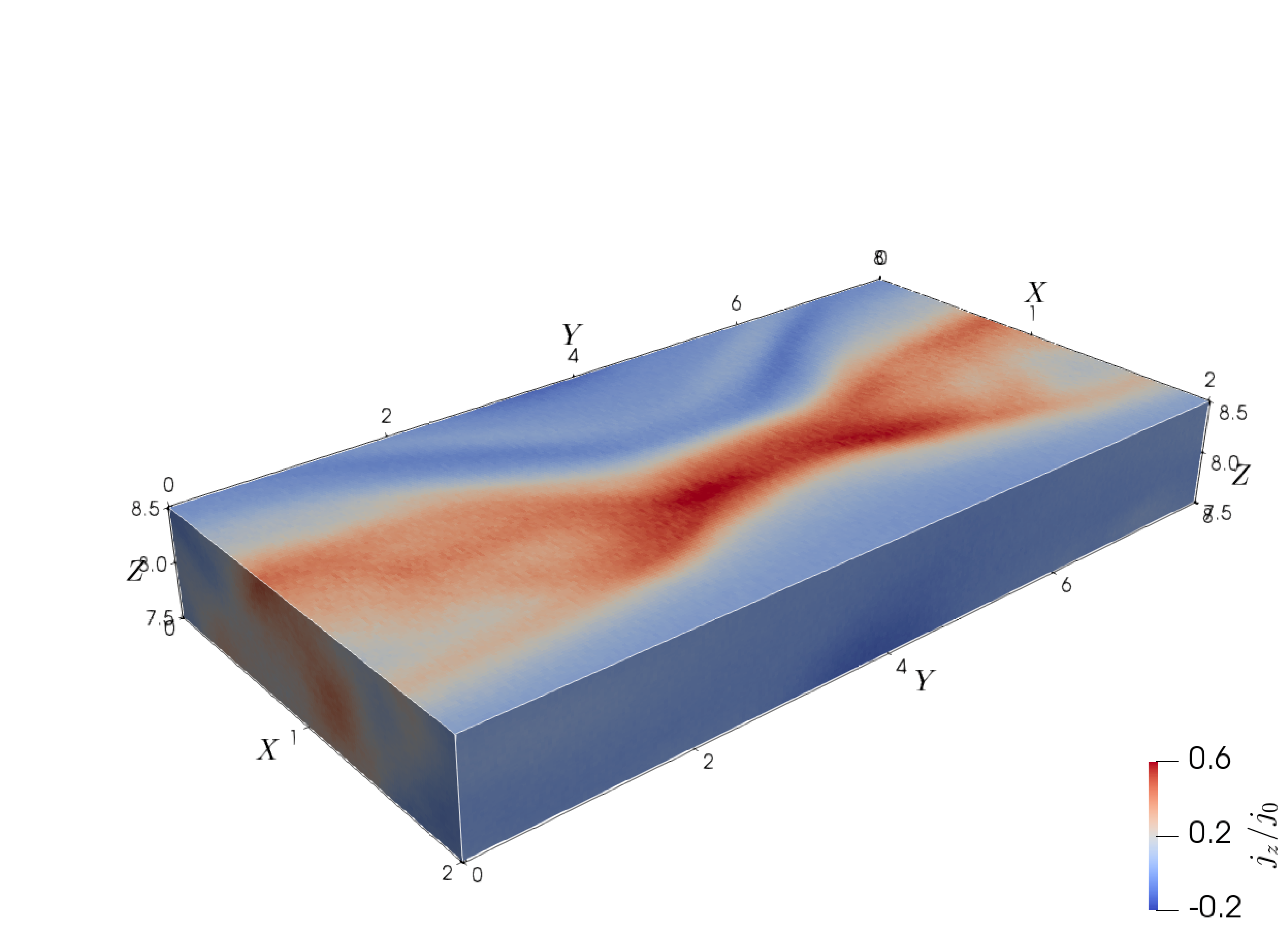}
        \caption[]{$j_z$ at $z=8\pm 0.5d_i$.}
    \end{subfigure}
    \centering
    \begin{subfigure}{0.46\textwidth}
        \centering
       \includegraphics[height=0.8\textwidth,width=\linewidth]{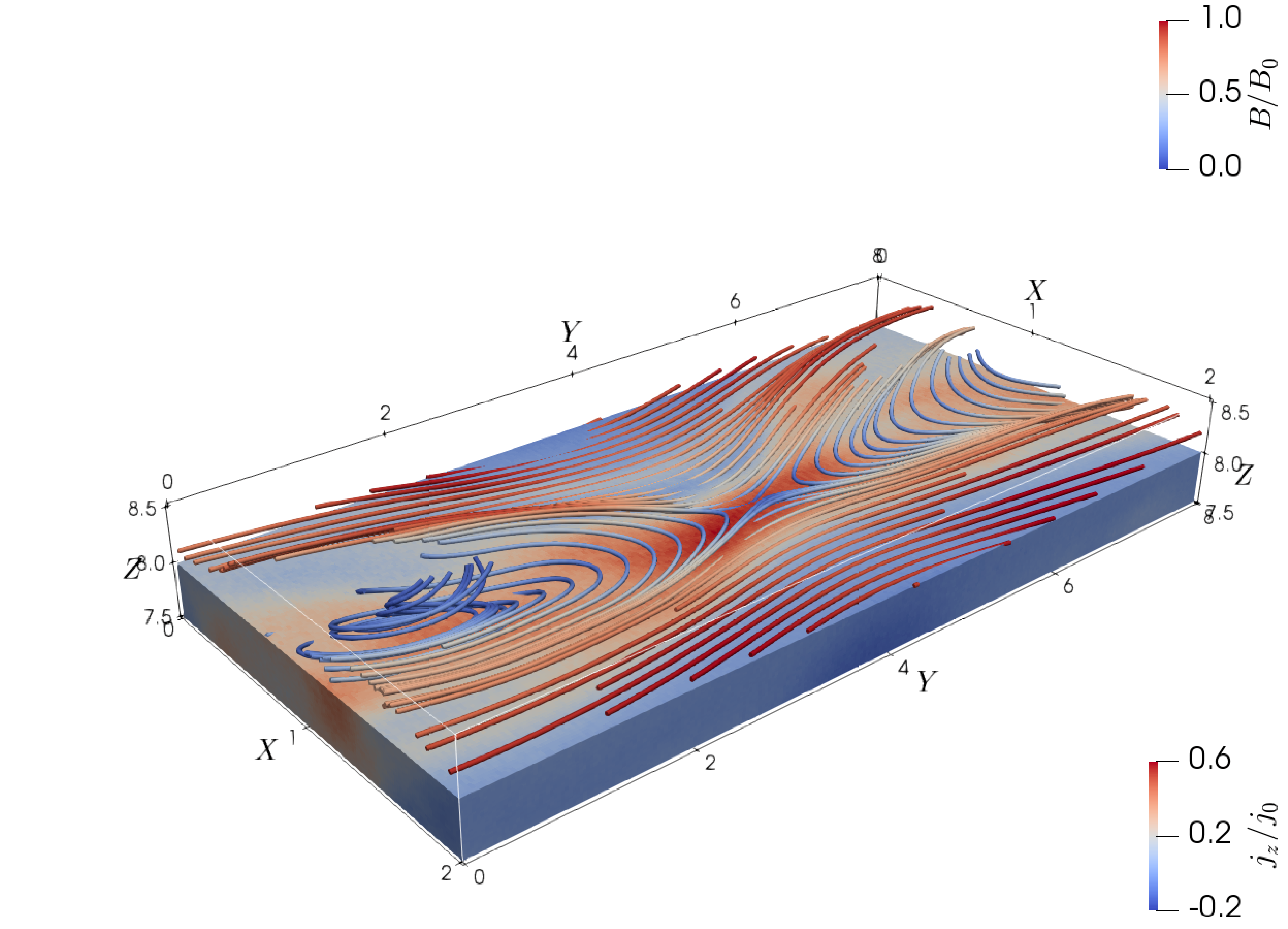}
        \caption[]{Magnetic field lines within $z=8\pm 0.5d_i$.}
    \end{subfigure}
    \centering
    \begin{subfigure}{0.46\textwidth}
        \centering
       \includegraphics[height=0.8\textwidth,width=\linewidth]{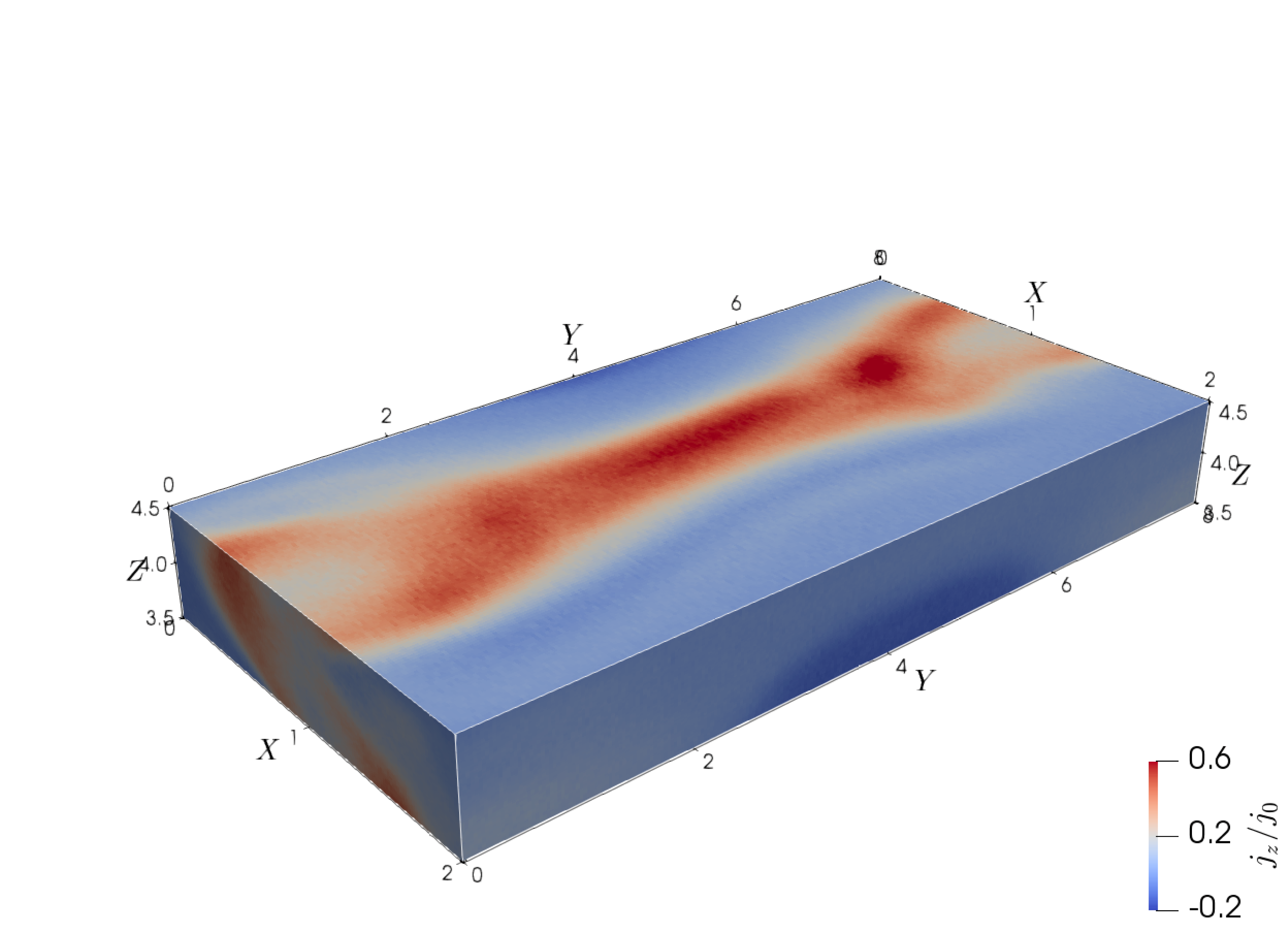}
        \caption[]{$j_z$ at $z=4\pm 0.5d_i$.}
    \end{subfigure}
    \centering
    \begin{subfigure}{0.46\textwidth}
        \centering
       \includegraphics[height=0.8\textwidth,width=\linewidth]{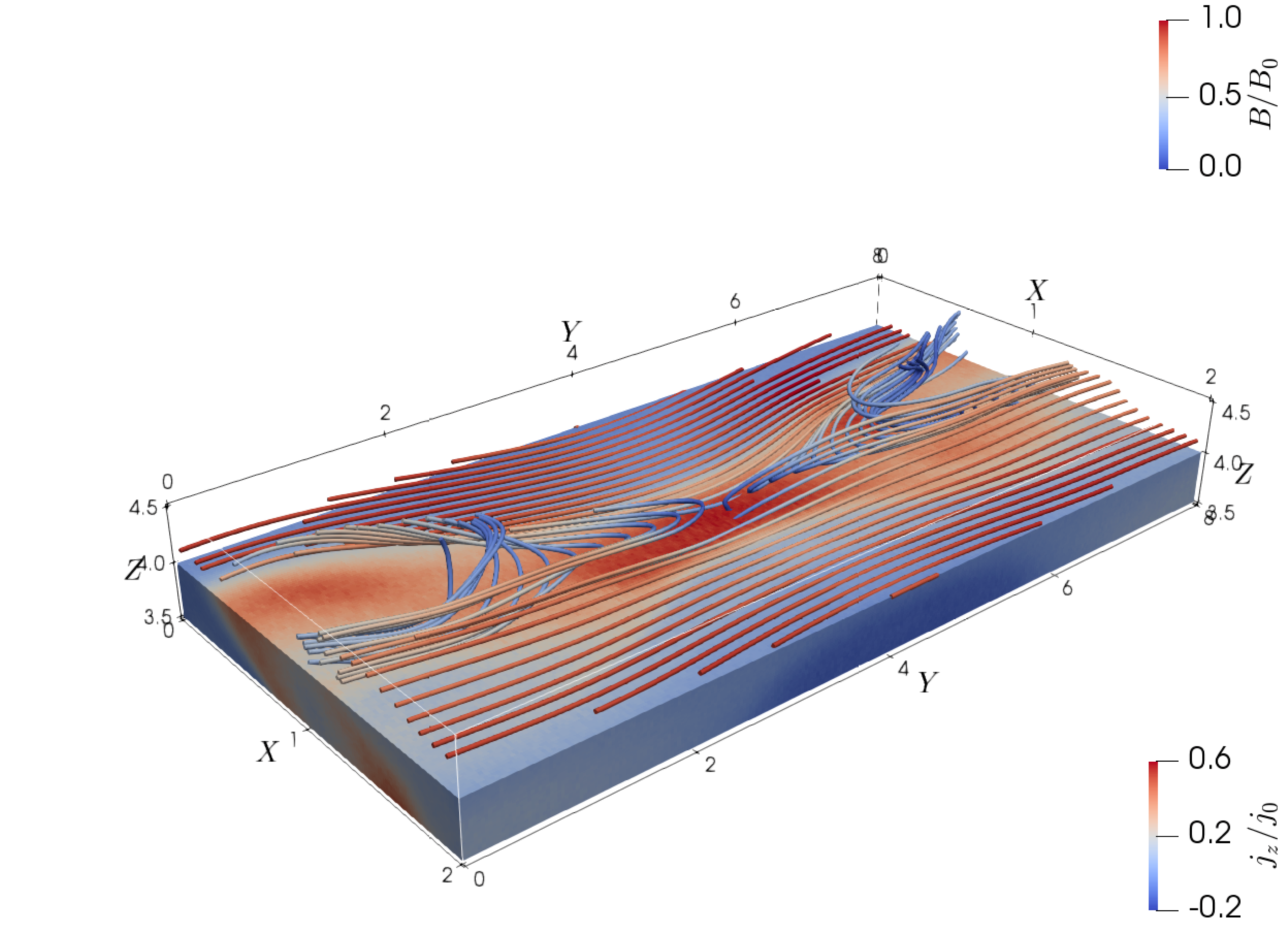}
        \caption[]{Magnetic field lines within $z=4\pm 0.5d_i$.}
    \end{subfigure}
    \caption[]{The out-of-plane current density $j_z$ (left column: a, c, e) and corresponding magnetic field lines (right column: b, d, f) within current sheet at $z=12\pm0.5d_i,\ 8\pm0.5d_i,\ 4\pm0.5d_i$ (from top to bottom), respectively, at $t=5.25\ \Omega_{ci}^{-1}$ for Run1. The current density $j_z$ is normalized by $j_0=en_0v_{the}$, each magnetic field line is multicolored by the normalized strength of magnetic filed $B/B_0$ (with $B=|\boldsymbol{B}|$) where it runs. The unit of the three axes is the ion skin depth $d_i$.}
    \label{fig:jz_3d}
\end{figure*}

As mentioned above, we focused on one of the current sheets in each simulation in this study. \myreffig{fig:jz_3d} (a,c,e) show the 3D spatial configuration of the out-of-plane current density $j_z$ within the ranges $z=12\pm 0.5d_i,\ 8\pm 0.5d_i,\ 4\pm 0.5d_i$, respectively, at $t=5.25\ \Omega_{ci}^{-1}$ for Run1 (the antiparallel magnetic reconnection case with $b_g=0$).
They roughly have a similar structure on each reconnection plane, namely, the X-point varies around $x=d_i,\ y=4d_i$ to some degree and the O-point fluctuate around $x=d_i,\ y=0\sim 2d_i$ and $x=d_i,\ y=6\sim 8d_i$, respectively.
\myreffig{fig:jz_3d}(b,d,f) show the corresponding 3D spatial structure of magnetic field lines near the reconnection plane at $z=12d_i,\ 8d_i,\ 4d_i$ respectively.
The magnetic field lines around the O points show a spiral-like form. At $z=8d_i$, the magnetic field lines at $x=d_i,\ y=8d_i$ are roughly onto the reconnection plane.
While in the diffusion and separatrices regions, the magnetic field lines at different $z$-heights roughly run within the $x-y$ plane, with little variation on the $z-$direction. This implies that on each reconnection plane, the topology of magnetic field lines in most regions especially in the inflow and separatrix regions behaves like the typical 2D reconnection structure.
There are variations in the topology of magnetic field lines along the $z$ direction, but they can still be roughly approximated by a 2D
reconnection structure. Those variations can be seen in \myreffig{fig:jz-xz} showing the structure of the current density component $j_z$ in the out-of-the-reconnection plane direction, which modulates the variations of the magnetic field at the calculated plane (near the X-point).
Similar results are observed in other simulations.
In the following, EVDFs are estimated on each reconnection plane at different $z$-heights. At each location (point) on a reconnection plane, EVDFs are estimated by velocities of electrons within a spherical region of radius $\le0.1d_i$ centered at this point in the local velocity reference frame determined by \myrefeq{eq:local_frame}.

\begin{figure*}
    \centering
    \begin{subfigure}{0.39\textwidth}
        \centering
       \includegraphics[height=0.8\textwidth,width=\linewidth]{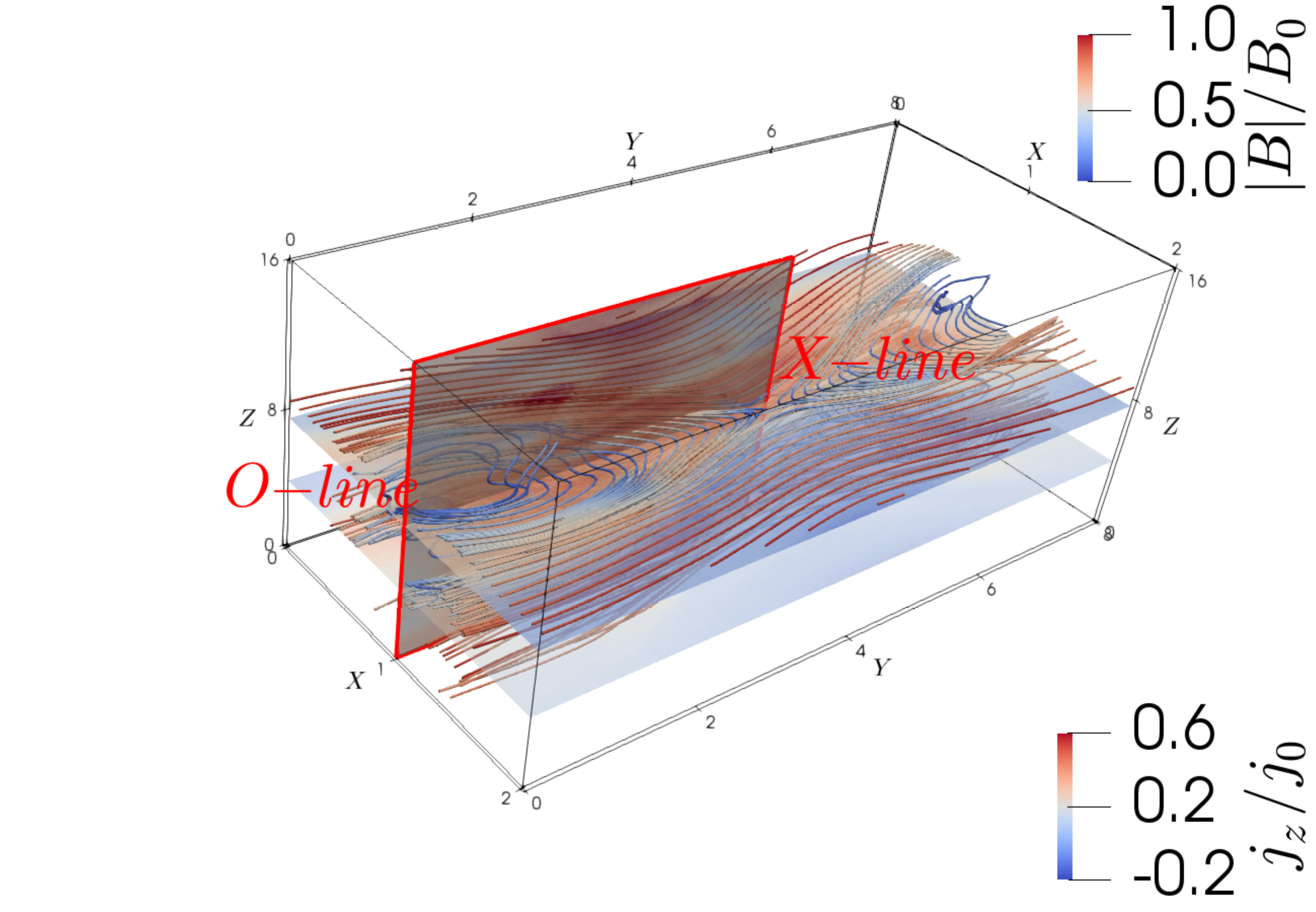}
        \caption[]{}
    \end{subfigure}
    \centering
    \begin{subfigure}{0.59\textwidth}
        \centering
       \includegraphics[height=1.0\textwidth,width=\linewidth]{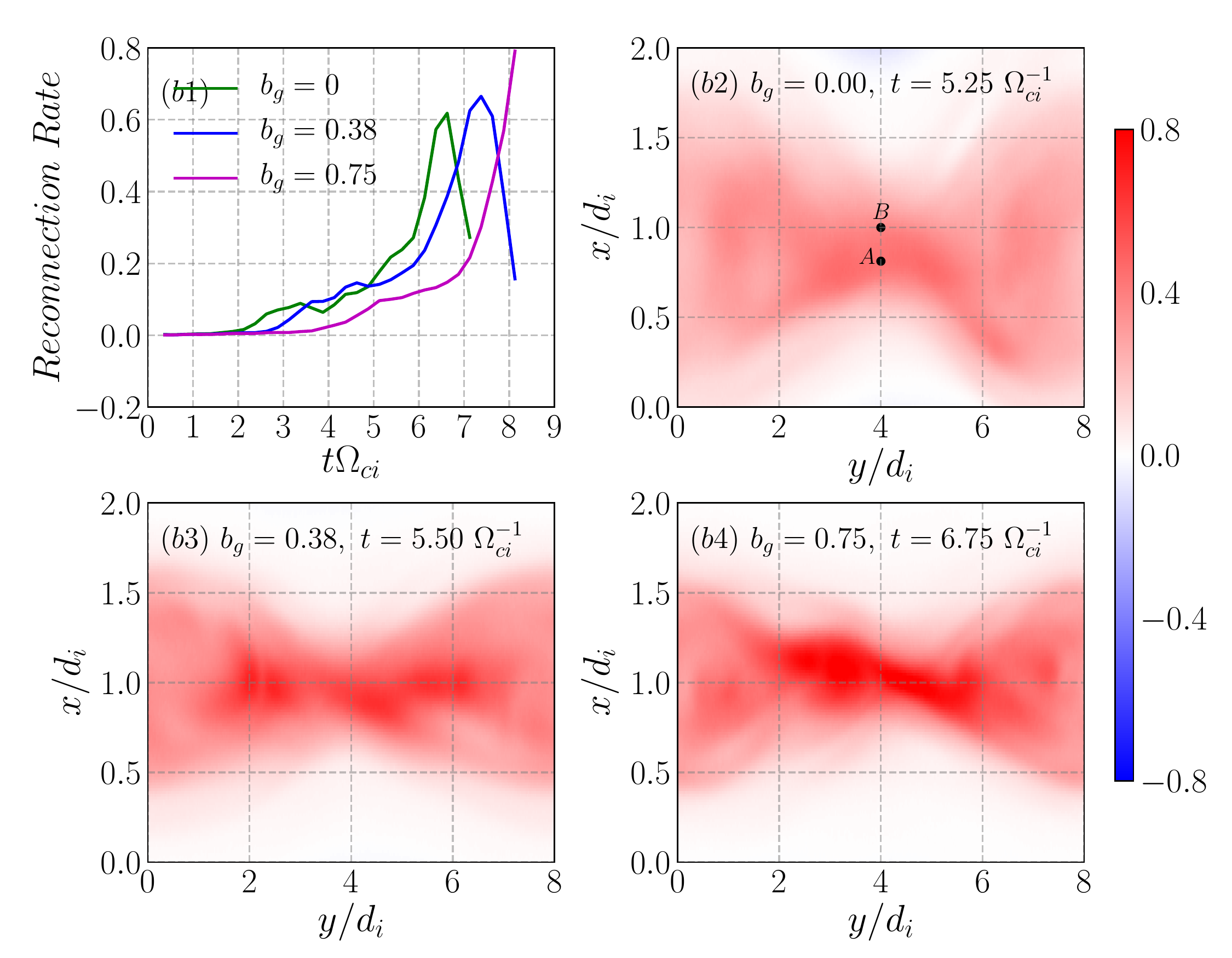}
        \caption[]{}
    \end{subfigure}
    \caption[]{(a) Illustration of the rectangular plane (delimited by the red box) between X-line,  O-line and boundaries of the simulation box where the magnetic flux across. The reconnection rate is calculated by the time derivative of the reconnected magnetic flux passing through this plane $\Psi=\int \vec{B}\cdot d\vec{S}$. (b1) Reconnection rate  $d\Psi/dt$ (solid curves). (b2-4) The out-of-plane current density $j_z$ at $z=8\ d_i$ for Run1 at $t=5.25\ \Omega_{ci}^{-1}$, Run2 at $t=5.5\ \Omega_{ci}^{-1}$ and Run3 at $t=6.75\ \Omega_{ci}^{-1}$ displayed on the $x-y$ reconnection plane, respectively. The reconnection rate is normalized by $B_{\infty}v_A/c$, with $v_A$ the Alfv\'en speed, and the current density $j_z$ is normalized by $j_0=en_0v_{the}$.}
    \label{fig:reconnection_rate}
\end{figure*}

\myreffig{fig:reconnection_rate} (b1) shows the reconnection rates of our three PIC magnetic reconnection simulations in various guide fields.
This quantity characterizes the reconnection efficiency and is numerically calculated by the time rate of the reconnected magnetic flux passing through the rectangular plane (see the rectangular plane delimited by the red box in \myreffig{fig:reconnection_rate}(a)) between the X-line,  O-line and boundaries of the simulation box, namely, $d\Psi/dt$ (see solid curves in \myreffig{fig:reconnection_rate}(b1)). Note that for this calculation we assumed a standard 2D reconnection geometry for the location of the X and O points. This ignores the variation in the out-of-plane direction and the more complicated structure of the magnetic field due to the wavy structure of the current sheet in the out-of-plane direction (see \myreffig{fig:jz-xz}). But in average (along the z-direction) the X and O points are approximately located at the points required by our calculations of the reconnection rate as explained above.

Due to the initial perturbation and reconnected magnetic flux available in the simulation box, reconnection saturates at $t=6.5\Omega_{ci}^{-1}$ for Run1, $t=7.5\Omega_{ci}^{-1}$ for Run2 and after $t=8.2\Omega_{ci}^{-1}$ for Run3 respectively.
Those times are relatively short in units of the inverse cyclotron times in comparison with other simulations.
This mainly has to do with the relatively small simulation box size along the $x$ direction and the consequent small available magnetic flux to be converted by magnetic reconnection. The larger the simulation box the larger the available magnetic flux which implies that magnetic reconnection can be sustained for longer times.
A second reason is the relatively thin initial current sheet. That enhances the growth rates of the tearing mode instability triggering magnetic reconnection. As a result simulations of magnetic reoconection will take less time to evolve and to reach saturation in comparison with equivalent simulations using an initially thicker current sheet.

The maxima of reconnection rates are larger than $0.6\ B_{\infty}v_A/c$, with $v_A=B_{\infty}/\sqrt{4\pi n_im_i}$ the Alfv\'en speed. The results show that a stronger guide field delays the onset of reconnection. This effect has already been discussed extensively for a wide variety of magnetic reconnection condition, being attributed to the Hall effect of magnetized electrons interacting with the ions \citep{Horiuchi1997,Ricci2004}.
Note that the obtained peak reconnection values are larger than the typical reconnection values, i.e., 0.1 in normalized units of $B_{\infty}v_A/c$ (in CGS units). Those values are commonly observed \citep[][]{Cassak2017,Liu2017}. Our larger values can be attributed to numerical effects due to the reduced size of the simulations box and the interaction of the second current sheet (not shown in those plots).

In this study, we concentrated on EVDFs generated by reconnection near the time when reconnection rates reach values about and slightly above $\sim0.1$ in normalized units. For example, at time $t=5.0-5.5\Omega_{ci}^{-1}$ for Run1, $t=5.5-6.0\Omega_{ci}^{-1}$ for Run2 and $t=6.75-7.25\Omega_{ci}^{-1}$ for Run3.
The out-of-reconnection-plane current density $j_z$ at $z=8\ d_i$ of Run1 at $t=5.25\ \Omega_{ci}^{-1}$, Run2 at $t=5.5\ \Omega_{ci}^{-1}$ and Run3 at $t=6.75\ \Omega_{ci}^{-1}$ in the $x-y$ plane are shown in \myreffig{fig:reconnection_rate} (b2, b3, b4), respectively.

\subsection{Non-thermal EVDFs: Harris and background electron populations}\label{sec:harris_background} 

As discussed in \myrefsec{sec:simulation}, each simulation is initialized with two electron populations, one establishes the Harris current sheet equilibrium and the other one with a constant density background (see \myrefeq{eq:cs_harris_ne}). We find that the Harris electrons contribute more to the formation of non-thermal electron beams, while background electrons are mainly Maxwellian distributed at each point on the reconnection plane.

We should note the relative contribution of the Harris population to the total electron distribution function depends on time scales of reconnection, which are reduced in the case of small domains sizes, like in our simulations. We would expect that at later times fresh plasma from the inflow region, belonging to the background population will have a more important contribution. But a quantitative assessment of this effect would require a larger parametric study involving not only different simulation sizes, but also eventually different types and amplitudes of the initial perturbation which also play a role in the onset of reconnection.
In addition, we should also mention that the boundary conditions also influence the contribution of different populations. In a simulation with periodic boundary conditions like ours, the Harris population is pushed to the outflow regions which eventually reenter the simulation domain, so that the Harris population can always be found inside of the current sheet. In a simulation with open boundary conditions in the outflow region, a large part of the Harris population will eventually escape, with a consequent larger proportion of background electrons coming from the inflow plasma.

\begin{figure}
    \centering
   \includegraphics[width=0.99\linewidth]{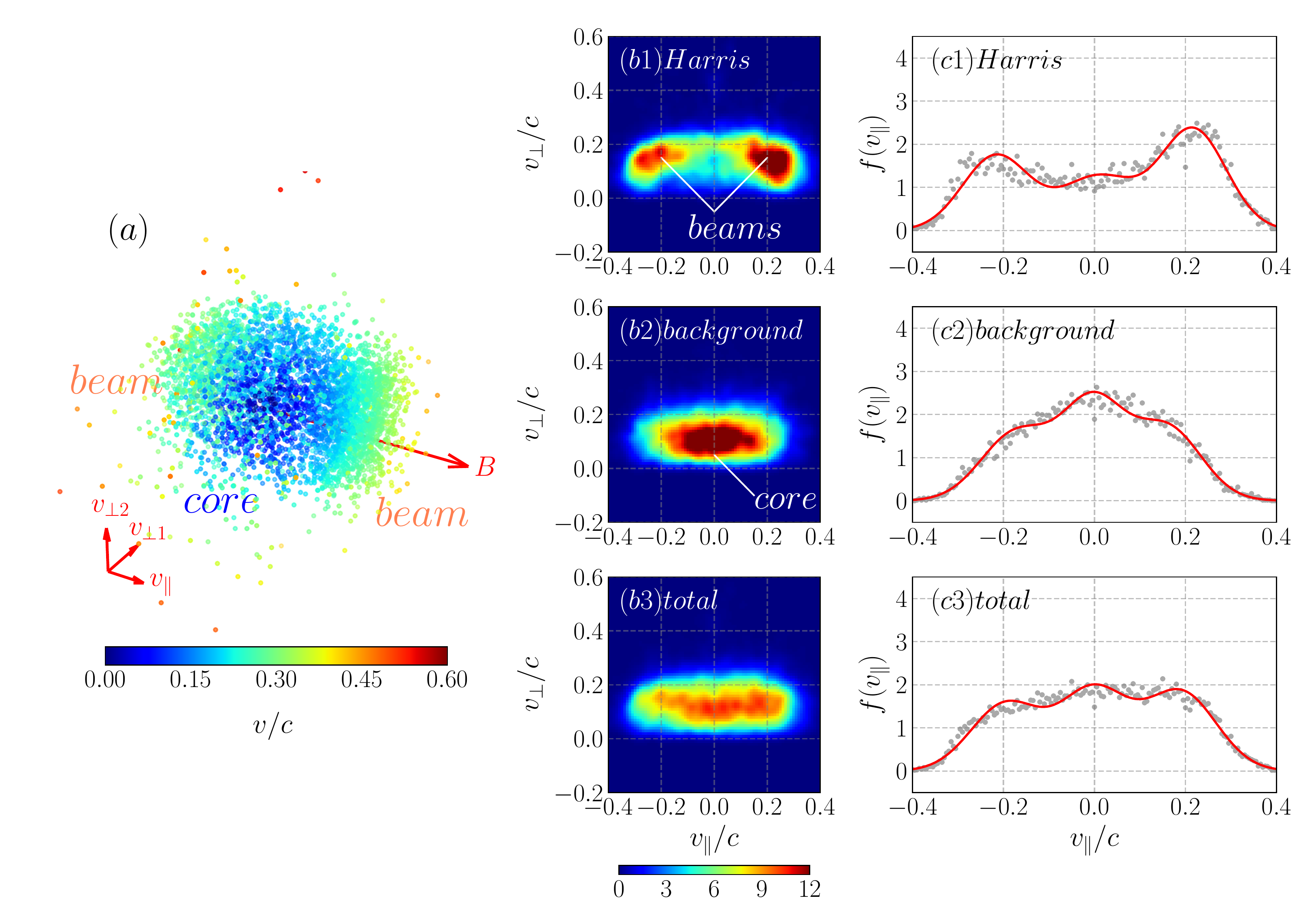}
    \caption[]{Velocity distribution of electrons within a spherical region of radius $\le0.1d_i$ centered at point A (see \myreffig{fig:reconnection_rate} (a2)) on the reconnection plane at $z=8d_i$ and time $t=5.25\Omega_{ci}^{-1}$ for Run1. (a) Distribution of total electrons in the 3D velocity space (scatter plot). Each dot indicates an electron and it is multicolored by the speed $v=\sqrt{v_{\parallel}^2+v_{\perp}^2}$ of the electron. (b1-b3) 2D EVDFs $f(v_{\parallel},v_{\perp})$ of Harris, background and total electrons in $v_{\parallel}-v_{\perp}$ plane respectively. (c1-c3) 1D EVDFs $f(v_{\parallel})$ (the red curves) of Harris, background and total electrons respectively. The gray dots show the frequency of particle number per bin in velocity space (for a bin width of $w=0.005c$).}
    \label{fig:beam_3d_vpara}
\end{figure}

\myreffig{fig:beam_3d_vpara} shows the distribution of electrons in the velocity space. The electrons are collected from a spherical region of radius $\le0.1d_i$ centered at point A (see \myreffig{fig:reconnection_rate} (a2)) in the diffusion region on the reconnection plane at $z=8d_i$ and $t=5.25\Omega_{ci}^{-1}$ for the antiparallel magnetic reconnection Run1.
\myreffig{fig:beam_3d_vpara}(a) shows the distribution (scatter plot) of the total electrons in the 3D velocity space. Each dot indicates an electron and it is multicolored by the speed $v=\sqrt{v_{\parallel}^2+v_{\perp}^2}$ of the electron. The total electrons can be separated into two parts: the thermal core (see dots with cold colors) and the non-thermal electrons (see dots with warm colors) which resemble beams along the field-aligned direction parallel to $\boldsymbol{B}$. 
Indeed, those beams can be seen in the 2D EVDF of the Harris electrons (see \myreffig{fig:beam_3d_vpara}(b1)) as two separate enhancements centered approximately at $v_{\parallel}=\pm 0.2c,\ v_{\perp}=0.15c$. The corresponding 1D parallel EVDF $f(v_{\parallel})$, which are obtained by integrating the 2D EVDFs along the perpendicular direction (see \myreffig{fig:beam_3d_vpara}(c1)), also shows locations of two separate beams at about $v_{\parallel}=\pm 0.2c$ in the parallel direction. 
The 2D EVDF of background electrons (see \myreffig{fig:beam_3d_vpara}(b2)) are mainly Maxwellian distributed around $v_{\parallel}=0,\ v_{\perp}=0$ in the $v_{\parallel}-v_{\perp}$ plane, and its 1D EVDF $f(v_{\parallel})$ (see \myreffig{fig:beam_3d_vpara}(c2)) shows a nearly bell-shaped distribution function around $v_{\parallel}=0$.
In this way, the 1D EVDF of total electrons (as a sum of both Harris and background electrons) shows three separated maxima with comparable values along the parallel direction, i.e., two from the Harris electrons at $v_{\parallel}=\pm 0.2c$ and one from the background electrons at $v_{\parallel}=0$ (see \myreffig{fig:beam_3d_vpara}(c3)).

\begin{figure}
    \centering
   \includegraphics[width=0.99\linewidth]{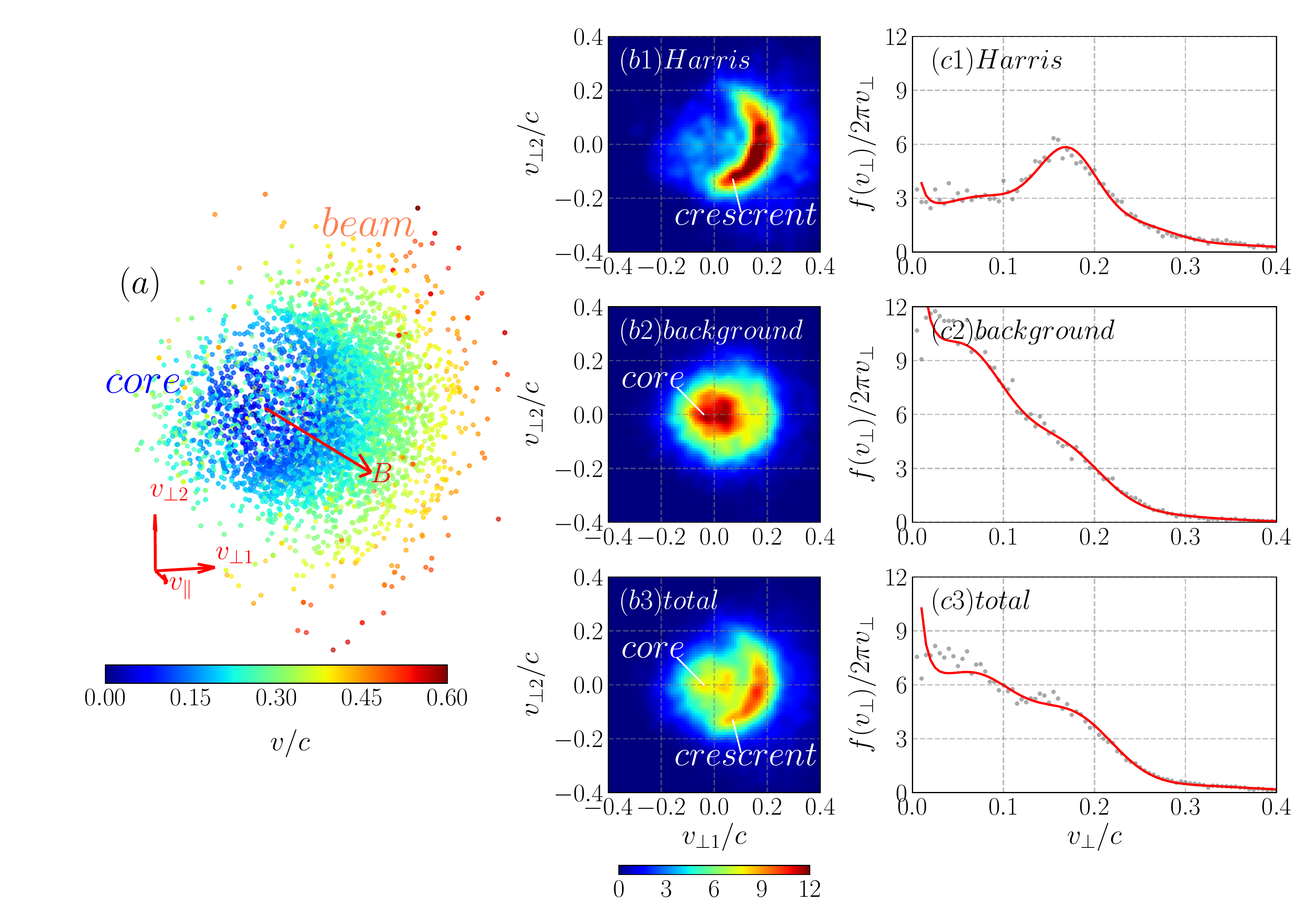}
    \caption[]{Velocity distribution of electrons within a spherical region of radius $\le0.1d_i$ centered at point B (see \myreffig{fig:reconnection_rate} (a2)) on the reconnection plane at $z=8d_i$ and $t=5.25\Omega_{ci}^{-1}$ for Run1. (a) Distribution of total electrons in the 3D velocity space. (b1-b3) 2D EVDFs $f(v_{\perp 1},v_{\perp 2})$ of Harris, background and total electrons in $v_{\perp 1}-v_{\perp 2}$ plane respectively. (c1-c3) 1D EVDFs $f(v_{\perp})/2\pi v_{\perp}$ (the red curves) of Harris, background and total electrons, respectively. Other quantities are same to those shown in \myreffig{fig:beam_3d_vpara}.}
   \label{fig:beam_3d_vperp}
\end{figure}

\myreffig{fig:beam_3d_vperp} shows the velocity distribution of electrons collected from a spherical region of radius $\le0.1d_i$ centered at point B (see \myreffig{fig:reconnection_rate} (a2)) a bit off the X point into the inflow region  on the reconnection plane at $z=8d_i$ and $t=5.25\Omega_{ci}^{-1}$ for Run1.
\myreffig{fig:beam_3d_vperp}(a) shows the distribution of total electrons in the 3D velocity space: the thermal core (see dots with cold colors) and non-thermal electron beams off the local magnetic field $\boldsymbol{B}$ axis and to the positive $v_{\perp1}$ direction (see dots with warm colors). 
The beam is better seen in the 2D EVDF $f(v_{\perp1},v_{\perp2})$ of Harris electrons shown in the $v_{\perp1}-v_{\perp2}$ plane (see \myreffig{fig:beam_3d_vperp}(b1)), which are obtained by integrating along the parallel direction. The Harris electrons population shows a crescent-shaped feature in the $v_{\perp1}-v_{\perp2}$ plane, while the azimuthally integrated 1D EVDF $f(v_{\perp})/2\pi v_{\perp}$ shows that a local maximum occurs at $v_{\perp}\approx 0.16c$ (see \myreffig{fig:beam_3d_vperp}(c1)). The latter implies a positive gradient in the perpendicular EVDF. 
On the other hand the bulk flow speed of the background electrons is near zero and these electrons exhibit a Maxwellian distribution (see 2D EVDF in \myreffig{fig:beam_3d_vperp}(b2) and its 1D EVDF in \myreffig{fig:beam_3d_vperp}(c2)). 
As a result, the resulting total 2D EVDF shows both thermal core and the crescent-shape feature (see \myreffig{fig:beam_3d_vperp}(b3)), while its 1D EVDF, on the other hand, shows no positive gradient in the $f(v_{\perp})/2\pi v_{\perp}$ (see \myreffig{fig:beam_3d_vperp}(c3)).

Note that at the points A and B, even though positive gradients exist in the 1D EVDFs calculated from Harris electrons (see \myreffig{fig:beam_3d_vpara}(c1) and \myreffig{fig:beam_3d_vperp}(c1)), the gradients are not present in the 1D EVDFs from total electrons (see \myreffig{fig:beam_3d_vpara}(c3) and \myreffig{fig:beam_3d_vperp}(c3)). This is because the background electrons are mainly thermally distributed and contribute more significantly to the total electron density than the Harris electrons, which has a significant influence on the formation of positive gradients in 1D EVDFs of total electrons. As a result, the locations where sources of free energy are available is a subset of those where non-thermal electron beams exist. 

The results show that the Harris electron population tend to be the population responsible for the non-thermal features in the total EVDFs (in both parallel and perpendicular direction to the local magnetic field), while the background population deviates slightly from a Maxwellian distribution.
In this study, we concentrated on the non-thermal electron beams which can offer sources of free energy in the form of positive gradients in their EVDFs as a necessary conditions for micro-instabilities. For this purpose, we identify and focus on two types of EVDFs: (1) those with a positive velocity gradient in 1D EVDFs from the Harris electron population, which is useful to identify the formation mechanism of non-thermal electron beams. (2) those with a positive gradient in the 1D EVDF from the total electrons (as a sum of Harris and background electrons), which determines the existence of parallel/perpendicular sources of free energy.

\subsection{Harris electron population EVDFs}

\begin{figure}
    \centering
   \includegraphics[width=0.99\linewidth]{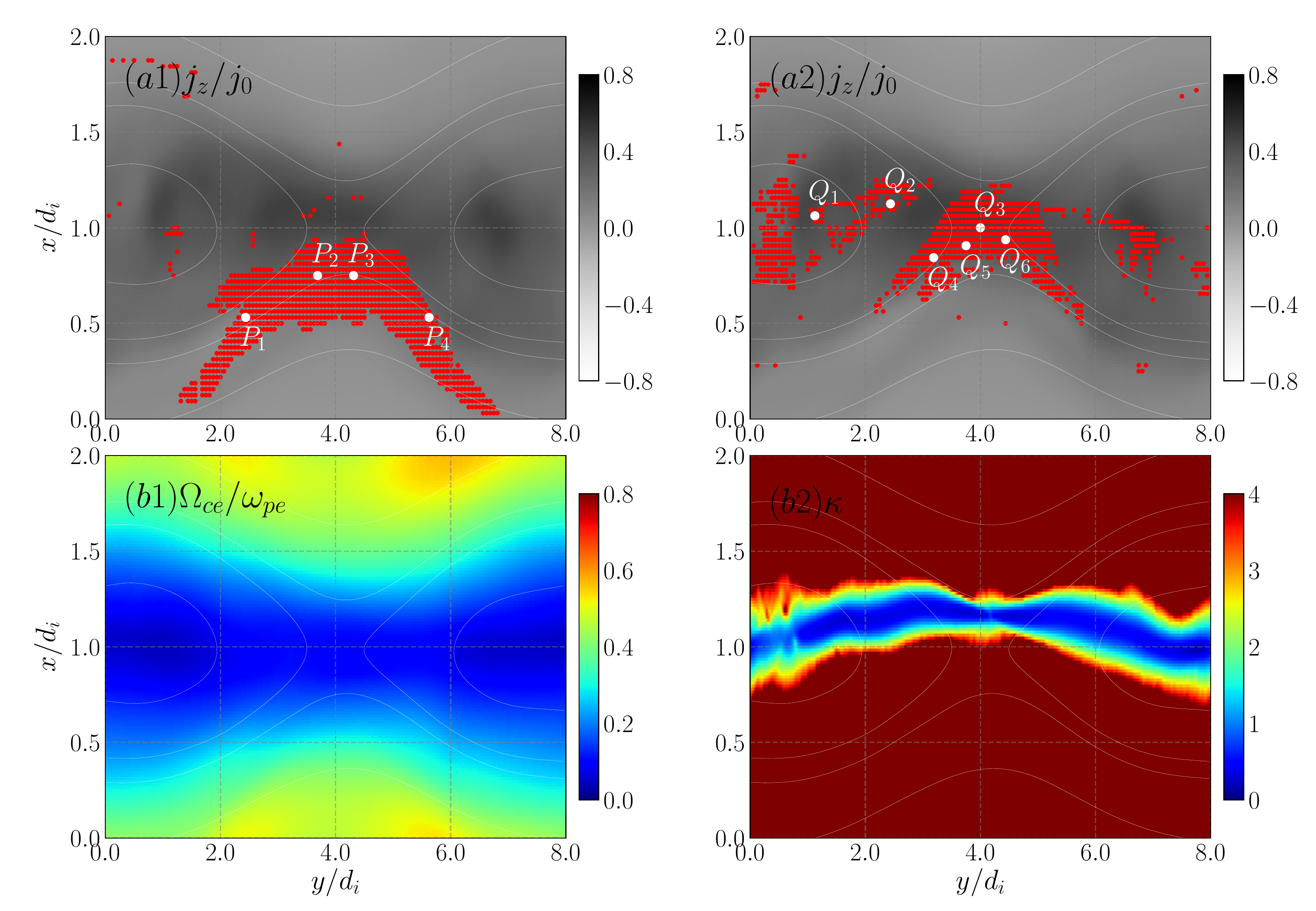}
    \caption[]{Distribution of non-thermal electron beam (red dot) in the directions parallel (a1) and perpendicular (a2) to the local magnetic field on the plane $z=8d_i$ at $t=5.25\Omega_{ci}^{-1}$ for Run1. The non-thermal electron beam is determined by the existence of positive velocity gradient(s) in EVDFs of the Harris electron population at each location on the reconnection plane. The background quantity in gray-scale is the out-of-reconnection-plane current density $j_z$ normalized by $j_0$. (b1) The ratio of electron cyclotron frequency to plasma frequency $\Omega_{ce}/\omega_{pe}$ and (b2) the curvature parameter $\kappa$ calculated by \myrefeq{eq:kappa_parameter}. Magnetic field lines (white curves) are overlaid, which are computed from the out-of-plane vector potential to a first order approximation.}
    \label{fig:source_quant}
\end{figure}

\myreffig{fig:source_quant} shows the distribution of non-thermal electron beams, which is determined by the existence of positive velocity gradient(s) in 1D EVDFs of the Harris electron population in the field-aligned (denoted by red dots in \myreffig{fig:source_quant}(a1)) and perpendicular (denoted by red dots in \myreffig{fig:source_quant}(a1)) directions to the local magnetic field separately, as well the ratio of electron cyclotron frequency (\myreffig{fig:source_quant}(b1)) to plasma frequency $\Omega_{ce}/\omega_{pe}$ and the curvature parameter $\kappa$ (\myreffig{fig:source_quant}(b2)) calculated by \myrefeq{eq:kappa_parameter} on the plane $z=8d_i$ at $t=5.25\Omega_{ci}^{-1}$ for Run1. 
In this case, EVDFs with a positive parallel velocity gradient in the Harris electron population mainly form in the diffusion region below the X point and two bottom-branches of separatrices, while the corresponding perpendicular velocity gradients are generally distributed in the diffusion region and outflow region near the midplane (roughly located at $x=d_i$) of reconnection plane. 
The cyclotron to plasma frequency ratio $\Omega_{ce}/\omega_{pe}$ is generally greater than $0.5$ in the inflow regions, while it is less than $0.5$ elsewhere. In the separatrices the ratio $\Omega_{ce}/\omega_{pe}\approx 0.4\sim 0.5$. 

In the diffusion region and outflow region near the midplane, the curvature parameter $\kappa$ is roughly about $1$, namely, $\kappa\approx 1$. This implies electrons here are mainly unmagnetized (or weakly magnetized) and non-adiabatic, they mainly contribute to the formation of EVDFs with a positive perpendicular velocity gradient. While EVDFs with a positive parallel velocity gradient are mainly due to magnetized electrons with $\kappa\ge 3$ in the separatrices.

\begin{figure}
    \centering
    \includegraphics[width=0.7\linewidth]{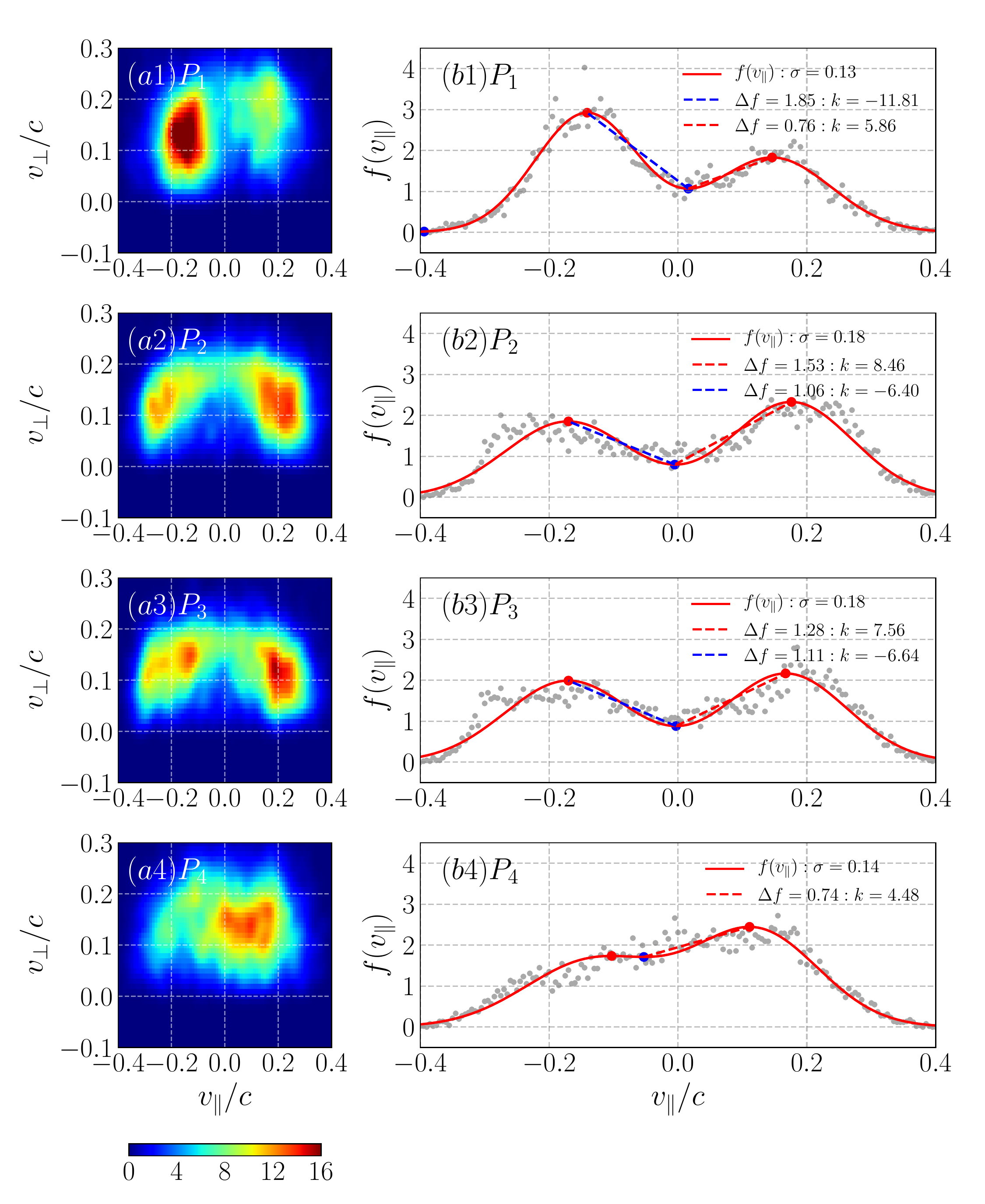}
    \caption[]{2D EVDFs in the $v_{\parallel}-v_{\perp}$ plane (left column) and 1D EVDFs $f(v_{\parallel})$ (right column) at four different points (see points $P_i\ (i=1,2,3,4)$ in \myreffig{fig:source_quant}(a1)) on the reconnection plane at $z=8d_i$ and time at $t=5.25\Omega_{ci}^{-1}$ for Run1.
    Those EVDFs were calculated with only the Harris electron population.
    The gray dots denote frequency of electron per bin in velocity space $f_i\ (i=1,2,\dots)$ (for a bin width $w=0.005c$), and the red solid curve indicate 1D EVDF $f(v_{\parallel})$ estimated by the BGMM method.
    $\sigma$ represents an estimation of the uncertainty between the data and  fitted curve.
    The red points indicate the local maximum and the blue dots denote local minimum of 1D EVDFs, respectively.
    The  positive velocity gradients or slopes $k$ are indicated with dashed lines, while $\Delta f$ is their (vertical) difference.
    The slope $k$ of the 1D EVDFs $f(v_{\parallel})$ has units of $c^{-1}$), while that of $f(v_{\perp})/2\pi v_{\perp}$ has units of $c^{-2}$. The effective slope $k_{\rm eff}$ of $f(v_{\perp})$ (unit: $c^{-1}$) is also shown.}
    \label{fig:evdf_para}
\end{figure}

\myreffig{fig:evdf_para} shows 2D EVDFs in the $v_{\parallel}-v_{\perp}$ plane as well as the associated integrated 1D EVDFs $f(v_{\parallel})$ at four different points (i.e., points $P_i\ (i=1,2,3,4)$ in \myreffig{fig:source_quant}(a1)). These EVDFs are derived from Harris electrons within a spherical region of radius $\le0.1d_i$ at each location on the reconnection plane $z=8d_i$ and time $t=5.25\Omega_{ci}^{-1}$ for Run1. 
Let us discuss firstly the EVDFs at points $P_2$ and $P_3$ in the diffusion region. \myreffig{fig:evdf_para} (a2) shows that at point $P_2$ (see \myreffig{fig:source_quant}(a1)) two counter-propagating beams with parallel bulk flow speed $\approx v_{\parallel}\approx\pm 0.18c$ in the $v_{\parallel}-v_{\perp}$ plane, while \myreffig{fig:evdf_para} (b2) clearly shows corresponding 1D EVDF $f(v_{\parallel})$ having two peaks located at $v_{\parallel}\approx\pm 0.18c$.
A rough estimation of the uncertainty of the fitting is given  by the standard deviation $\sigma$ between the data $f_i$ and the fitted curve $f(v_{\parallel})$, i.e.,
\begin{equation}\label{sigma_fitting}
\sigma=\sqrt{\sum\limits_i \left(f_i-f(v_i)\right)^2/N}
\end{equation}
For this case it is $\sigma=0.18c^{-1}$.
The 1D EVDF clearly has positive gradients $v_{\parallel}\cdot\partial f/\partial v_{\parallel}>0$ in the ranges $v_{\parallel}=[-0.18,0]c$ and $v_{\parallel}=[0,0.18]c$, respectively.
The slopes of those gradients  $\partial f/\partial v_{\parallel}$, calculated
by fitting all the data points between the corresponding local maxima to the local minima, are  $k=-6.4$ and $k=8.46$ (in units of $c^{-1}$), respectively.
The difference (along the vertical axis) for theses two gradients is about $\Delta f=1.53c^{-1}$ and $1.06c^{-1}$, respectively.
Those quantities, obtained from the fitting data, should be compared with the standard deviation Eq.~\ref{sigma_fitting}
in order to assess how significant are they compared to the input data.
We can clearly see that they are significantly larger by at least a factor of 5, so this
confirms the reliability of our fittings and in particular the calculated gradients.
This is of course only a simple estimation with significant drawbacks that will be
discussed later.

All this implies this EVDF offer sources of free energy which could possibly cause counter-streaming instabilities, provided that the background population does not significantly contribute to the total electron density.
Similar results are observed in the EVDFs at point $P_3$ (see \myreffig{fig:evdf_para} (a3,b3)).

Let us focus now on the EVDFs in the separatrix region farther away from the diffusion region. At the point $P_1$ (see \myreffig{fig:source_quant}(a1)), \myreffig{fig:evdf_para} (a1) shows an electron beam formed at $v_{\parallel}=-0.18c,\ v_{\perp}=[0.1,0.15]c$ and another beam formed at t $v_{\parallel}=0.16c,\ v_{\perp}=0.2c$. 
Their 1D EVDF has peaks located at about $v_{\parallel}=-0.18c$ and $v_{\parallel}=0.16c$ (see \myreffig{fig:evdf_para}(b1)). The value differences (along vertical axis) for theses two gradients is about $\Delta f=1.85c^{-1}$ and $0.76c^{-1}$ separately, they are more than the deviation $\sigma=0.13c^{-1}$. Their velocity gradient slopes are $k=-11.81$ and $k=5.86$ (in units of $c^{-1}$) respectively.
At point $P_4$ (see \myreffig{fig:source_quant}(a1)), \myreffig{fig:evdf_para} (a4) exhibits an electron beam located at $v_{\parallel}=0.1c$ and $v_{\perp}=0.15c$, while its 1D EVDF features a peak located at about $v_{\parallel}=0.15c$ (see \myreffig{fig:evdf_para}(b4)). The value difference (along vertical axis) for this gradient is about $\Delta f=0.74c^{-1}$ being more than the deviation $\sigma=0.14c^{-1}$.
Since this EVDF has a positive gradient at $v_{\parallel}=0.15c$, it offers a source of free energy which could possibly cause bump-on-tail instabilities, as long as the background population does not significantly contribute to the total electron density.

\begin{figure}
    \centering
   \includegraphics[width=0.99\linewidth]{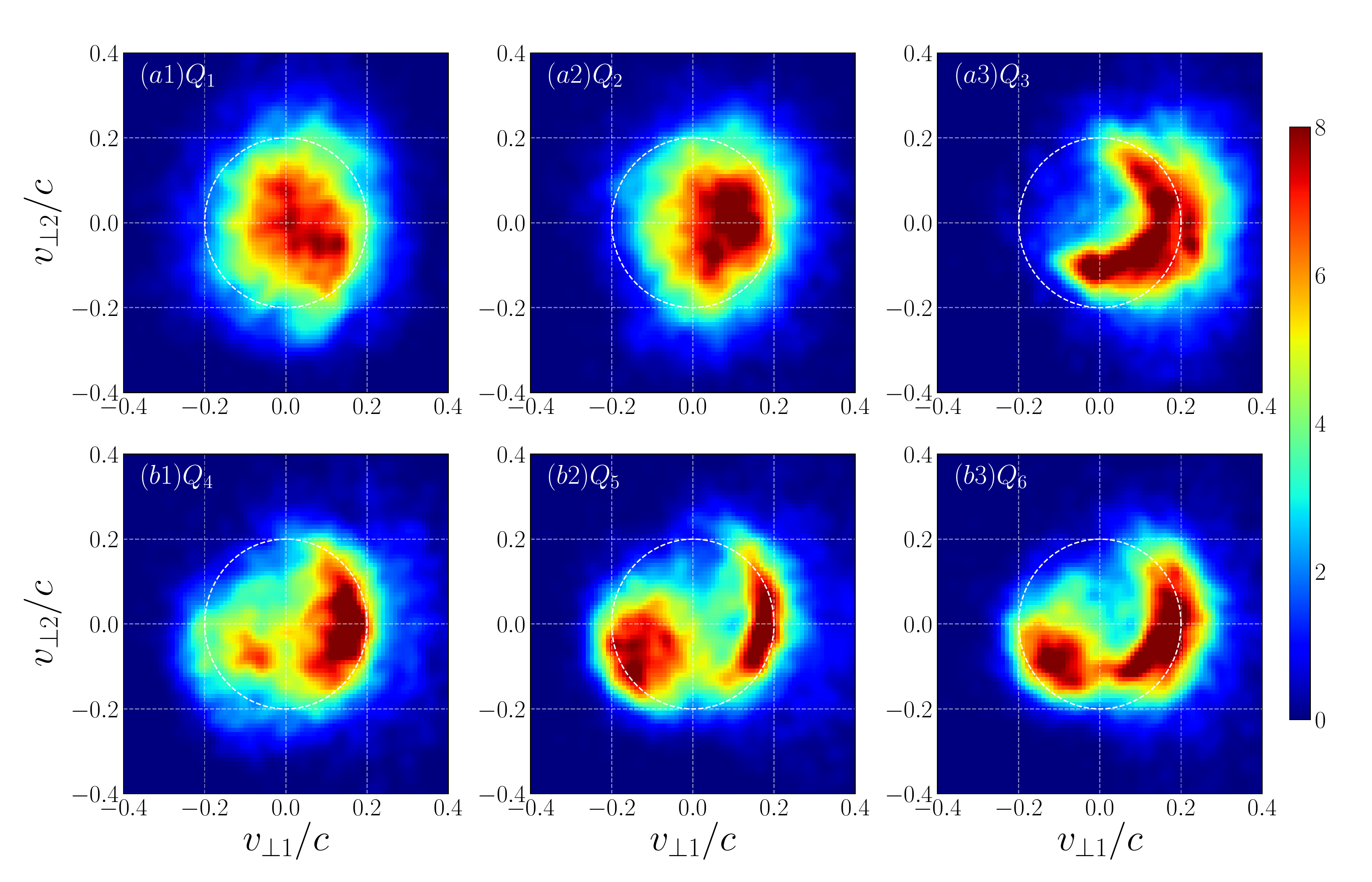}
    \caption[]{2D EVDFs in the $v_{\perp1}-v_{\perp2}$ plane at six different points (see points $Q_i\ (i=1,2,...,6)$ in \myreffig{fig:source_quant}(a2)) on the reconnection plane $z=8d_i$ at $t=5.25\Omega_{ci}^{-1}$ for Run1. Those EVDFs were calculated  with only the Harris electron population. The dashed white circle centered at the $v_{\perp1}=0,v_{\perp2}=0$ has radius of $0.2c$.}
    \label{fig:evdf_perp}
\end{figure}

\myreffig{fig:evdf_perp} shows 2D EVDFs in the $v_{\perp1}-v_{\perp2}$ plane, these electrons are calculated from the Harris electron population at six different points (i.e., points $Q_i(i=1,2\dots,6)$ in \myreffig{fig:source_quant}(a2)) on the reconnection plane $z=8d_i$ and at time $t=5.25\Omega_{ci}^{-1}$ for Run1.
The points located near and below the diffusion region, i.e., $Q_3,\ Q_4,\ Q_5$ and $Q_6$ (see \myreffig{fig:source_quant}(a2)) feature perpendicular crescent-shaped EVDFs, which are partial rings centered at $v_{\perp1}=0,v_{\perp2}=0$ with a radius of about $0.2c$. 
Note that the EVDFs at the points $Q_4,\ Q_5$ and $Q_6$ (see \myreffig{fig:evdf_perp}(b1,b2,b3)) feature an additional population in the quadrant $v_{\perp1}<0,v_{\perp2}<0$. This implies their origin is different from the main partial ring population. Because those points are located in regions with $\kappa\ge3$, it is plausible to think their origin is due to the magnetized and adiabatic electrons taking gyrotropic electron motion. 
These EVDFs can offer perpendicular sources of free energy to cause ECMIs since  positive velocity gradients exist in their corresponding integrated 1D perpendicular EVDFs $f(v_{\perp})$, assuming that the contribution of the background electron population to the total electron density is not significant.
The EVDFs in the outflow region, i.e., at points $Q_1$ and $Q_2$ (see \myreffig{fig:source_quant}(a2)) are dominated by a thermal core (see \myreffig{fig:evdf_perp}(a1,a2)). This is possibly attributed to the thermalization of non-thermal electron beams generated near the X point as they move away from the diffusion region into the outflow region. 
The perpendicular EVDF at point $Q_2$ has a positive gradient in its 1D EVDF $f(v_{\perp})$ and thus it could offer a source of free energy to cause ECMIs, provided again that the background electron population does not significantly contribute to the total electron density. 
However, the perpendicular EVDFs at point $Q_1$ are unable to offer sources of free energy since the electrons are mainly Maxwellian distributed around $v_{\perp1}=0,v_{\perp2}=0$ with a thermal spread width of about $0.1c$.

\subsection{Identification of sources of free energy}

\begin{figure*}
    \centering
    \begin{subfigure}{0.35\textwidth}
        \centering
       \includegraphics[height=0.8\textwidth,width=\linewidth]{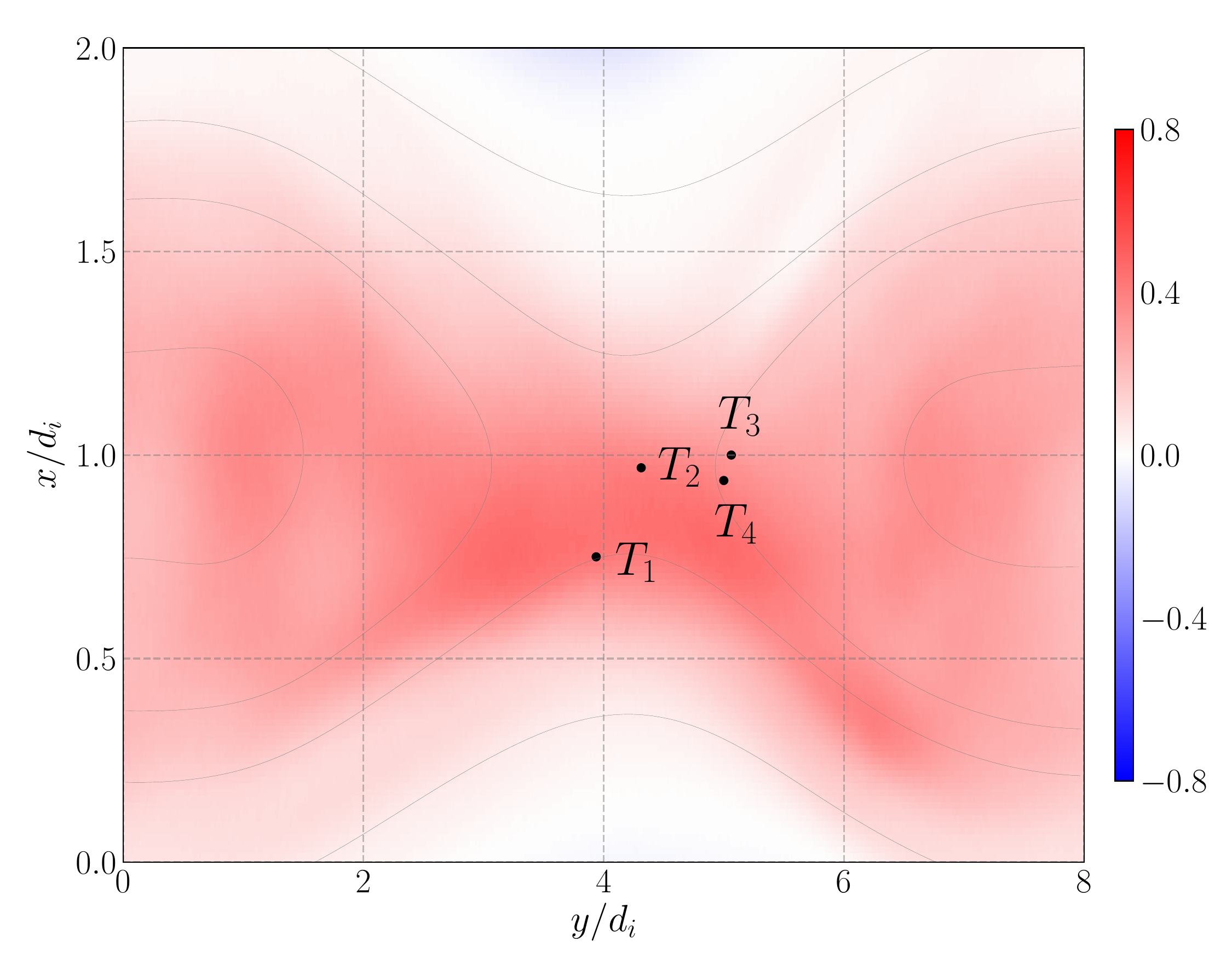}
        \caption[]{}
    \end{subfigure}
    \centering
    \begin{subfigure}{0.64\textwidth}
        \centering
       \includegraphics[height=1.0\textwidth,width=\linewidth]{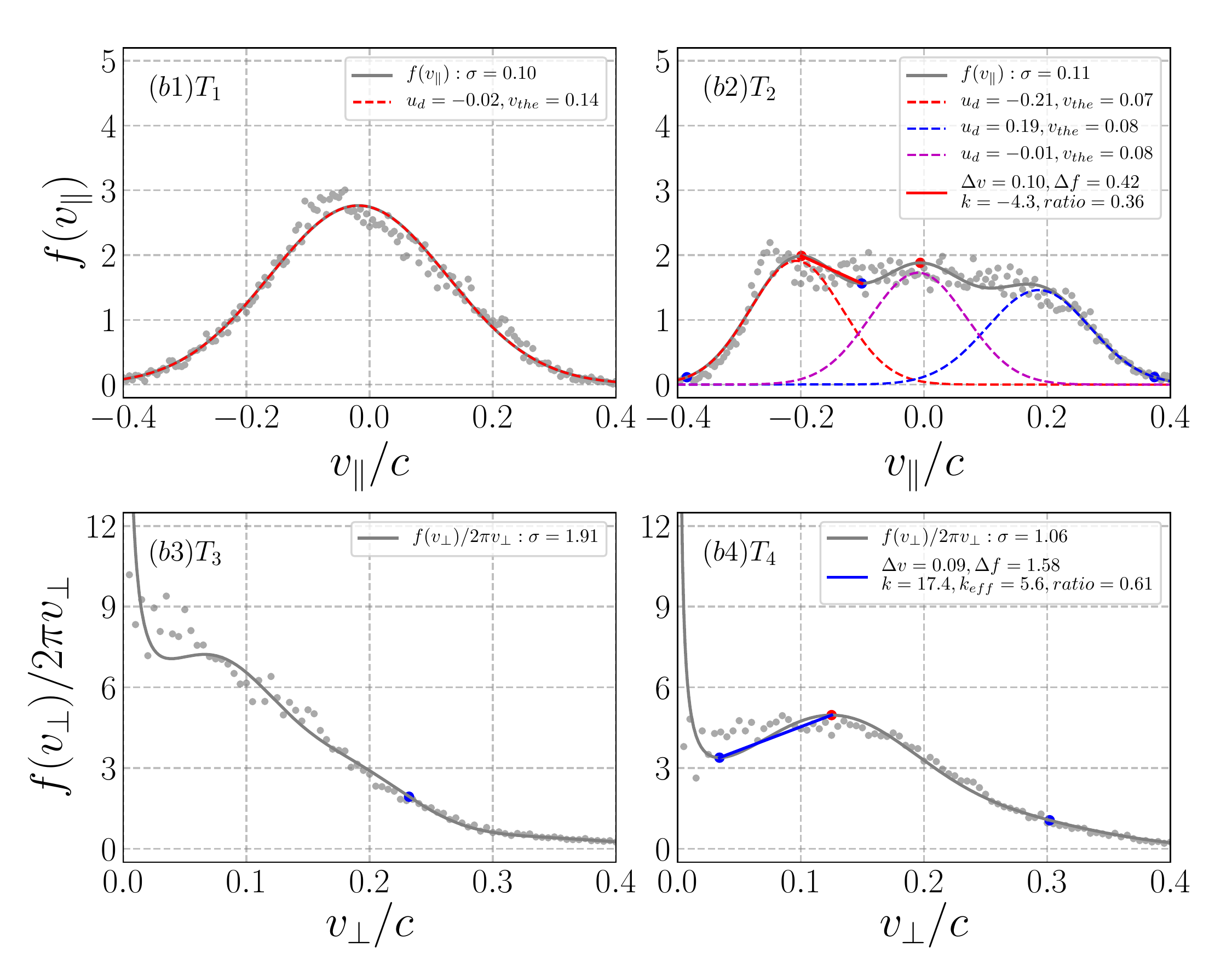}
        \caption[]{}
    \end{subfigure}
    \caption[]{(a) The out-of-plane current density $j_z$ (normalized by $j_0=en_0v_{the}$) and four points $T_i\ (i=1,2,3,4)$ on the reconnection plane at $z=8d_i$ and time $t=5.25\Omega_{ci}^{-1}$ for Run1.
    (b) Typical parallel/perpendicular 1D EVDFs at points $T_i\ (i=1,2,3,4)$ on the reconnection plane.
    The 1D EVDFs are estimated from the total electrons within a spherical region with radius $0.1d_i$ at each point.
    The gray dots denote frequency of electron per bin in velocity space $f_i\ (i=1,2,\dots)$ (using the same bin width as in Fig~\ref{fig:evdf_para}, i.e., $w=0.005c$) and the gray curve indicate 1D EVDFa $f(v_{\parallel})$ estimated by the BGMM method.
    The parallel 1D EVDF is decomposed into different Gaussian components by the BGMM method (see dashed curves in panel (b1,b2)) and the density ratio of each fitted Maxwellian to the total distributions are estimated.
    Other quantities and legend are same as Fig~\ref{fig:evdf_para}.}
    \label{fig:evdf_category}
\end{figure*}

\myreffig{fig:evdf_category}(b) shows examples of typical parallel and perpendicular 1D EVDFs at four different points (i.e., $T_i\ (i=1,2,3,4)$ in \myreffig{fig:evdf_category}(a)) in the diffusion region and separatrices of the current sheet on the reconnection plane at $z=8d_i$ and $t=5.25\Omega_{ci}^{-1}$ for Run1. The 1D EVDF at each point $T_i\ (i=1,2,3,4)$ is calculated from the total electrons within a spherical region of radius $\le0.1d_i$ centered at the point on the reconnection plane.
\myreffig{fig:evdf_category}(b1) shows the 1D parallel EVDF $f(v_{\parallel})$ at the point $T_1$. The EVDF is practically a Maxwellian distribution with a mean velocity $v_{\parallel}\approx -0.01c$. The fitting by the BGMM method allows to determine the width of this thermal EVDF as $v_{the}\approx 0.14c$. There is no positive velocity gradient in the 1D EVDF and thus it cannot offer any source of free energy.
\myreffig{fig:evdf_category}(b2) shows the 1D parallel EVDF $f(v_{\parallel})$ at point $T_2$.
Its standard deviation (an estimation of the uncertainty of the fitting) from the fitted curve, using the definition of Eq.~\ref{sigma_fitting} is $\sigma=0.11c^{-1}$.
Besides a peak centered at $v_{\parallel}=0$, there are other two peaks located at $v_{\parallel}\approx\pm 0.2c$. These two peaks indicate two possible non-thermal field-aligned electron beams.
The left-most beam, centered at  $v_{\parallel}\sim-0.2c$ constitutes $36\%$ of the total electron population.
It features a positive velocity gradient with slope $k=-4.3$ (in units of $c^{-1}$, see the red fitting slopes).
Its associated difference (along the vertical axis) is $\Delta f=0.42c^{-1}$ is greater than the standard deviation $\sigma=0.11c^{-1}$,
but not by the same (larger) amount as in Fig.~\ref{fig:evdf_para}.
Although this could indicate the reliability of the fitting and the calculated gradient or slope,
it is important to notice that the standard deviation of the input data
is not homogeneous, but small at the tails of the VDFs
and larger near the maxima, where the slopes are actually calculated.
So it is necessary to be cautious with the interpretation of the goodness of fit of our calculations,
keeping in mind that the determined uncertainties provide just a very rough estimation.
We have also carried out similar calculations for different bin widths of the input VDF data (not shown here).
From this we can conclude that the deviations between the input data and the GMM fittings
decrease as the bin width of the input data increases, and it is good enough
in such a way that the slopes of the positive velocity gradients
can be sufficiently distinguished, in the sense that their vertical differences ($\Delta f$)
are larger that the (global) standard deviation between the data and the fitting.
More work would be needed to make more accurate statements about those uncertainties.

This all implies that source of free energy, i.e., $v_{\parallel}\cdot \partial f/\partial v_{\parallel}>0$, could be available to eventually cause streaming-like instabilities.
\myreffig{fig:evdf_category}(b3) shows the 1D perpendicular EVDF $f(v_{\perp})/2\pi v_{\perp}$ at point $T_3$.
The EVDF has a maximum near $v_{\perp}=0$ and monotonically decreases to zero.
There is no source of free energy.
\myreffig{fig:evdf_category}(b4) shows the 1D perpendicular EVDF $f(v_{\perp})/2\pi v_{\perp}$ at the point $T_4$.
The standard deviation between data and fitting from Eq.~\ref{sigma_fitting} is about $\sigma=1.06c^{-2}$.
There is a positive velocity gradient with a slope of $k=17.4$ (in unit of $c^{-2}$) in $f(v_{\perp})/2\pi v_{\perp}$ between $v_{\perp}\approx 0.05c$ and $v_{\perp}\approx 0.16c$.
Its associated vertical difference is $\Delta f=1.58c^{-2}$, which is just barely about $\sigma=1.06c^{-2}$.
This indicates that the uncertainty of this gradient with respect to the input data is relatively large,
so that the positive velocity gradient is not that robust and prone to noise, but it exists.
The associated effective slope in $f(v_{\perp})$ in $f(v_{\perp})$ can be estimated as $k_{eff}=5.6$ (in unit of $c^{-1}$). We apply the BGMM method to fit the 1D EVDF $f(v_{\perp})/2\pi v_{\perp}$, the non-thermal beam constitutes about $61\%$ of the total electron population.
As a result, perpendicular sources of free energy could be available to cause ECMIs.

\subsection{Distribution of sources of free energy}

We systematically evaluate positive velocity gradients in the 1D parallel and perpendicular EVDFs, i.e., $f(v_{\parallel})$ and $f(v_{\perp})/2\pi v_{\perp}$, at locations on each reconnection plane to determine sources of free energy formed in magnetic reconnection.
At each point, the 1D EVDFs are calculated from total electrons within a spherical region of radius $\le 0.1d_i$ centered at this point.
In this way, not only the spatial distribution of the resulting EVDFs with positive velocity gradient(s) and thus the possible sources of free energy for micro-instabilities can be visualized, but it also allows to partly understand their formation mechanism by comparing them with the spatial distributions of electromagnetic fields and other quantities.

\begin{figure*}
    \centering
   \includegraphics[width=0.99\linewidth]{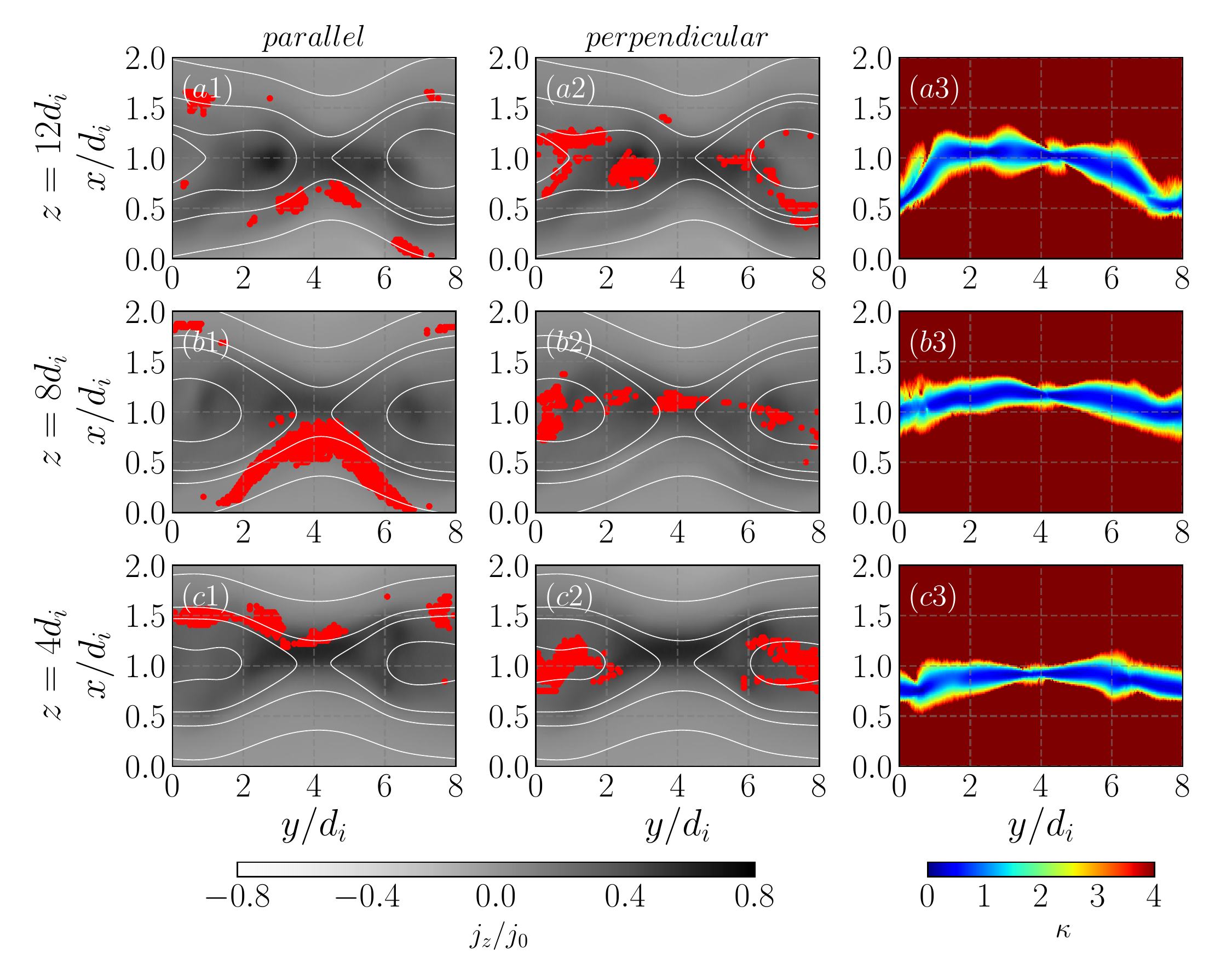}
    \caption[]{Distribution of sources of free energy (red dot) along the parallel (left column) and perpendicular (middle column) direction to the local magnetic field, the curvature parameter $\kappa$ (right column) estimated by \myrefeq{eq:kappa_parameter} on the reconnection plane $z=12,\ 8,\ 4d_i$ (from top to bottom), respectively, at time $t=5.25\Omega_{ci}^{-1}$ for Run1. The sources of free energy is determined by the existence of positive gradient(s) in EVDFs of total electron population at each location on the reconnection plane. The background quantity in gray-scale is the corresponding out-of-plane current density $j_z$ normalized by $j_0=en_0v_{the}$. Magnetic field lines (white curves) are overlaid.}
    \label{fig:source_Nz}
\end{figure*}

\myreffig{fig:source_Nz} shows the distribution of possible sources of free energy, i.e., EVDFs with positive gradients in the parallel and perpendicular direction to the local magnetic field separately along the out-of-plane direction at $z=12d_i,\ 8d_i,\ 4d_i$ respectively, at time $t=5.25\ \Omega_{ci}^{-1}$ for Run1. Along the field-aligned direction, sources of free energy at $z=12d_i$ are distributed in the left-top and right-bottom separatrix branches, as well as in the diffusion region below the X point (see \myreffig{fig:source_Nz}(a1)). 
Most of the sources of free energy at $z=8d_i$ are observed in the diffusion region below the X point and bottom separatrix regions along the magnetic field lines (see \myreffig{fig:source_Nz}(b1)). \myreffig{fig:source_Nz}(c1) shows sources of free energy at $z=4d_i$ mainly form in the diffusion region above the X point and top separatrix regions along the magnetic field lines.
In the direction perpendicular to local magnetic field, sources of free energy at $z=4d_i$ are mainly distributed in the outflow regions near the midplane of reconnection at different heights (see \myreffig{fig:source_Nz}(c2)), while those at $z=12d_i$ and $z=8d_i$ are observed in the diffusion region and outflow regions near the midplane (see \myreffig{fig:source_Nz}(a2,b2)).
\myreffig{fig:source_Nz}(a3,b3,c3) shows the curvature parameter $\kappa$ estimated according to \myrefeq{eq:kappa_parameter} at each reconnection plane $z=12d_i,\ 8d_i,\ 4d_i$ for Run1 respectively.
The quantity of $\kappa\le 3$ mainly appears in the diffusion and outflow regions near the midplane (i.e., $x\approx d_i$). 
Along the field-aligned direction, sources of free energy are mainly observed in the region with $\kappa\ge 3$, this implies the formation of parallel sources of free energy is mainly due to magnetized and adiabatic electrons. In the perpendicular direction to the local magnetic field, sources of free energy are generally observed in the regions with $\kappa\le 3$, this implies that the formation of perpendicular sources of free energy is mainly due to unmagnetized and non-adiabatic electrons.

\begin{figure}[h!]
    \centering
    \includegraphics[width=0.8\linewidth]{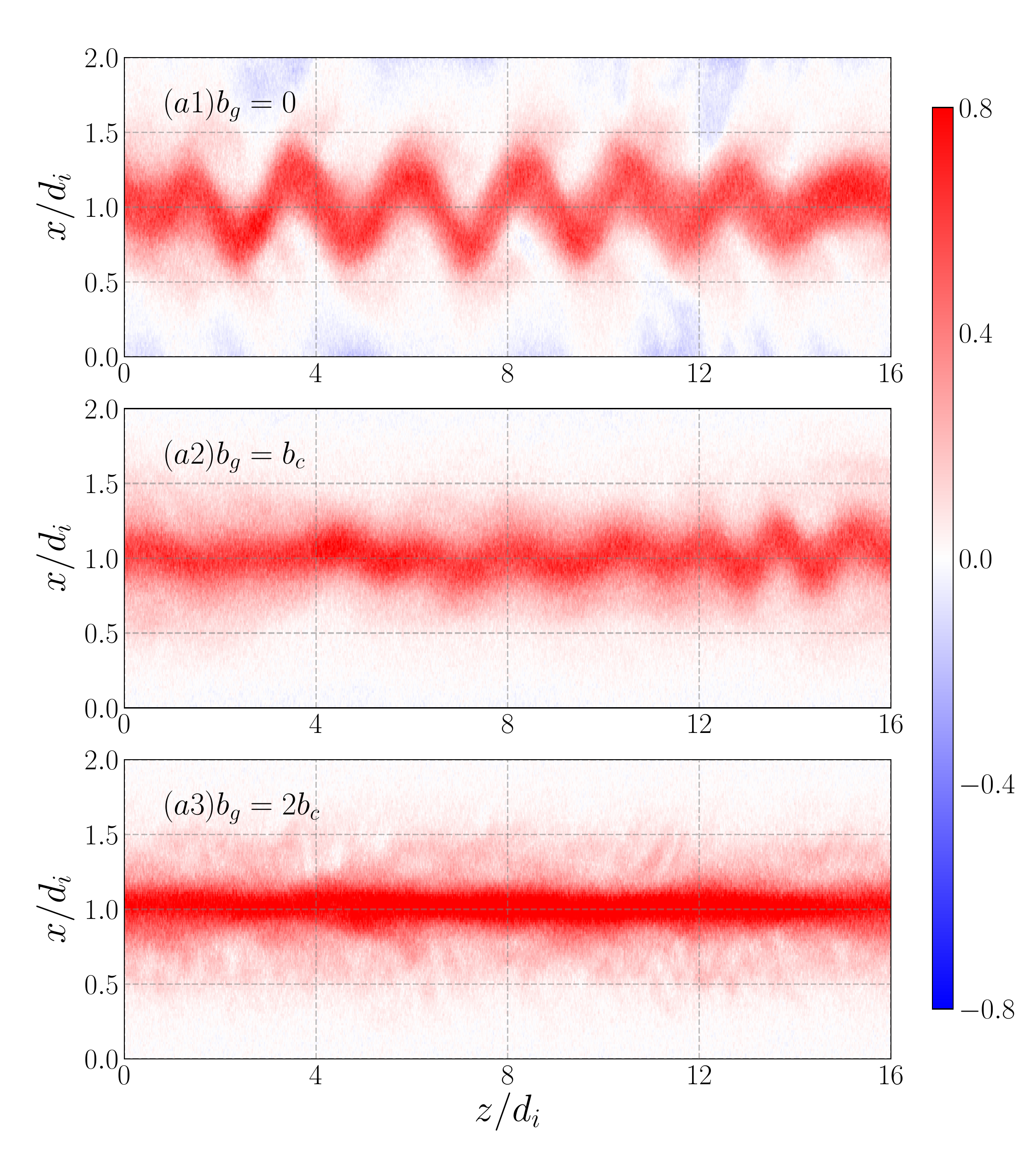}
    \caption[]{Current density component $j_z$ on the plane $y=4d_i$ (at the current sheet center and spanning the out-of-the-reconnection plane direction $z$ and the direction across the magnetic field gradients $x$) at time $t=5.25\Omega_{ci}^{-1}$ for Run1, Run2 and Run3 respectively. Here the current density $j_z$ is normalized by $j_0=en_0v_{the}$.}
    \label{fig:jz-xz}
\end{figure}

\myreffig{fig:jz-xz} shows the variation of the current density component $j_z$ on the plane $y=4d_i$ (at the current sheet) in order to show the out-of-plane variation. For antiparallel magnetic reonncetion Run1 (\myreffig{fig:jz-xz}(a1)), the current density structure is wavy along the $z$ direction, probably due to the density gradients at the edge of the current sheet driving the so-called lower-hybrid drift instability. We observed that this current sheet "kinking" weakens with increasing guide field (see \myreffig{fig:jz-xz}(a2,a3)). \myreffig{fig:jz-xz}(a3) shows reconnection with a stronger guide field actually makes the current sheet structure to behave more 2D-like.
This kinking of the current sheet modulates the distribution of sources of free energy (distribution functions with positive gradients) along the z-direction. Such a modulation would be absent in 2D reconnection, because there is not such a kinking instability in a 2D geometry. This current sheet oscillation causes a complicated flow pattern with alternate and different shear and counter-streaming bulk plasma flows on different $z-$planes. And this, in turn, explains the asymmetry of the distribution of sources of free energy, in particular, in antiparallel magnetic reconnection Run1 (see \myreffig{fig:source_Nz}(left)).

\begin{figure*}
    \centering
   \includegraphics[width=0.99\linewidth]{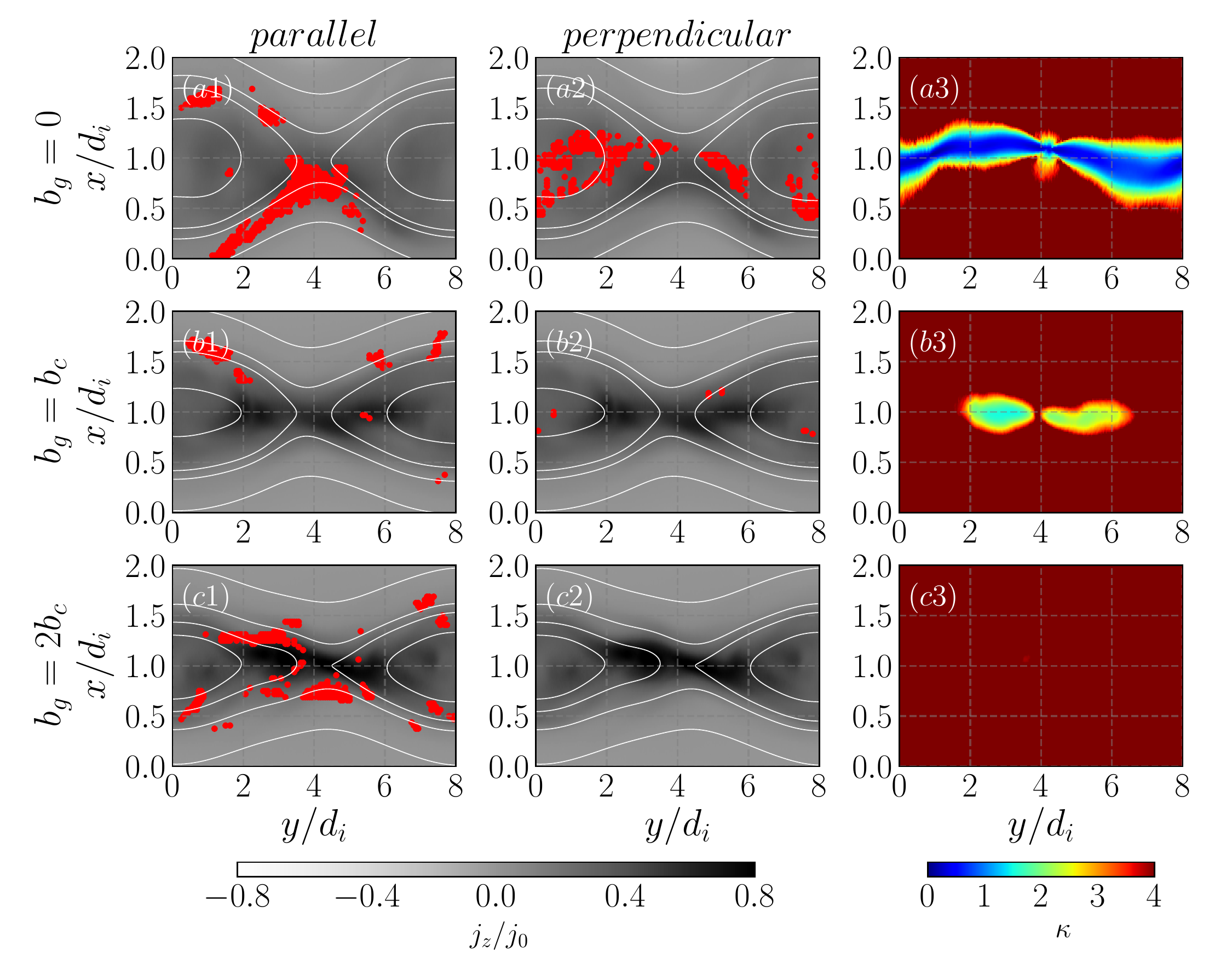}
    \caption[]{Distribution of sources of free energy along the parallel (left column) and perpendicular (middle column) direction to the local magnetic field, and the curvature parameter $\kappa$ estimated by \myrefeq{eq:kappa_parameter} (right column) in the reconnection plane $z=8d_i$ for Run1 ($b_g=0,\ t=5.25\Omega_{ci}^{-1}$), Run2 ($b_g=b_c,\ t=5.5\Omega_{ci}^{-1}$), Run3 ($b_g=2b_c,\ t=6.75\Omega_{ci}^{-1}$), respectively. Other quantities are same to those in \myreffig{fig:source_Nz}.}
    \label{fig:source_bg}
\end{figure*}

\myreffig{fig:source_bg} shows the distribution of sources of free energy, i.e., EVDFs with positive gradients along the parallel and perpendicular directions separately on the reconnection plane at $z=8d_i$ for Run1 ($b_g=0,\ t=5.25\Omega_{ci}^{-1}$), Run2 ($b_g=b_c,\ t=5.5\Omega_{ci}^{-1}$) and Run3 ($b_g=2b_c,\ t=6.75\Omega_{ci}^{-1}$), respectively.
It is found that the parallel (field-aligned) sources of free energy generally appear in the diffusion region near the X point and separatrices. For the antiparallel reconnection Run1 (see \myreffig{fig:source_bg}(a1)) and the strong guide field reconnection Run3 (see \myreffig{fig:source_bg}(c1)), sources of free energy mainly form in the diffusion region off the X point and separatrices, while sources of free energy are mainly located in the top two separatrices for weak guide field reconnection Run2 (see \myreffig{fig:source_bg}(b1)).
Along the perpendicular direction, sources of free energy are generally distributed in the diffusion region and the outflow regions near the midplane for the antiparallel reconnection Run1 (see \myreffig{fig:source_bg}(a2)). However, few sources of free energy form on the reconnection plane for Run2 (see \myreffig{fig:source_bg}(b2)) and Run3 (see \myreffig{fig:source_bg}(c2)). This implies the guide field strength has a significantly negative influence on the formation of the sources of free energy in the direction perpendicular to the local magnetic field: as the guide field strength increases, the spatial extent of perpendicular sources of free energy decreases significantly. 
The curvature parameter $\kappa$ is also estimated based on \myrefeq{eq:kappa_parameter} at reconnection plane $z=8d_i$ for simulations of magnetic reconnection in different guide fields strengths (see \myreffig{fig:source_bg}(a3,b3,c3)). 
Along the field-aligned direction, sources of free energy are mainly due to magnetized and adiabatic electrons with $\kappa\ge 3$ in the separatrices. 
While in the perpendicular direction to the local magnetic field, sources of free energy are generally observed in the regions with $\kappa\le 3$, this implies formation of perpendicular sources of free energy is mainly due to the unmagnetized and non-adiabatic electrons. 
For the antiparallel reconnection Run1, the curvature parameter $\kappa$ is less than $3$ near the midplane (i.e., $x\approx d_i$) on reconnection plane, and the perpendicular sources of free energy are mainly generated in the region with $\kappa\le 3$. However, for the reconnection cases with finite guide-field Run2 and Run3, the value of the curvature parameter is practically always larger than $3$ everywhere. In these cases, few perpendicular sources of free energy are observed, which proves a significantly negative correlation between the electron magnetization and perpendicular sources of free energy.

\section{Conclusions and discussions}\label{sec:conclusion}

We carried out 3D PIC code simulations of magnetic reconnection in order to investigate the formation of non-thermal electron beams and EVDFs for plasma conditions suitable to the solar corona.
We focused on the EVDF features that can be considered as necessary conditions for microscopic plasma instabilities that could play a role in radio emission processes.
This requires the identification of possible source of free energy, namely, positive velocity space gradient(s) in the EVDFs in the direction parallel and perpendicular to the local magnetic field, respectively.
For this sake we implemented a machine learning algorithm to fit the reduced 1D EVDFs and searched for their velocity gradient(s).
The effects of a guide field on those feature of non-thermal electron beams were also investigated.

Our results are summarized as follows:
\begin{itemize}
    \item Possible parallel (or field-aligned) sources of free energy for streaming-like instabilities, are mainly generated in the separatrices and diffusion region. This kind of source of free energy is determined by positive gradient(s) in the 1D parallel EVDF, i.e., $v_{\parallel}\cdot \partial f(v_{\parallel})/v_{\parallel}>0$. Those EVDFs are a necessary condition for the instabilities leading to the plasma emission mechanism, which  cause waves at the plasma frequency and possibly their harmonics \citep[][]{Yao2021}.
    \item Possible perpendicular sources of free energy for cyclotron maser instabilities are mainly formed in the diffusion and outflow region near the midplane of reconnection.
	This kind of source of free energy is determined by the positive gradient in the perpendicular 1D EVDF, i.e., $v_{\perp}\cdot f(v_{\perp})>0$. These non-thermal electrons are non-adiabatic and their EVDF is mostly characterized by a crescent-shaped feature. Those EVDFs represent a necessary condition that can cause ECMIs and generate waves at the harmonics of the electron cyclotron frequency \citep[][]{Yao2021}.
    \item As the strength of guide field increases, 1D EVDFs with positive velocity gradient(s) in perpendicular direction to the local magnetic field are less likely to be generated by reconnection.
\end{itemize}

\citet{Dupuis2020} introduced a machine learning method to detect the formation of non-thermal electron beams by comparing the 2D EVDFs, i.e., $f(v_{\parallel},v_{\perp})$ and $f(v_{\perp1},v_{\perp2})$, generated by 2D magnetic reconnection with Maxwellian distributed EVDFs. In this study, we extended this method to 1D EVDFs, i.e., $f(v_{\parallel})$ and $f(v_{\perp})$ generated by 3D magnetic reconnection. We investigated the positive velocity gradient(s) in the 1D parallel and perpendicular EVDFs in order to determine whether sources of free energy are available at each locations in the reconnection region.

Our results depend on the reliability of the fitting, for which an estimation of its accuracy was not performed.
Only a very rough estimate of the  uncertainties of the EVDFs fitting by the BGMM method were carried out, in particular to assess the reliability of the positive velocity gradients compared to those fittings.
This does not guarantee the accuracy our results, but it provides some kind of positive evidence.
A more rigorous goodness-of-fit statistics still needs to be performed to guarantee the reliability of our method but that is deferred to a future work.

An investigation of microscopic plasma instabilities, such as the streaming-like instabilities and ECMIs, due to those nonthermal EVDFs are beyond the scope of this study. We can only speculate here that it is unlikely that this kind of instabilities can be detected in these simulations, because positive gradients in EVDFs are relatively weak, the possible micro-instabilities caused by them are thus relatively weak in comparison with other macroscopic instabilities and plasma flows. 
We plan to investigate the micro-instabilities and properties of resulting wave emission due to non-thermal EVDFs featuring possible sources of free energy in a forthcoming publication, in which instabilities associated to those EVDFs are separately investigated and they could become more relevant.

A 3D magnetic reconnection configuration is essential for our purposes since EVDFs evolve differently comparing to their counterparts in 2D magnetic reconnection. For example, it is expected that EVDFs with positive gradients in the out-of-plane direction (i.e., along $v_z$ direction) will release their energy via streaming-instabilities via unstable waves with a $k_z$ vector. This process is not possible in a 2D configuration.
This is particularly relevant for reconnection under the influence of a (relatively strong) guide field, because the total magnetic field nearly points to the $z-$direction (at least near the reconnection midplane), the reconnection electric field will accelerate electrons along this direction. So the evolution of those electron distribution functions will be clearly different in 3D and 2D reconnetion.

The formation mechanism of those non-thermal electron beams and resulting velocity distribution functions is also valuable to investigate in a future work. In particular, for the field-aligned non-thermal electron beams, the reconnection electric field and parallel electric field have significant influence on the electron motion and acceleration \citep[][]{Dahlin2016,Drake2003,Egedal2016a}. For the non-thermal electron beams that forms in the direction perpendicular to the local magnetic field, there are several proposed mechanisms (as those discussed in the introduction), including the $\boldsymbol{E}\times\boldsymbol{B}$ or gradient-$B$ drifts.
It is still not clear which process dominantly affects the electron acceleration in our 3D kinetic magnetic reconnection simulations. For this purpose, a study of dynamic orbits of electrons and ions, their interaction with reconnection electromagnetic fields and their dependence on the strength of guide field is necessary, which is beyond the scope of this work.

\section{Acknowledgements}

We gratefully acknowledge the developers of the ACRONYM code, the \textit{Verein zur F\"orderung kinetischer Plasmasimulationen e.V.} and the financial support by the German Science Foundation (DFG), projects MU-4255/1-1 and BU 777/15-1.
We also gratefully acknowledge the possibility of using the computing resources of the Max Planck Computing and Data Facility (MPCDF, formerly known as RZG) at Garching and of the Max-Planck-Institute for Solar System Research at G\"ottingen as well as of the Technical University Berlin, Germany.
We also thank the referees for their comments and suggestions that allowed us to improve the presentation of our results.

\appendix
\par

\section{The Bayesian Gaussian mixture model} 
\label{app:BGMM}

The application of the Bayesian Gaussian mixture model (BGMM) \citep[][]{Bishop2006} on electron velocity distribution function (EVDF) is briefly described in this appendix.

The velocity distribution function of electrons can be fitted by the Gaussian mixture model (GMM) \citep[][]{Bishop2006}, namely, a linear superposition of $K$ Gaussians, as follows:
\begin{align}
    f(\boldsymbol{v}|\boldsymbol{\Phi})&=\sum_{k=1}^K A_k\mathcal{N}(\boldsymbol{v}|\boldsymbol{\mu}_k,\boldsymbol{\Sigma}_k)
    \label{eq:GMM_PDF}
\end{align}
here $\boldsymbol{v}$ is the velocity vector. $A_k$ corresponds to the weight of the $k$-th component of the Gaussian distribution. $A_k\in \left[0,1\right]$ is also called the mixing coefficient, it is an effective probability and thus the summation of $A_k$ over $k$ is 1, i.e., $\displaystyle\sum_k A_k=1$.

The multivariate Gaussian distribution $\mathcal{N}$ parameterized by the mean vector $\boldsymbol{\mu}_k$ and the covariance matrix $\boldsymbol{\Sigma}_k$ is defined as
\begin{align}
    \mathcal{N}(\boldsymbol{v}|\boldsymbol{\mu}_k,\boldsymbol{\Sigma}_k)=\frac{1}{\left(2\pi\right)^{3/2}}\big|\boldsymbol{\Sigma}_k\big|^{-1/2}\cdot\exp\left[-\frac{1}{2}\left(\boldsymbol{v}-\boldsymbol{\mu}_k\right)^T\boldsymbol{\Sigma}_k^{-1}\left(\boldsymbol{v}-\boldsymbol{\mu}_k\right)\right]
\end{align}
the mixture parameter $\boldsymbol{\Phi}=\{A_1,A_2,...A_K,\boldsymbol{\mu}_1,\boldsymbol{\mu}_2,...,\boldsymbol{\mu}_K,\boldsymbol{\Sigma}_1,\boldsymbol{\Sigma}_2,...,\boldsymbol{\Sigma}_K\}$ includes all the parameters of the GMM. 

The quantity $|A_i|^2$ represents physically the intensity of the $k$-th component of a Gaussian distribution, and $\boldsymbol{\mu}_k$ and $\boldsymbol{\Sigma}_k$ indicate the bulk flow velocity and thermal speed respectively.

Introducing a $K$-dimensional binary random latent variable $\boldsymbol{z}\in\mathcal{R}$. $\boldsymbol{z}$ has a $1$-of-$K$ representation, namely, in which a particular element of $\boldsymbol{z}$ is equal to 1 and other elements is zero, i.e., $z_k\in \{0,1\}$, and the sum of all $z_k$ is 1, i.e., $\displaystyle\sum_kz_k=1$.

The marginal distribution with respect to $\boldsymbol{z}$ is specified by the mixing coefficients $A_k$, namely
\begin{align}
    f(z_k=1)=A_k
\end{align}

Considering the $1$-of-$K$ representation of the latent variable $\boldsymbol{z}$, the distribution in terms of $\boldsymbol{z}$ thus can be written into the following form:
\begin{align}
    f(\boldsymbol{z})=\prod_k A_k^{z_k}
\end{align}

Given a particle value of the latent variable $\boldsymbol{z}$, for example, the $kth$ element of $\boldsymbol{z}$. The condition distribution with respect to the variable $\boldsymbol{v}$ obeys a Gaussian in the following form:
\begin{align}
    f(\boldsymbol{v}|z_k=1)=\mathcal{N}(\boldsymbol{v}|\boldsymbol{\mu}_k,\boldsymbol{\Sigma}_k)
\end{align}
the condition distribution in terms of variable $\boldsymbol{v}$ with a given $\boldsymbol{z}$ thus is
\begin{align}
    f\left(\boldsymbol{v}|\boldsymbol{z}\right)=\prod_k \mathcal{N}(\boldsymbol{v}|\boldsymbol{\mu}_k,\boldsymbol{\Sigma}_k)^{z_k}
\end{align}

The joint distribution function of variables $\boldsymbol{x}$ and $\boldsymbol{z}$ is obtained by the product rule as follows:
\begin{align}
    f\left(\boldsymbol{v},\boldsymbol{z}\right)=f\left(\boldsymbol{z}\right)f\left(\boldsymbol{v}|\boldsymbol{z}\right)
    \label{eq:GMM_PDF_joint}
\end{align}

Then the marginal distribution function with respect to variable $\boldsymbol{v}$ \myrefeq{eq:GMM_PDF} is obtained by summing (or integrating) the joint distribution function \myrefeq{eq:GMM_PDF_joint} over all possible states of the latent variable $\boldsymbol{z}$, namely
\begin{align}
    f\left(\boldsymbol{v}\right)=\sum_{\boldsymbol{z}}f\left(\boldsymbol{v},\boldsymbol{z}\right)=\prod_k A_k\mathcal{N}(\boldsymbol{v}|\boldsymbol{\mu}_k,\boldsymbol{\Sigma}_k)
    \label{eq:GMM_PDF_marginal}
\end{align}

It is helpful to estimate the parameters of the Gaussian mixture involved the variable $\boldsymbol{z}$ by the above marginal distribution function \myrefeq{eq:GMM_PDF_marginal}.

Given a set of data of the velocity, namely, $\boldsymbol{v}_1,\boldsymbol{v}_2,\cdots,\boldsymbol{v}_n$. Thus each data $\boldsymbol{v}_i$ is determined by a specific value of the latent variable $z_k$. 
It is convenient to introduce the condition probability with respect to variable $\boldsymbol{z}$ using the Bayes's theorem \citep[][]{Bishop2006}, namely
\begin{align}
    \gamma_{ki}=f\left(z_k=1|\boldsymbol{v}_i\right)=\frac{f(z_k=1)f(\boldsymbol{v}_i|z_k=1)}{f(\boldsymbol{v}_i)}=\frac{A_k\mathcal{N}(\boldsymbol{v}_i|\boldsymbol{\mu}_k,\boldsymbol{\Sigma}_k)}{\sum\limits_kA_k\mathcal{N}(\boldsymbol{v}_i|\boldsymbol{\mu}_k,\boldsymbol{\Sigma}_k)}
    \label{eq:GMM_responsibility}
\end{align}
The mixing probability $A_k$ is thus the prior probability of $z_k=1$, and the quantity $\gamma_{ki}$ can be treated as the corresponding posterior probability for a particular observed $\boldsymbol{v}$.

Assuming the observed data points $\boldsymbol{v}_1,\boldsymbol{v}_2,\cdots,\boldsymbol{v}_n$ are independent from the distribution, the log likelihood of the Gaussian mixture \myrefeq{eq:GMM_PDF} can be constructed by the data $\boldsymbol{v}$ as follows:
\begin{align}
    \mathcal{L}\left(\boldsymbol{v}|\boldsymbol{\Phi}\right)=\sum_i\ln \left[\sum\limits_kA_k\mathcal{N}(\boldsymbol{v}_i|\boldsymbol{\mu}_k,\boldsymbol{\Sigma}_k)\right]
    \label{eq:likelihood_log}
\end{align}

The parameters of the mixture Gaussian model are thus corresponding to a solution of the maximum likelihood, i.e., $\max\ \mathcal{L}\left(\boldsymbol{v}|\boldsymbol{\Phi}\right)$.

Letting the gradient of the likelihood function to be zero with respect to the mean $\boldsymbol{\mu}_k$, the covariance $\boldsymbol{\Sigma}_k$ and the weight (mixture property) $A_k$, yields, for arbitrary $k\in \left[1,2,\cdots,K\right]$, namely,
\begin{align}
    \boldsymbol{\mu}_k&=\qquad\qquad\frac{\sum\limits_i \gamma_{ki}\boldsymbol{v}_i}{\sum\limits_i \gamma_{ki}}\label{eq:GMM_mean}\\
    \boldsymbol{\Sigma}_k&=\frac{\sum\limits_i \gamma_{ki}\left(\boldsymbol{v}_i-\boldsymbol{\mu}_k\right)\left(\boldsymbol{v}_i-\boldsymbol{\mu}_k\right)^T}{\sum\limits_i \gamma_{ki}}\label{eq:GMM_covariance}\\
    A_k&=\qquad\qquad\frac{1}{n}\sum\limits_i\gamma_{ki}\label{eq:GMM_weight}
\end{align}

Given the observed data of $\boldsymbol{v}$, the posterior probability $\gamma_{ki}$ appears in the expressions above, thus it is viewed as the responsibility of the $kth$ component relate to the observation $\boldsymbol{v}$ \citep[][]{Bishop2006}.

Provided a Gaussian mixture model, an expectation-maximization algorithm \citep[or EM algorithm][]{Dempster1977,McLachlan2008} is developed to maximize the likelihood expression $\mathcal{L}$ as follows:
\begin{itemize}
    \item Step 1. Initializing the mean $\boldsymbol{\mu}_k$, the covariance $\boldsymbol{\Sigma}_k$ and the mixing probability $A_k$. Evaluating the initial value of the log likelihood $\mathcal{L}$.
    \item Step 2. Expectation step (E-step): computing the responsibility $\gamma_{ki}$ according to \myrefeq{eq:GMM_responsibility}.
    \item Step 3. Maximum step (M-step): re-estimating the value of the mean $\boldsymbol{\mu}_k$, the covariance $\boldsymbol{\Sigma}_k$ and the weight (mixture property) $A_k$ using \myrefeq{eq:GMM_mean}-\myrefeq{eq:GMM_weight} based on the $\gamma_{ki}$ obtained in the E step.
    \item Step 4. Estimating the log likelihood $\mathcal{L}$ using \myrefeq{eq:likelihood_log}. Then checking the convergence criterion, if the convergence criterion is not satisfied, then go to the E-step iteratively.
\end{itemize}

In order to compensate for the overfitting of GMM models, it is necessary to assess the appropriate number of sub-populations from the $K$-component Gaussian mixture models based on the model selection technique named Bayesian information criterion (BIC) \citep[][]{Burnham2004}, namely
\begin{align}
    BIC=K\cdot\ln N-2\ln \mathcal{L}
\end{align}
where $N$ is the sample size of the observation $\boldsymbol{v}$, $K$ denotes $K$ components Gaussian mixture model obtained from training of the data by the EM algorithm mentioned above.
By introducing this penalty term associated with the number of parameters of the GMM model, this model selection method can automatically determine the appropriate number of sub-populations \citep[][]{Bishop2006}.


\begin{thebibliography}{106}%
    \makeatletter
    \providecommand \@ifxundefined [1]{%
     \@ifx{#1\undefined}
    }%
    \providecommand \@ifnum [1]{%
     \ifnum #1\expandafter \@firstoftwo
     \else \expandafter \@secondoftwo
     \fi
    }%
    \providecommand \@ifx [1]{%
     \ifx #1\expandafter \@firstoftwo
     \else \expandafter \@secondoftwo
     \fi
    }%
    \providecommand \natexlab [1]{#1}%
    \providecommand \enquote  [1]{``#1''}%
    \providecommand \bibnamefont  [1]{#1}%
    \providecommand \bibfnamefont [1]{#1}%
    \providecommand \citenamefont [1]{#1}%
    \providecommand \href@noop [0]{\@secondoftwo}%
    \providecommand \href [0]{\begingroup \@sanitize@url \@href}%
    \providecommand \@href[1]{\@@startlink{#1}\@@href}%
    \providecommand \@@href[1]{\endgroup#1\@@endlink}%
    \providecommand \@sanitize@url [0]{\catcode `\\12\catcode `\$12\catcode
      `\&12\catcode `\#12\catcode `\^12\catcode `\_12\catcode `\%12\relax}%
    \providecommand \@@startlink[1]{}%
    \providecommand \@@endlink[0]{}%
    \providecommand \url  [0]{\begingroup\@sanitize@url \@url }%
    \providecommand \@url [1]{\endgroup\@href {#1}{\urlprefix }}%
    \providecommand \urlprefix  [0]{URL }%
    \providecommand \Eprint [0]{\href }%
    \providecommand \doibase [0]{http://dx.doi.org/}%
    \providecommand \selectlanguage [0]{\@gobble}%
    \providecommand \bibinfo  [0]{\@secondoftwo}%
    \providecommand \bibfield  [0]{\@secondoftwo}%
    \providecommand \translation [1]{[#1]}%
    \providecommand \BibitemOpen [0]{}%
    \providecommand \bibitemStop [0]{}%
    \providecommand \bibitemNoStop [0]{.\EOS\space}%
    \providecommand \EOS [0]{\spacefactor3000\relax}%
    \providecommand \BibitemShut  [1]{\csname bibitem#1\endcsname}%
    \let\auto@bib@innerbib\@empty
    \bibitem [{\citenamefont {Birn}\ and\ \citenamefont {Priest}(2007)}]{Birn2007}%
      \BibitemOpen
      \bibfield  {author} {\bibinfo {author} {\bibfnamefont {J.}~\bibnamefont
      {Birn}}\ and\ \bibinfo {author} {\bibfnamefont {E.~R.}\ \bibnamefont
      {Priest}},\ }\href {\doibase 10.1017/CBO9780511536151} {\emph {\bibinfo
      {title} {{Reconnection of Magnetic Fields: magnetohydrodynamics and
      collisionless theory and observations}}}},\ edited by\ \bibinfo {editor}
      {\bibfnamefont {J.}~\bibnamefont {Birn}}\ and\ \bibinfo {editor}
      {\bibfnamefont {E.~R.}\ \bibnamefont {Priest}}\ (\bibinfo  {publisher}
      {Cambridge University Press},\ \bibinfo {address} {Cambridge},\ \bibinfo
      {year} {2007})\BibitemShut {NoStop}%
    \bibitem [{\citenamefont {Mann}, \citenamefont {Warmuth},\ and\ \citenamefont
      {Aurass}(2009)}]{Mann2009}%
      \BibitemOpen
      \bibfield  {author} {\bibinfo {author} {\bibfnamefont {G.}~\bibnamefont
      {Mann}}, \bibinfo {author} {\bibfnamefont {A.}~\bibnamefont {Warmuth}}, \
      and\ \bibinfo {author} {\bibfnamefont {H.}~\bibnamefont {Aurass}},\ }\href
      {\doibase 10.1051/0004-6361:200810099} {\bibfield  {journal} {\bibinfo
      {journal} {Astron. Astrophys.}\ }\textbf {\bibinfo {volume} {494}},\ \bibinfo
      {pages} {669} (\bibinfo {year} {2009})}\BibitemShut {NoStop}%
    \bibitem [{\citenamefont {Treumann}\ and\ \citenamefont
      {Baumjohann}(2013)}]{Treumann2013b}%
      \BibitemOpen
      \bibfield  {author} {\bibinfo {author} {\bibfnamefont {R.~A.}\ \bibnamefont
      {Treumann}}\ and\ \bibinfo {author} {\bibfnamefont {W.}~\bibnamefont
      {Baumjohann}},\ }\href {\doibase 10.3389/fphy.2013.00031} {\bibfield
      {journal} {\bibinfo  {journal} {Front. Phys.}\ }\textbf {\bibinfo {volume}
      {1}},\ \bibinfo {pages} {1} (\bibinfo {year} {2013})}\BibitemShut {NoStop}%
    \bibitem [{\citenamefont {Demars}\ and\ \citenamefont
      {Schunk}(1990)}]{Demars1990}%
      \BibitemOpen
      \bibfield  {author} {\bibinfo {author} {\bibfnamefont {H.}~\bibnamefont
      {Demars}}\ and\ \bibinfo {author} {\bibfnamefont {R.}~\bibnamefont
      {Schunk}},\ }\href {\doibase 10.1016/0032-0633(90)90018-L} {\bibfield
      {journal} {\bibinfo  {journal} {Planet. Space Sci.}\ }\textbf {\bibinfo
      {volume} {38}},\ \bibinfo {pages} {1091} (\bibinfo {year}
      {1990})}\BibitemShut {NoStop}%
    \bibitem [{\citenamefont {Marsch}\ and\ \citenamefont
      {Bourouaine}(2011)}]{Marsch2011}%
      \BibitemOpen
      \bibfield  {author} {\bibinfo {author} {\bibfnamefont {E.}~\bibnamefont
      {Marsch}}\ and\ \bibinfo {author} {\bibfnamefont {S.}~\bibnamefont
      {Bourouaine}},\ }\href {\doibase 10.5194/angeo-29-2089-2011} {\bibfield
      {journal} {\bibinfo  {journal} {Ann. Geophys.}\ }\textbf {\bibinfo {volume}
      {29}},\ \bibinfo {pages} {2089} (\bibinfo {year} {2011})}\BibitemShut
      {NoStop}%
    \bibitem [{\citenamefont {Pierrard}\ \emph {et~al.}(2010)\citenamefont
      {Pierrard}, \citenamefont {Voitenko}, \citenamefont {Maksimovic},
      \citenamefont {Issautier}, \citenamefont {Meyer-Vernet}, \citenamefont
      {Moncuquet},\ and\ \citenamefont {Pantellini}}]{Pierrard2010}%
      \BibitemOpen
      \bibfield  {author} {\bibinfo {author} {\bibfnamefont {V.}~\bibnamefont
      {Pierrard}}, \bibinfo {author} {\bibfnamefont {Y.}~\bibnamefont {Voitenko}},
      \bibinfo {author} {\bibfnamefont {M.}~\bibnamefont {Maksimovic}}, \bibinfo
      {author} {\bibfnamefont {K.}~\bibnamefont {Issautier}}, \bibinfo {author}
      {\bibfnamefont {N.}~\bibnamefont {Meyer-Vernet}}, \bibinfo {author}
      {\bibfnamefont {M.}~\bibnamefont {Moncuquet}}, \ and\ \bibinfo {author}
      {\bibfnamefont {F.}~\bibnamefont {Pantellini}},\ }in\ \href {\doibase
      10.1063/1.3395812} {\emph {\bibinfo {booktitle} {AIP Conf. Proc.}}},\ Vol.\
      \bibinfo {volume} {1216}\ (\bibinfo {year} {2010})\ pp.\ \bibinfo {pages}
      {102--105}\BibitemShut {NoStop}%
    \bibitem [{\citenamefont {Vaisberg}\ \emph {et~al.}(2004)\citenamefont
      {Vaisberg}, \citenamefont {Avanov}, \citenamefont {Moore},\ and\
      \citenamefont {Smirnov}}]{Vaisberg2004}%
      \BibitemOpen
      \bibfield  {author} {\bibinfo {author} {\bibfnamefont {O.~L.}\ \bibnamefont
      {Vaisberg}}, \bibinfo {author} {\bibfnamefont {L.~A.}\ \bibnamefont
      {Avanov}}, \bibinfo {author} {\bibfnamefont {T.~E.}\ \bibnamefont {Moore}}, \
      and\ \bibinfo {author} {\bibfnamefont {V.~N.}\ \bibnamefont {Smirnov}},\
      }\href {\doibase 10.5194/angeo-22-213-2004} {\bibfield  {journal} {\bibinfo
      {journal} {Ann. Geophys.}\ }\textbf {\bibinfo {volume} {22}},\ \bibinfo
      {pages} {213} (\bibinfo {year} {2004})}\BibitemShut {NoStop}%
    \bibitem [{\citenamefont {Perri}\ \emph {et~al.}(2020)\citenamefont {Perri},
      \citenamefont {Perrone}, \citenamefont {Yordanova}, \citenamefont
      {Sorriso-Valvo}, \citenamefont {Paterson}, \citenamefont {Gershman},
      \citenamefont {Giles}, \citenamefont {Pollock}, \citenamefont {Dorelli},
      \citenamefont {Avanov}, \citenamefont {Lavraud}, \citenamefont {Saito},
      \citenamefont {Nakamura}, \citenamefont {Fischer}, \citenamefont
      {Baumjohann}, \citenamefont {Plaschke}, \citenamefont {Narita}, \citenamefont
      {Magnes}, \citenamefont {Russell}, \citenamefont {Strangeway}, \citenamefont
      {Contel}, \citenamefont {Khotyaintsev},\ and\ \citenamefont
      {Valentini}}]{Perri2020}%
      \BibitemOpen
      \bibfield  {author} {\bibinfo {author} {\bibfnamefont {S.}~\bibnamefont
      {Perri}}, \bibinfo {author} {\bibfnamefont {D.}~\bibnamefont {Perrone}},
      \bibinfo {author} {\bibfnamefont {E.}~\bibnamefont {Yordanova}}, \bibinfo
      {author} {\bibfnamefont {L.}~\bibnamefont {Sorriso-Valvo}}, \bibinfo {author}
      {\bibfnamefont {W.~R.}\ \bibnamefont {Paterson}}, \bibinfo {author}
      {\bibfnamefont {D.~J.}\ \bibnamefont {Gershman}}, \bibinfo {author}
      {\bibfnamefont {B.~L.}\ \bibnamefont {Giles}}, \bibinfo {author}
      {\bibfnamefont {C.~J.}\ \bibnamefont {Pollock}}, \bibinfo {author}
      {\bibfnamefont {J.~C.}\ \bibnamefont {Dorelli}}, \bibinfo {author}
      {\bibfnamefont {L.~A.}\ \bibnamefont {Avanov}}, \bibinfo {author}
      {\bibfnamefont {B.}~\bibnamefont {Lavraud}}, \bibinfo {author} {\bibfnamefont
      {Y.}~\bibnamefont {Saito}}, \bibinfo {author} {\bibfnamefont
      {R.}~\bibnamefont {Nakamura}}, \bibinfo {author} {\bibfnamefont
      {D.}~\bibnamefont {Fischer}}, \bibinfo {author} {\bibfnamefont
      {W.}~\bibnamefont {Baumjohann}}, \bibinfo {author} {\bibfnamefont
      {F.}~\bibnamefont {Plaschke}}, \bibinfo {author} {\bibfnamefont
      {Y.}~\bibnamefont {Narita}}, \bibinfo {author} {\bibfnamefont
      {W.}~\bibnamefont {Magnes}}, \bibinfo {author} {\bibfnamefont {C.~T.}\
      \bibnamefont {Russell}}, \bibinfo {author} {\bibfnamefont {R.~J.}\
      \bibnamefont {Strangeway}}, \bibinfo {author} {\bibfnamefont {O.~L.}\
      \bibnamefont {Contel}}, \bibinfo {author} {\bibfnamefont {Y.}~\bibnamefont
      {Khotyaintsev}}, \ and\ \bibinfo {author} {\bibfnamefont {F.}~\bibnamefont
      {Valentini}},\ }\href {\doibase 10.1017/S0022377820000021} {\bibfield
      {journal} {\bibinfo  {journal} {J. Plasma Phys.}\ ,\ \bibinfo {pages} {1}}
      (\bibinfo {year} {2020})},\ \Eprint {http://arxiv.org/abs/1905.09466}
      {arXiv:1905.09466} \BibitemShut {NoStop}%
    \bibitem [{\citenamefont {Runov}\ \emph {et~al.}(2021)\citenamefont {Runov},
      \citenamefont {Grandin}, \citenamefont {Palmroth}, \citenamefont {Battarbee},
      \citenamefont {Ganse}, \citenamefont {Hietala}, \citenamefont {Hoilijoki},
      \citenamefont {Kilpua}, \citenamefont {Pfau-Kempf}, \citenamefont
      {Toledo-Redondo}, \citenamefont {Turc},\ and\ \citenamefont
      {Turner}}]{Runov2021}%
      \BibitemOpen
      \bibfield  {author} {\bibinfo {author} {\bibfnamefont {A.}~\bibnamefont
      {Runov}}, \bibinfo {author} {\bibfnamefont {M.}~\bibnamefont {Grandin}},
      \bibinfo {author} {\bibfnamefont {M.}~\bibnamefont {Palmroth}}, \bibinfo
      {author} {\bibfnamefont {M.}~\bibnamefont {Battarbee}}, \bibinfo {author}
      {\bibfnamefont {U.}~\bibnamefont {Ganse}}, \bibinfo {author} {\bibfnamefont
      {H.}~\bibnamefont {Hietala}}, \bibinfo {author} {\bibfnamefont
      {S.}~\bibnamefont {Hoilijoki}}, \bibinfo {author} {\bibfnamefont
      {E.}~\bibnamefont {Kilpua}}, \bibinfo {author} {\bibfnamefont
      {Y.}~\bibnamefont {Pfau-Kempf}}, \bibinfo {author} {\bibfnamefont
      {S.}~\bibnamefont {Toledo-Redondo}}, \bibinfo {author} {\bibfnamefont
      {L.}~\bibnamefont {Turc}}, \ and\ \bibinfo {author} {\bibfnamefont
      {D.}~\bibnamefont {Turner}},\ }\href {\doibase 10.5194/angeo-39-599-2021}
      {\bibfield  {journal} {\bibinfo  {journal} {Ann. Geophys.}\ }\textbf
      {\bibinfo {volume} {39}},\ \bibinfo {pages} {599} (\bibinfo {year}
      {2021})}\BibitemShut {NoStop}%
    \bibitem [{\citenamefont {Lin}(1998)}]{Lin1998}%
      \BibitemOpen
      \bibfield  {author} {\bibinfo {author} {\bibfnamefont {R.~P.}\ \bibnamefont
      {Lin}},\ }in\ \href {\doibase 10.1007/978-94-011-4762-0_4} {\emph {\bibinfo
      {booktitle} {Adv. Compos. Explor. Mission}}},\ Vol.~\bibinfo {volume} {86}\
      (\bibinfo  {publisher} {Springer Netherlands},\ \bibinfo {address}
      {Dordrecht},\ \bibinfo {year} {1998})\ pp.\ \bibinfo {pages}
      {61--78}\BibitemShut {NoStop}%
    \bibitem [{\citenamefont {Maksimovic}\ \emph {et~al.}(1999)\citenamefont
      {Maksimovic}, \citenamefont {Pierrard}, \citenamefont {Lemaire},\ and\
      \citenamefont {Larson}}]{Maksimovic2008b}%
      \BibitemOpen
      \bibfield  {author} {\bibinfo {author} {\bibfnamefont {M.}~\bibnamefont
      {Maksimovic}}, \bibinfo {author} {\bibfnamefont {V.}~\bibnamefont
      {Pierrard}}, \bibinfo {author} {\bibfnamefont {J.}~\bibnamefont {Lemaire}}, \
      and\ \bibinfo {author} {\bibfnamefont {D.}~\bibnamefont {Larson}},\ }in\
      \href {\doibase 10.1063/1.58759} {\emph {\bibinfo {booktitle} {AIP Conf.
      Proc.}}},\ Vol.\ \bibinfo {volume} {104}\ (\bibinfo  {publisher} {AIP},\
      \bibinfo {year} {1999})\ pp.\ \bibinfo {pages} {267--268}\BibitemShut
      {NoStop}%
    \bibitem [{\citenamefont {Burch}\ \emph {et~al.}(2016)\citenamefont {Burch},
      \citenamefont {Torbert}, \citenamefont {Phan}, \citenamefont {Chen},
      \citenamefont {Moore}, \citenamefont {Ergun}, \citenamefont {Eastwood},
      \citenamefont {Gershman}, \citenamefont {Cassak}, \citenamefont {Argall},
      \citenamefont {Wang}, \citenamefont {Hesse}, \citenamefont {Pollock},
      \citenamefont {Giles}, \citenamefont {Nakamura}, \citenamefont {Mauk},
      \citenamefont {Fuselier}, \citenamefont {Russell}, \citenamefont
      {Strangeway}, \citenamefont {Drake}, \citenamefont {Shay}, \citenamefont
      {Khotyaintsev}, \citenamefont {Lindqvist}, \citenamefont {Marklund},
      \citenamefont {Wilder}, \citenamefont {Young}, \citenamefont {Torkar},
      \citenamefont {Goldstein}, \citenamefont {Dorelli}, \citenamefont {Avanov},
      \citenamefont {Oka}, \citenamefont {Baker}, \citenamefont {Jaynes},
      \citenamefont {Goodrich}, \citenamefont {Cohen}, \citenamefont {Turner},
      \citenamefont {Fennell}, \citenamefont {Blake}, \citenamefont {Clemmons},
      \citenamefont {Goldman}, \citenamefont {Newman}, \citenamefont {Petrinec},
      \citenamefont {Trattner}, \citenamefont {Lavraud}, \citenamefont {Reiff},
      \citenamefont {Baumjohann}, \citenamefont {Magnes}, \citenamefont {Steller},
      \citenamefont {Lewis}, \citenamefont {Saito}, \citenamefont {Coffey},\ and\
      \citenamefont {Chandler}}]{Burch2016}%
      \BibitemOpen
      \bibfield  {author} {\bibinfo {author} {\bibfnamefont {J.~L.}\ \bibnamefont
      {Burch}}, \bibinfo {author} {\bibfnamefont {R.~B.}\ \bibnamefont {Torbert}},
      \bibinfo {author} {\bibfnamefont {T.~D.}\ \bibnamefont {Phan}}, \bibinfo
      {author} {\bibfnamefont {L.-J.}\ \bibnamefont {Chen}}, \bibinfo {author}
      {\bibfnamefont {T.~E.}\ \bibnamefont {Moore}}, \bibinfo {author}
      {\bibfnamefont {R.~E.}\ \bibnamefont {Ergun}}, \bibinfo {author}
      {\bibfnamefont {J.~P.}\ \bibnamefont {Eastwood}}, \bibinfo {author}
      {\bibfnamefont {D.~J.}\ \bibnamefont {Gershman}}, \bibinfo {author}
      {\bibfnamefont {P.~A.}\ \bibnamefont {Cassak}}, \bibinfo {author}
      {\bibfnamefont {M.~R.}\ \bibnamefont {Argall}}, \bibinfo {author}
      {\bibfnamefont {S.}~\bibnamefont {Wang}}, \bibinfo {author} {\bibfnamefont
      {M.}~\bibnamefont {Hesse}}, \bibinfo {author} {\bibfnamefont {C.~J.}\
      \bibnamefont {Pollock}}, \bibinfo {author} {\bibfnamefont {B.~L.}\
      \bibnamefont {Giles}}, \bibinfo {author} {\bibfnamefont {R.}~\bibnamefont
      {Nakamura}}, \bibinfo {author} {\bibfnamefont {B.~H.}\ \bibnamefont {Mauk}},
      \bibinfo {author} {\bibfnamefont {S.~A.}\ \bibnamefont {Fuselier}}, \bibinfo
      {author} {\bibfnamefont {C.~T.}\ \bibnamefont {Russell}}, \bibinfo {author}
      {\bibfnamefont {R.~J.}\ \bibnamefont {Strangeway}}, \bibinfo {author}
      {\bibfnamefont {J.~F.}\ \bibnamefont {Drake}}, \bibinfo {author}
      {\bibfnamefont {M.~A.}\ \bibnamefont {Shay}}, \bibinfo {author}
      {\bibfnamefont {Y.~V.}\ \bibnamefont {Khotyaintsev}}, \bibinfo {author}
      {\bibfnamefont {P.-A.}\ \bibnamefont {Lindqvist}}, \bibinfo {author}
      {\bibfnamefont {G.}~\bibnamefont {Marklund}}, \bibinfo {author}
      {\bibfnamefont {F.~D.}\ \bibnamefont {Wilder}}, \bibinfo {author}
      {\bibfnamefont {D.~T.}\ \bibnamefont {Young}}, \bibinfo {author}
      {\bibfnamefont {K.}~\bibnamefont {Torkar}}, \bibinfo {author} {\bibfnamefont
      {J.}~\bibnamefont {Goldstein}}, \bibinfo {author} {\bibfnamefont {J.~C.}\
      \bibnamefont {Dorelli}}, \bibinfo {author} {\bibfnamefont {L.~A.}\
      \bibnamefont {Avanov}}, \bibinfo {author} {\bibfnamefont {M.}~\bibnamefont
      {Oka}}, \bibinfo {author} {\bibfnamefont {D.~N.}\ \bibnamefont {Baker}},
      \bibinfo {author} {\bibfnamefont {A.~N.}\ \bibnamefont {Jaynes}}, \bibinfo
      {author} {\bibfnamefont {K.~A.}\ \bibnamefont {Goodrich}}, \bibinfo {author}
      {\bibfnamefont {I.~J.}\ \bibnamefont {Cohen}}, \bibinfo {author}
      {\bibfnamefont {D.~L.}\ \bibnamefont {Turner}}, \bibinfo {author}
      {\bibfnamefont {J.~F.}\ \bibnamefont {Fennell}}, \bibinfo {author}
      {\bibfnamefont {J.~B.}\ \bibnamefont {Blake}}, \bibinfo {author}
      {\bibfnamefont {J.}~\bibnamefont {Clemmons}}, \bibinfo {author}
      {\bibfnamefont {M.}~\bibnamefont {Goldman}}, \bibinfo {author} {\bibfnamefont
      {D.}~\bibnamefont {Newman}}, \bibinfo {author} {\bibfnamefont {S.~M.}\
      \bibnamefont {Petrinec}}, \bibinfo {author} {\bibfnamefont {K.~J.}\
      \bibnamefont {Trattner}}, \bibinfo {author} {\bibfnamefont {B.}~\bibnamefont
      {Lavraud}}, \bibinfo {author} {\bibfnamefont {P.~H.}\ \bibnamefont {Reiff}},
      \bibinfo {author} {\bibfnamefont {W.}~\bibnamefont {Baumjohann}}, \bibinfo
      {author} {\bibfnamefont {W.}~\bibnamefont {Magnes}}, \bibinfo {author}
      {\bibfnamefont {M.}~\bibnamefont {Steller}}, \bibinfo {author} {\bibfnamefont
      {W.}~\bibnamefont {Lewis}}, \bibinfo {author} {\bibfnamefont
      {Y.}~\bibnamefont {Saito}}, \bibinfo {author} {\bibfnamefont
      {V.}~\bibnamefont {Coffey}}, \ and\ \bibinfo {author} {\bibfnamefont
      {M.}~\bibnamefont {Chandler}},\ }\href {\doibase 10.1126/science.aaf2939}
      {\bibfield  {journal} {\bibinfo  {journal} {Science}\ }\textbf {\bibinfo
      {volume} {352}},\ \bibinfo {pages} {aaf2939} (\bibinfo {year}
      {2016})}\BibitemShut {NoStop}%
    \bibitem [{\citenamefont {Burch}\ and\ \citenamefont
      {Phan}(2016)}]{Burch2016b}%
      \BibitemOpen
      \bibfield  {author} {\bibinfo {author} {\bibfnamefont {J.~L.}\ \bibnamefont
      {Burch}}\ and\ \bibinfo {author} {\bibfnamefont {T.~D.}\ \bibnamefont
      {Phan}},\ }\href {\doibase 10.1002/2016GL069787} {\bibfield  {journal}
      {\bibinfo  {journal} {Geophys. Res. Lett.}\ }\textbf {\bibinfo {volume}
      {43}},\ \bibinfo {pages} {8327} (\bibinfo {year} {2016})}\BibitemShut
      {NoStop}%
    \bibitem [{\citenamefont {Ergun}\ \emph {et~al.}(2018)\citenamefont {Ergun},
      \citenamefont {Goodrich}, \citenamefont {Wilder}, \citenamefont {Ahmadi},
      \citenamefont {Holmes}, \citenamefont {Eriksson}, \citenamefont {Stawarz},
      \citenamefont {Nakamura}, \citenamefont {Genestreti}, \citenamefont {Hesse},
      \citenamefont {Burch}, \citenamefont {Torbert}, \citenamefont {Phan},
      \citenamefont {Schwartz}, \citenamefont {Eastwood}, \citenamefont
      {Strangeway}, \citenamefont {{Le Contel}}, \citenamefont {Russell},
      \citenamefont {Argall}, \citenamefont {Lindqvist}, \citenamefont {Chen},
      \citenamefont {Cassak}, \citenamefont {Giles}, \citenamefont {Dorelli},
      \citenamefont {Gershman}, \citenamefont {Leonard}, \citenamefont {Lavraud},
      \citenamefont {Retino}, \citenamefont {Matthaeus},\ and\ \citenamefont
      {Vaivads}}]{Ergun2018}%
      \BibitemOpen
      \bibfield  {author} {\bibinfo {author} {\bibfnamefont {R.~E.}\ \bibnamefont
      {Ergun}}, \bibinfo {author} {\bibfnamefont {K.~A.}\ \bibnamefont {Goodrich}},
      \bibinfo {author} {\bibfnamefont {F.~D.}\ \bibnamefont {Wilder}}, \bibinfo
      {author} {\bibfnamefont {N.}~\bibnamefont {Ahmadi}}, \bibinfo {author}
      {\bibfnamefont {J.~C.}\ \bibnamefont {Holmes}}, \bibinfo {author}
      {\bibfnamefont {S.}~\bibnamefont {Eriksson}}, \bibinfo {author}
      {\bibfnamefont {J.~E.}\ \bibnamefont {Stawarz}}, \bibinfo {author}
      {\bibfnamefont {R.}~\bibnamefont {Nakamura}}, \bibinfo {author}
      {\bibfnamefont {K.~J.}\ \bibnamefont {Genestreti}}, \bibinfo {author}
      {\bibfnamefont {M.}~\bibnamefont {Hesse}}, \bibinfo {author} {\bibfnamefont
      {J.~L.}\ \bibnamefont {Burch}}, \bibinfo {author} {\bibfnamefont {R.~B.}\
      \bibnamefont {Torbert}}, \bibinfo {author} {\bibfnamefont {T.~D.}\
      \bibnamefont {Phan}}, \bibinfo {author} {\bibfnamefont {S.~J.}\ \bibnamefont
      {Schwartz}}, \bibinfo {author} {\bibfnamefont {J.~P.}\ \bibnamefont
      {Eastwood}}, \bibinfo {author} {\bibfnamefont {R.~J.}\ \bibnamefont
      {Strangeway}}, \bibinfo {author} {\bibfnamefont {O.}~\bibnamefont {{Le
      Contel}}}, \bibinfo {author} {\bibfnamefont {C.~T.}\ \bibnamefont {Russell}},
      \bibinfo {author} {\bibfnamefont {M.~R.}\ \bibnamefont {Argall}}, \bibinfo
      {author} {\bibfnamefont {P.}~\bibnamefont {Lindqvist}}, \bibinfo {author}
      {\bibfnamefont {L.~J.}\ \bibnamefont {Chen}}, \bibinfo {author}
      {\bibfnamefont {P.~A.}\ \bibnamefont {Cassak}}, \bibinfo {author}
      {\bibfnamefont {B.~L.}\ \bibnamefont {Giles}}, \bibinfo {author}
      {\bibfnamefont {J.~C.}\ \bibnamefont {Dorelli}}, \bibinfo {author}
      {\bibfnamefont {D.}~\bibnamefont {Gershman}}, \bibinfo {author}
      {\bibfnamefont {T.~W.}\ \bibnamefont {Leonard}}, \bibinfo {author}
      {\bibfnamefont {B.}~\bibnamefont {Lavraud}}, \bibinfo {author} {\bibfnamefont
      {A.}~\bibnamefont {Retino}}, \bibinfo {author} {\bibfnamefont
      {W.}~\bibnamefont {Matthaeus}}, \ and\ \bibinfo {author} {\bibfnamefont
      {A.}~\bibnamefont {Vaivads}},\ }\href {\doibase 10.1002/2018GL076993}
      {\bibfield  {journal} {\bibinfo  {journal} {Geophys. Res. Lett.}\ }\textbf
      {\bibinfo {volume} {45}},\ \bibinfo {pages} {3338} (\bibinfo {year}
      {2018})}\BibitemShut {NoStop}%
    \bibitem [{\citenamefont {Hoshino}, \citenamefont {Hiraide},\ and\
      \citenamefont {Mukai}(2001)}]{Hoshino2001}%
      \BibitemOpen
      \bibfield  {author} {\bibinfo {author} {\bibfnamefont {M.}~\bibnamefont
      {Hoshino}}, \bibinfo {author} {\bibfnamefont {K.}~\bibnamefont {Hiraide}}, \
      and\ \bibinfo {author} {\bibfnamefont {T.}~\bibnamefont {Mukai}},\ }\href
      {\doibase 10.1186/BF03353282} {\bibfield  {journal} {\bibinfo  {journal}
      {Earth Planets Space}\ }\textbf {\bibinfo {volume} {53}},\ \bibinfo {pages}
      {627} (\bibinfo {year} {2001})}\BibitemShut {NoStop}%
    \bibitem [{\citenamefont {Drake}\ \emph {et~al.}(2003)\citenamefont {Drake},
      \citenamefont {Swisdak}, \citenamefont {Cattell}, \citenamefont {Shay},
      \citenamefont {Rogers},\ and\ \citenamefont {Zeiler}}]{Drake2003}%
      \BibitemOpen
      \bibfield  {author} {\bibinfo {author} {\bibfnamefont {J.}~\bibnamefont
      {Drake}}, \bibinfo {author} {\bibfnamefont {M.}~\bibnamefont {Swisdak}},
      \bibinfo {author} {\bibfnamefont {C.}~\bibnamefont {Cattell}}, \bibinfo
      {author} {\bibfnamefont {M.}~\bibnamefont {Shay}}, \bibinfo {author}
      {\bibfnamefont {B.}~\bibnamefont {Rogers}}, \ and\ \bibinfo {author}
      {\bibfnamefont {A.}~\bibnamefont {Zeiler}},\ }\href {\doibase
      10.1126/science.1080333} {\bibfield  {journal} {\bibinfo  {journal}
      {Science}\ }\textbf {\bibinfo {volume} {299}},\ \bibinfo {pages} {873}
      (\bibinfo {year} {2003})}\BibitemShut {NoStop}%
    \bibitem [{\citenamefont {Che}\ \emph {et~al.}(2010)\citenamefont {Che},
      \citenamefont {Drake}, \citenamefont {Swisdak},\ and\ \citenamefont
      {Yoon}}]{Che2010}%
      \BibitemOpen
      \bibfield  {author} {\bibinfo {author} {\bibfnamefont {H.}~\bibnamefont
      {Che}}, \bibinfo {author} {\bibfnamefont {J.~F.}\ \bibnamefont {Drake}},
      \bibinfo {author} {\bibfnamefont {M.}~\bibnamefont {Swisdak}}, \ and\
      \bibinfo {author} {\bibfnamefont {P.~H.}\ \bibnamefont {Yoon}},\ }\href
      {\doibase 10.1029/2010GL043608} {\bibfield  {journal} {\bibinfo  {journal}
      {Geophys. Res. Lett.}\ }\textbf {\bibinfo {volume} {37}},\ \bibinfo {pages}
      {L11105} (\bibinfo {year} {2010})}\BibitemShut {NoStop}%
    \bibitem [{\citenamefont {Ng}\ \emph {et~al.}(2011)\citenamefont {Ng},
      \citenamefont {Egedal}, \citenamefont {Le}, \citenamefont {Daughton},\ and\
      \citenamefont {Chen}}]{Ng2011}%
      \BibitemOpen
      \bibfield  {author} {\bibinfo {author} {\bibfnamefont {J.}~\bibnamefont
      {Ng}}, \bibinfo {author} {\bibfnamefont {J.}~\bibnamefont {Egedal}}, \bibinfo
      {author} {\bibfnamefont {A.}~\bibnamefont {Le}}, \bibinfo {author}
      {\bibfnamefont {W.}~\bibnamefont {Daughton}}, \ and\ \bibinfo {author}
      {\bibfnamefont {L.-J.}\ \bibnamefont {Chen}},\ }\href {\doibase
      10.1103/PhysRevLett.106.065002} {\bibfield  {journal} {\bibinfo  {journal}
      {Phys. Rev. Lett.}\ }\textbf {\bibinfo {volume} {106}},\ \bibinfo {pages}
      {065002} (\bibinfo {year} {2011})}\BibitemShut {NoStop}%
    \bibitem [{\citenamefont {Fujimoto}(2014)}]{Fujimoto2014a}%
      \BibitemOpen
      \bibfield  {author} {\bibinfo {author} {\bibfnamefont {K.}~\bibnamefont
      {Fujimoto}},\ }\href {\doibase 10.1002/2014GL059893} {\bibfield  {journal}
      {\bibinfo  {journal} {Geophys. Res. Lett.}\ }\textbf {\bibinfo {volume}
      {41}},\ \bibinfo {pages} {2721} (\bibinfo {year} {2014})}\BibitemShut
      {NoStop}%
    \bibitem [{\citenamefont {Bessho}, \citenamefont {Chen},\ and\ \citenamefont
      {Hesse}(2016)}]{Bessho2016}%
      \BibitemOpen
      \bibfield  {author} {\bibinfo {author} {\bibfnamefont {N.}~\bibnamefont
      {Bessho}}, \bibinfo {author} {\bibfnamefont {L.-J.}\ \bibnamefont {Chen}}, \
      and\ \bibinfo {author} {\bibfnamefont {M.}~\bibnamefont {Hesse}},\ }\href
      {\doibase 10.1002/2016GL067886} {\bibfield  {journal} {\bibinfo  {journal}
      {Geophys. Res. Lett.}\ }\textbf {\bibinfo {volume} {43}},\ \bibinfo {pages}
      {1828} (\bibinfo {year} {2016})}\BibitemShut {NoStop}%
    \bibitem [{\citenamefont {Mu{\~{n}}oz}\ and\ \citenamefont
      {B{\"{u}}chner}(2016)}]{Munoz2016}%
      \BibitemOpen
      \bibfield  {author} {\bibinfo {author} {\bibfnamefont {P.~A.}\ \bibnamefont
      {Mu{\~{n}}oz}}\ and\ \bibinfo {author} {\bibfnamefont {J.}~\bibnamefont
      {B{\"{u}}chner}},\ }\href {\doibase 10.1063/1.4963773} {\bibfield  {journal}
      {\bibinfo  {journal} {Phys. Plasmas}\ }\textbf {\bibinfo {volume} {23}},\
      \bibinfo {pages} {102103} (\bibinfo {year} {2016})}\BibitemShut {NoStop}%
    \bibitem [{\citenamefont {Egedal}\ \emph
      {et~al.}(2016{\natexlab{a}})\citenamefont {Egedal}, \citenamefont {Le},
      \citenamefont {Daughton}, \citenamefont {Wetherton}, \citenamefont {Cassak},
      \citenamefont {Chen}, \citenamefont {Lavraud}, \citenamefont {Torbert},
      \citenamefont {Dorelli}, \citenamefont {Gershman},\ and\ \citenamefont
      {Avanov}}]{Egedal2016}%
      \BibitemOpen
      \bibfield  {author} {\bibinfo {author} {\bibfnamefont {J.}~\bibnamefont
      {Egedal}}, \bibinfo {author} {\bibfnamefont {A.}~\bibnamefont {Le}}, \bibinfo
      {author} {\bibfnamefont {W.}~\bibnamefont {Daughton}}, \bibinfo {author}
      {\bibfnamefont {B.}~\bibnamefont {Wetherton}}, \bibinfo {author}
      {\bibfnamefont {P.~A.}\ \bibnamefont {Cassak}}, \bibinfo {author}
      {\bibfnamefont {L.-J.}\ \bibnamefont {Chen}}, \bibinfo {author}
      {\bibfnamefont {B.}~\bibnamefont {Lavraud}}, \bibinfo {author} {\bibfnamefont
      {R.~B.}\ \bibnamefont {Torbert}}, \bibinfo {author} {\bibfnamefont
      {J.}~\bibnamefont {Dorelli}}, \bibinfo {author} {\bibfnamefont {D.~J.}\
      \bibnamefont {Gershman}}, \ and\ \bibinfo {author} {\bibfnamefont {L.~A.}\
      \bibnamefont {Avanov}},\ }\href {\doibase 10.1103/PhysRevLett.117.185101}
      {\bibfield  {journal} {\bibinfo  {journal} {Phys. Rev. Lett.}\ }\textbf
      {\bibinfo {volume} {117}},\ \bibinfo {pages} {185101} (\bibinfo {year}
      {2016}{\natexlab{a}})}\BibitemShut {NoStop}%
    \bibitem [{\citenamefont {Zenitani}\ and\ \citenamefont
      {Nagai}(2016)}]{Zenitani2016}%
      \BibitemOpen
      \bibfield  {author} {\bibinfo {author} {\bibfnamefont {S.}~\bibnamefont
      {Zenitani}}\ and\ \bibinfo {author} {\bibfnamefont {T.}~\bibnamefont
      {Nagai}},\ }\href {\doibase 10.1063/1.4963008} {\bibfield  {journal}
      {\bibinfo  {journal} {Phys. Plasmas}\ }\textbf {\bibinfo {volume} {23}},\
      \bibinfo {pages} {102102} (\bibinfo {year} {2016})}\BibitemShut {NoStop}%
    \bibitem [{\citenamefont {{\O}ieroset}\ \emph {et~al.}(2002)\citenamefont
      {{\O}ieroset}, \citenamefont {Lin}, \citenamefont {Phan}, \citenamefont
      {Larson},\ and\ \citenamefont {Bale}}]{Oieroset2002}%
      \BibitemOpen
      \bibfield  {author} {\bibinfo {author} {\bibfnamefont {M.}~\bibnamefont
      {{\O}ieroset}}, \bibinfo {author} {\bibfnamefont {R.~P.}\ \bibnamefont
      {Lin}}, \bibinfo {author} {\bibfnamefont {T.~D.}\ \bibnamefont {Phan}},
      \bibinfo {author} {\bibfnamefont {D.~E.}\ \bibnamefont {Larson}}, \ and\
      \bibinfo {author} {\bibfnamefont {S.~D.}\ \bibnamefont {Bale}},\ }\href
      {\doibase 10.1103/PhysRevLett.89.195001} {\bibfield  {journal} {\bibinfo
      {journal} {Phys. Rev. Lett.}\ }\textbf {\bibinfo {volume} {89}},\ \bibinfo
      {pages} {195001} (\bibinfo {year} {2002})}\BibitemShut {NoStop}%
    \bibitem [{\citenamefont {Dahlin}, \citenamefont {Drake},\ and\ \citenamefont
      {Swisdak}(2015)}]{Dahlin2015}%
      \BibitemOpen
      \bibfield  {author} {\bibinfo {author} {\bibfnamefont {J.~T.}\ \bibnamefont
      {Dahlin}}, \bibinfo {author} {\bibfnamefont {J.~F.}\ \bibnamefont {Drake}}, \
      and\ \bibinfo {author} {\bibfnamefont {M.}~\bibnamefont {Swisdak}},\ }\href
      {\doibase 10.1063/1.4933212} {\bibfield  {journal} {\bibinfo  {journal}
      {Phys. Plasmas}\ }\textbf {\bibinfo {volume} {22}},\ \bibinfo {pages}
      {100704} (\bibinfo {year} {2015})},\ \Eprint
      {http://arxiv.org/abs/1503.02218} {arXiv:1503.02218} \BibitemShut {NoStop}%
    \bibitem [{\citenamefont {Asano}\ \emph {et~al.}(2008)\citenamefont {Asano},
      \citenamefont {Nakamura}, \citenamefont {Shinohara}, \citenamefont
      {Fujimoto}, \citenamefont {Takada}, \citenamefont {Baumjohann}, \citenamefont
      {Owen}, \citenamefont {Fazakerley}, \citenamefont {Runov}, \citenamefont
      {Nagai}, \citenamefont {Lucek},\ and\ \citenamefont
      {R{\`{e}}me}}]{Asano2008}%
      \BibitemOpen
      \bibfield  {author} {\bibinfo {author} {\bibfnamefont {Y.}~\bibnamefont
      {Asano}}, \bibinfo {author} {\bibfnamefont {R.}~\bibnamefont {Nakamura}},
      \bibinfo {author} {\bibfnamefont {I.}~\bibnamefont {Shinohara}}, \bibinfo
      {author} {\bibfnamefont {M.}~\bibnamefont {Fujimoto}}, \bibinfo {author}
      {\bibfnamefont {T.}~\bibnamefont {Takada}}, \bibinfo {author} {\bibfnamefont
      {W.}~\bibnamefont {Baumjohann}}, \bibinfo {author} {\bibfnamefont {C.~J.}\
      \bibnamefont {Owen}}, \bibinfo {author} {\bibfnamefont {A.~N.}\ \bibnamefont
      {Fazakerley}}, \bibinfo {author} {\bibfnamefont {A.}~\bibnamefont {Runov}},
      \bibinfo {author} {\bibfnamefont {T.}~\bibnamefont {Nagai}}, \bibinfo
      {author} {\bibfnamefont {E.~A.}\ \bibnamefont {Lucek}}, \ and\ \bibinfo
      {author} {\bibfnamefont {H.}~\bibnamefont {R{\`{e}}me}},\ }\href {\doibase
      10.1029/2007JA012461} {\bibfield  {journal} {\bibinfo  {journal} {J. Geophys.
      Res. Sp. Phys.}\ }\textbf {\bibinfo {volume} {113}},\ \bibinfo {pages}
      {A01207} (\bibinfo {year} {2008})}\BibitemShut {NoStop}%
    \bibitem [{\citenamefont {Egedal}\ \emph {et~al.}(2015)\citenamefont {Egedal},
      \citenamefont {Daughton}, \citenamefont {Le},\ and\ \citenamefont
      {Borg}}]{Egedal2015}%
      \BibitemOpen
      \bibfield  {author} {\bibinfo {author} {\bibfnamefont {J.}~\bibnamefont
      {Egedal}}, \bibinfo {author} {\bibfnamefont {W.}~\bibnamefont {Daughton}},
      \bibinfo {author} {\bibfnamefont {A.}~\bibnamefont {Le}}, \ and\ \bibinfo
      {author} {\bibfnamefont {A.~L.}\ \bibnamefont {Borg}},\ }\href {\doibase
      10.1063/1.4933055} {\bibfield  {journal} {\bibinfo  {journal} {Phys.
      Plasmas}\ }\textbf {\bibinfo {volume} {22}},\ \bibinfo {pages} {101208}
      (\bibinfo {year} {2015})}\BibitemShut {NoStop}%
    \bibitem [{\citenamefont {Bessho}\ \emph {et~al.}(2017)\citenamefont {Bessho},
      \citenamefont {Chen}, \citenamefont {Hesse},\ and\ \citenamefont
      {Wang}}]{Bessho2017}%
      \BibitemOpen
      \bibfield  {author} {\bibinfo {author} {\bibfnamefont {N.}~\bibnamefont
      {Bessho}}, \bibinfo {author} {\bibfnamefont {L.-J.}\ \bibnamefont {Chen}},
      \bibinfo {author} {\bibfnamefont {M.}~\bibnamefont {Hesse}}, \ and\ \bibinfo
      {author} {\bibfnamefont {S.}~\bibnamefont {Wang}},\ }\href {\doibase
      10.1063/1.4989737} {\bibfield  {journal} {\bibinfo  {journal} {Phys.
      Plasmas}\ }\textbf {\bibinfo {volume} {24}},\ \bibinfo {pages} {072903}
      (\bibinfo {year} {2017})}\BibitemShut {NoStop}%
    \bibitem [{\citenamefont {Melrose}(1980)}]{Melrose1980}%
      \BibitemOpen
      \bibfield  {author} {\bibinfo {author} {\bibfnamefont {D.}~\bibnamefont
      {Melrose}},\ }\href {\doibase 10.1007/BF00212597} {\bibfield  {journal}
      {\bibinfo  {journal} {Space Sci. Rev.}\ }\textbf {\bibinfo {volume} {26}},\
      \bibinfo {pages} {3} (\bibinfo {year} {1980})}\BibitemShut {NoStop}%
    \bibitem [{\citenamefont {Melrose}(2017)}]{Melrose2017}%
      \BibitemOpen
      \bibfield  {author} {\bibinfo {author} {\bibfnamefont {D.~B.}\ \bibnamefont
      {Melrose}},\ }\href {\doibase 10.1007/s41614-017-0007-0} {\bibfield
      {journal} {\bibinfo  {journal} {Rev. Mod. Plasma Phys.}\ }\textbf {\bibinfo
      {volume} {1}},\ \bibinfo {pages} {5} (\bibinfo {year} {2017})}\BibitemShut
      {NoStop}%
    \bibitem [{\citenamefont {Ni}\ \emph {et~al.}(2020)\citenamefont {Ni},
      \citenamefont {Chen}, \citenamefont {Li}, \citenamefont {Zhang},
      \citenamefont {Ning}, \citenamefont {Kong}, \citenamefont {Wang},\ and\
      \citenamefont {Hosseinpour}}]{Ni2020}%
      \BibitemOpen
      \bibfield  {author} {\bibinfo {author} {\bibfnamefont {S.}~\bibnamefont
      {Ni}}, \bibinfo {author} {\bibfnamefont {Y.}~\bibnamefont {Chen}}, \bibinfo
      {author} {\bibfnamefont {C.}~\bibnamefont {Li}}, \bibinfo {author}
      {\bibfnamefont {Z.}~\bibnamefont {Zhang}}, \bibinfo {author} {\bibfnamefont
      {H.}~\bibnamefont {Ning}}, \bibinfo {author} {\bibfnamefont {X.}~\bibnamefont
      {Kong}}, \bibinfo {author} {\bibfnamefont {B.}~\bibnamefont {Wang}}, \ and\
      \bibinfo {author} {\bibfnamefont {M.}~\bibnamefont {Hosseinpour}},\ }\href
      {\doibase 10.3847/2041-8213/ab7750} {\bibfield  {journal} {\bibinfo
      {journal} {Astrophys. J.}\ }\textbf {\bibinfo {volume} {891}},\ \bibinfo
      {pages} {L25} (\bibinfo {year} {2020})}\BibitemShut {NoStop}%
    \bibitem [{\citenamefont {Aschwanden}(2002)}]{Aschwanden2002}%
      \BibitemOpen
      \bibfield  {author} {\bibinfo {author} {\bibfnamefont {M.~J.}\ \bibnamefont
      {Aschwanden}},\ }\href {\doibase 10.1023/A:1019712124366} {\bibfield
      {journal} {\bibinfo  {journal} {Space Sci. Rev.}\ }\textbf {\bibinfo {volume}
      {101}},\ \bibinfo {pages} {1} (\bibinfo {year} {2002})}\BibitemShut {NoStop}%
    \bibitem [{\citenamefont {Yan}, \citenamefont {Chen},\ and\ \citenamefont
      {Yu}(2015)}]{Yan2015}%
      \BibitemOpen
      \bibfield  {author} {\bibinfo {author} {\bibfnamefont {Y.}~\bibnamefont
      {Yan}}, \bibinfo {author} {\bibfnamefont {L.}~\bibnamefont {Chen}}, \ and\
      \bibinfo {author} {\bibfnamefont {S.}~\bibnamefont {Yu}},\ }\href {\doibase
      10.1017/S174392131600051X} {\bibfield  {journal} {\bibinfo  {journal} {Proc.
      Int. Astron. Union}\ }\textbf {\bibinfo {volume} {11}},\ \bibinfo {pages}
      {427} (\bibinfo {year} {2015})}\BibitemShut {NoStop}%
    \bibitem [{\citenamefont {Chen}\ \emph {et~al.}(2018)\citenamefont {Chen},
      \citenamefont {Yu}, \citenamefont {Battaglia}, \citenamefont {Farid},
      \citenamefont {Savcheva}, \citenamefont {Reeves}, \citenamefont {Krucker},
      \citenamefont {Bastian}, \citenamefont {Guo},\ and\ \citenamefont
      {Tassev}}]{Chen2018}%
      \BibitemOpen
      \bibfield  {author} {\bibinfo {author} {\bibfnamefont {B.}~\bibnamefont
      {Chen}}, \bibinfo {author} {\bibfnamefont {S.}~\bibnamefont {Yu}}, \bibinfo
      {author} {\bibfnamefont {M.}~\bibnamefont {Battaglia}}, \bibinfo {author}
      {\bibfnamefont {S.}~\bibnamefont {Farid}}, \bibinfo {author} {\bibfnamefont
      {A.}~\bibnamefont {Savcheva}}, \bibinfo {author} {\bibfnamefont {K.~K.}\
      \bibnamefont {Reeves}}, \bibinfo {author} {\bibfnamefont {S.}~\bibnamefont
      {Krucker}}, \bibinfo {author} {\bibfnamefont {T.~S.}\ \bibnamefont
      {Bastian}}, \bibinfo {author} {\bibfnamefont {F.}~\bibnamefont {Guo}}, \ and\
      \bibinfo {author} {\bibfnamefont {S.}~\bibnamefont {Tassev}},\ }\href
      {\doibase 10.3847/1538-4357/aadb89} {\bibfield  {journal} {\bibinfo
      {journal} {Astrophys. J.}\ }\textbf {\bibinfo {volume} {866}},\ \bibinfo
      {pages} {62} (\bibinfo {year} {2018})}\BibitemShut {NoStop}%
    \bibitem [{\citenamefont {Penrose}(1960)}]{Penrose1960}%
      \BibitemOpen
      \bibfield  {author} {\bibinfo {author} {\bibfnamefont {O.}~\bibnamefont
      {Penrose}},\ }\href {\doibase 10.1063/1.1706024} {\bibfield  {journal}
      {\bibinfo  {journal} {Phys. Fluids}\ }\textbf {\bibinfo {volume} {3}},\
      \bibinfo {pages} {258} (\bibinfo {year} {1960})}\BibitemShut {NoStop}%
    \bibitem [{\citenamefont {Sestero}\ and\ \citenamefont
      {Curatolo}(1971)}]{Sestero1971}%
      \BibitemOpen
      \bibfield  {author} {\bibinfo {author} {\bibfnamefont {A.}~\bibnamefont
      {Sestero}}\ and\ \bibinfo {author} {\bibfnamefont {M.}~\bibnamefont
      {Curatolo}},\ }\href@noop {} {\bibfield  {journal} {\bibinfo  {journal} {Il
      Nuovo Cimento B}\ }\textbf {\bibinfo {volume} {5}},\ \bibinfo {pages} {165}
      (\bibinfo {year} {1971})}\BibitemShut {NoStop}%
    \bibitem [{\citenamefont {Nicholson}\ \emph {et~al.}(1978)\citenamefont
      {Nicholson}, \citenamefont {Goldman}, \citenamefont {Hoyng},\ and\
      \citenamefont {Weatherall}}]{Nicholson1978}%
      \BibitemOpen
      \bibfield  {author} {\bibinfo {author} {\bibfnamefont {D.~R.}\ \bibnamefont
      {Nicholson}}, \bibinfo {author} {\bibfnamefont {M.~V.}\ \bibnamefont
      {Goldman}}, \bibinfo {author} {\bibfnamefont {P.}~\bibnamefont {Hoyng}}, \
      and\ \bibinfo {author} {\bibfnamefont {J.~C.}\ \bibnamefont {Weatherall}},\
      }\href {\doibase 10.1086/156296} {\bibfield  {journal} {\bibinfo  {journal}
      {Astrophys. J.}\ }\textbf {\bibinfo {volume} {223}},\ \bibinfo {pages} {605}
      (\bibinfo {year} {1978})}\BibitemShut {NoStop}%
    \bibitem [{\citenamefont {Melrose}(1985)}]{Melrose1985}%
      \BibitemOpen
      \bibfield  {author} {\bibinfo {author} {\bibfnamefont {D.}~\bibnamefont
      {Melrose}},\ }in\ \href@noop {} {\emph {\bibinfo {booktitle} {Sol. Radiophys.
      Stud. Emiss. from Sun Metre Wavelengths}}},\ Vol.~\bibinfo {volume} {17},\
      \bibinfo {editor} {edited by\ \bibinfo {editor} {\bibfnamefont
      {D.}~\bibnamefont {McLean}}\ and\ \bibinfo {editor} {\bibfnamefont
      {N.}~\bibnamefont {Labrum}}}\ (\bibinfo {year} {1985})\ pp.\ \bibinfo {pages}
      {177--210}\BibitemShut {NoStop}%
    \bibitem [{\citenamefont {Melrose}(1987)}]{Melrose1987}%
      \BibitemOpen
      \bibfield  {author} {\bibinfo {author} {\bibfnamefont {D.~B.}\ \bibnamefont
      {Melrose}},\ }\href {\doibase 10.1007/BF00145443} {\bibfield  {journal}
      {\bibinfo  {journal} {Sol. Phys.}\ }\textbf {\bibinfo {volume} {111}},\
      \bibinfo {pages} {89} (\bibinfo {year} {1987})}\BibitemShut {NoStop}%
    \bibitem [{\citenamefont {Yao}\ \emph {et~al.}(2021)\citenamefont {Yao},
      \citenamefont {Mu{\~{n}}oz}, \citenamefont {B{\"{u}}chner}, \citenamefont
      {Zhou},\ and\ \citenamefont {Liu}}]{Yao2021}%
      \BibitemOpen
      \bibfield  {author} {\bibinfo {author} {\bibfnamefont {X.}~\bibnamefont
      {Yao}}, \bibinfo {author} {\bibfnamefont {P.~A.}\ \bibnamefont
      {Mu{\~{n}}oz}}, \bibinfo {author} {\bibfnamefont {J.}~\bibnamefont
      {B{\"{u}}chner}}, \bibinfo {author} {\bibfnamefont {X.}~\bibnamefont {Zhou}},
      \ and\ \bibinfo {author} {\bibfnamefont {S.}~\bibnamefont {Liu}},\ }\href
      {\doibase 10.1017/S0022377821000076} {\bibfield  {journal} {\bibinfo
      {journal} {J. Plasma Phys.}\ }\textbf {\bibinfo {volume} {87}},\ \bibinfo
      {pages} {905870203} (\bibinfo {year} {2021})}\BibitemShut {NoStop}%
    \bibitem [{\citenamefont {Ginzburg}\ and\ \citenamefont
      {Zhelezniakov}(1958)}]{Ginzburg1958}%
      \BibitemOpen
      \bibfield  {author} {\bibinfo {author} {\bibfnamefont {V.~L.}\ \bibnamefont
      {Ginzburg}}\ and\ \bibinfo {author} {\bibfnamefont {V.~V.}\ \bibnamefont
      {Zhelezniakov}},\ }\href@noop {} {\bibfield  {journal} {\bibinfo  {journal}
      {Sov. Astron.}\ }\textbf {\bibinfo {volume} {2}},\ \bibinfo {pages} {653}
      (\bibinfo {year} {1958})}\BibitemShut {NoStop}%
    \bibitem [{\citenamefont {Melrose}(1970{\natexlab{a}})}]{Melrose1970}%
      \BibitemOpen
      \bibfield  {author} {\bibinfo {author} {\bibfnamefont {D.}~\bibnamefont
      {Melrose}},\ }\href {\doibase 10.1071/PH700871} {\bibfield  {journal}
      {\bibinfo  {journal} {J. Phys.}\ }\textbf {\bibinfo {volume} {23}},\ \bibinfo
      {pages} {871} (\bibinfo {year} {1970}{\natexlab{a}})}\BibitemShut {NoStop}%
    \bibitem [{\citenamefont {Melrose}(1970{\natexlab{b}})}]{Melrose1970a}%
      \BibitemOpen
      \bibfield  {author} {\bibinfo {author} {\bibfnamefont {D.}~\bibnamefont
      {Melrose}},\ }\href {\doibase 10.1071/PH700885} {\bibfield  {journal}
      {\bibinfo  {journal} {J. Phys.}\ }\textbf {\bibinfo {volume} {23}},\ \bibinfo
      {pages} {885} (\bibinfo {year} {1970}{\natexlab{b}})}\BibitemShut {NoStop}%
    \bibitem [{\citenamefont {Reid}\ and\ \citenamefont
      {Ratcliffe}(2014)}]{Reid2014}%
      \BibitemOpen
      \bibfield  {author} {\bibinfo {author} {\bibfnamefont {H.~A.~S.}\
      \bibnamefont {Reid}}\ and\ \bibinfo {author} {\bibfnamefont {H.}~\bibnamefont
      {Ratcliffe}},\ }\href {\doibase 10.1088/1674-4527/14/7/003} {\bibfield
      {journal} {\bibinfo  {journal} {Res. Astron. Astrophys.}\ }\textbf {\bibinfo
      {volume} {14}},\ \bibinfo {pages} {773} (\bibinfo {year} {2014})}\BibitemShut
      {NoStop}%
    \bibitem [{\citenamefont {Henri}\ \emph {et~al.}(2019)\citenamefont {Henri},
      \citenamefont {Sgattoni}, \citenamefont {Briand}, \citenamefont {Amiranoff},\
      and\ \citenamefont {Riconda}}]{Henri2019}%
      \BibitemOpen
      \bibfield  {author} {\bibinfo {author} {\bibfnamefont {P.}~\bibnamefont
      {Henri}}, \bibinfo {author} {\bibfnamefont {A.}~\bibnamefont {Sgattoni}},
      \bibinfo {author} {\bibfnamefont {C.}~\bibnamefont {Briand}}, \bibinfo
      {author} {\bibfnamefont {F.}~\bibnamefont {Amiranoff}}, \ and\ \bibinfo
      {author} {\bibfnamefont {C.}~\bibnamefont {Riconda}},\ }\href {\doibase
      10.1029/2018JA025707} {\bibfield  {journal} {\bibinfo  {journal} {J. Geophys.
      Res.}\ }\textbf {\bibinfo {volume} {124}},\ \bibinfo {pages} {1475} (\bibinfo
      {year} {2019})}\BibitemShut {NoStop}%
    \bibitem [{\citenamefont {Yi}, \citenamefont {Yoon},\ and\ \citenamefont
      {Ryu}(2007)}]{Yi2007}%
      \BibitemOpen
      \bibfield  {author} {\bibinfo {author} {\bibfnamefont {S.}~\bibnamefont
      {Yi}}, \bibinfo {author} {\bibfnamefont {P.~H.}\ \bibnamefont {Yoon}}, \ and\
      \bibinfo {author} {\bibfnamefont {C.-M.}\ \bibnamefont {Ryu}},\ }\href
      {\doibase 10.1063/1.2424556} {\bibfield  {journal} {\bibinfo  {journal}
      {Phys. Plasmas}\ }\textbf {\bibinfo {volume} {14}},\ \bibinfo {pages}
      {013301} (\bibinfo {year} {2007})}\BibitemShut {NoStop}%
    \bibitem [{\citenamefont {Rhee}\ \emph {et~al.}(2009)\citenamefont {Rhee},
      \citenamefont {Ryu}, \citenamefont {Woo}, \citenamefont {Kaang},
      \citenamefont {Yi},\ and\ \citenamefont {Yoon}}]{Rhee2009}%
      \BibitemOpen
      \bibfield  {author} {\bibinfo {author} {\bibfnamefont {T.}~\bibnamefont
      {Rhee}}, \bibinfo {author} {\bibfnamefont {C.-M.}\ \bibnamefont {Ryu}},
      \bibinfo {author} {\bibfnamefont {M.}~\bibnamefont {Woo}}, \bibinfo {author}
      {\bibfnamefont {H.~H.}\ \bibnamefont {Kaang}}, \bibinfo {author}
      {\bibfnamefont {S.}~\bibnamefont {Yi}}, \ and\ \bibinfo {author}
      {\bibfnamefont {P.~H.}\ \bibnamefont {Yoon}},\ }\href {\doibase
      10.1088/0004-637X/694/1/618} {\bibfield  {journal} {\bibinfo  {journal}
      {Astrophys. J.}\ }\textbf {\bibinfo {volume} {694}},\ \bibinfo {pages} {618}
      (\bibinfo {year} {2009})}\BibitemShut {NoStop}%
    \bibitem [{\citenamefont {Ganse}\ \emph {et~al.}(2012)\citenamefont {Ganse},
      \citenamefont {Kilian}, \citenamefont {Vainio},\ and\ \citenamefont
      {Spanier}}]{Ganse2012}%
      \BibitemOpen
      \bibfield  {author} {\bibinfo {author} {\bibfnamefont {U.}~\bibnamefont
      {Ganse}}, \bibinfo {author} {\bibfnamefont {P.}~\bibnamefont {Kilian}},
      \bibinfo {author} {\bibfnamefont {R.}~\bibnamefont {Vainio}}, \ and\ \bibinfo
      {author} {\bibfnamefont {F.}~\bibnamefont {Spanier}},\ }\href {\doibase
      10.1007/s11207-012-0077-7} {\bibfield  {journal} {\bibinfo  {journal} {Sol.
      Phys.}\ }\textbf {\bibinfo {volume} {280}},\ \bibinfo {pages} {551} (\bibinfo
      {year} {2012})}\BibitemShut {NoStop}%
    \bibitem [{\citenamefont {Thurgood}\ and\ \citenamefont
      {Tsiklauri}(2015)}]{Thurgood2015}%
      \BibitemOpen
      \bibfield  {author} {\bibinfo {author} {\bibfnamefont {J.~O.}\ \bibnamefont
      {Thurgood}}\ and\ \bibinfo {author} {\bibfnamefont {D.}~\bibnamefont
      {Tsiklauri}},\ }\href {\doibase 10.1051/0004-6361/201527079} {\bibfield
      {journal} {\bibinfo  {journal} {Astron. Astrophys.}\ }\textbf {\bibinfo
      {volume} {584}},\ \bibinfo {pages} {A83} (\bibinfo {year}
      {2015})}\BibitemShut {NoStop}%
    \bibitem [{\citenamefont {Dahlin}, \citenamefont {Drake},\ and\ \citenamefont
      {Swisdak}(2016)}]{Dahlin2016}%
      \BibitemOpen
      \bibfield  {author} {\bibinfo {author} {\bibfnamefont {J.~T.}\ \bibnamefont
      {Dahlin}}, \bibinfo {author} {\bibfnamefont {J.~F.}\ \bibnamefont {Drake}}, \
      and\ \bibinfo {author} {\bibfnamefont {M.}~\bibnamefont {Swisdak}},\ }\href
      {\doibase 10.1063/1.4972082} {\bibfield  {journal} {\bibinfo  {journal}
      {Phys. Plasmas}\ }\textbf {\bibinfo {volume} {23}},\ \bibinfo {pages}
      {120704} (\bibinfo {year} {2016})}\BibitemShut {NoStop}%
    \bibitem [{\citenamefont {Goldman}, \citenamefont {Newman},\ and\ \citenamefont
      {Pritchett}(2008)}]{Goldman2008}%
      \BibitemOpen
      \bibfield  {author} {\bibinfo {author} {\bibfnamefont {M.~V.}\ \bibnamefont
      {Goldman}}, \bibinfo {author} {\bibfnamefont {D.~L.}\ \bibnamefont {Newman}},
      \ and\ \bibinfo {author} {\bibfnamefont {P.}~\bibnamefont {Pritchett}},\
      }\href {\doibase 10.1029/2008GL035608} {\bibfield  {journal} {\bibinfo
      {journal} {Geophys. Res. Lett.}\ }\textbf {\bibinfo {volume} {35}},\ \bibinfo
      {pages} {L22109} (\bibinfo {year} {2008})}\BibitemShut {NoStop}%
    \bibitem [{\citenamefont {Goldman}\ \emph {et~al.}(2014)\citenamefont
      {Goldman}, \citenamefont {Newman}, \citenamefont {Lapenta}, \citenamefont
      {Andersson}, \citenamefont {Gosling}, \citenamefont {Eriksson}, \citenamefont
      {Markidis}, \citenamefont {Eastwood},\ and\ \citenamefont
      {Ergun}}]{Goldman2014}%
      \BibitemOpen
      \bibfield  {author} {\bibinfo {author} {\bibfnamefont {M.}~\bibnamefont
      {Goldman}}, \bibinfo {author} {\bibfnamefont {D.~L.}\ \bibnamefont {Newman}},
      \bibinfo {author} {\bibfnamefont {G.}~\bibnamefont {Lapenta}}, \bibinfo
      {author} {\bibfnamefont {L.}~\bibnamefont {Andersson}}, \bibinfo {author}
      {\bibfnamefont {J.~T.}\ \bibnamefont {Gosling}}, \bibinfo {author}
      {\bibfnamefont {S.}~\bibnamefont {Eriksson}}, \bibinfo {author}
      {\bibfnamefont {S.}~\bibnamefont {Markidis}}, \bibinfo {author}
      {\bibfnamefont {J.~P.}\ \bibnamefont {Eastwood}}, \ and\ \bibinfo {author}
      {\bibfnamefont {R.}~\bibnamefont {Ergun}},\ }\href {\doibase
      10.1103/PhysRevLett.112.145002} {\bibfield  {journal} {\bibinfo  {journal}
      {Phys. Rev. Lett.}\ }\textbf {\bibinfo {volume} {112}},\ \bibinfo {pages}
      {145002} (\bibinfo {year} {2014})}\BibitemShut {NoStop}%
    \bibitem [{\citenamefont {Divin}\ \emph {et~al.}(2012)\citenamefont {Divin},
      \citenamefont {Lapenta}, \citenamefont {Markidis}, \citenamefont {Newman},\
      and\ \citenamefont {Goldman}}]{Divin2012}%
      \BibitemOpen
      \bibfield  {author} {\bibinfo {author} {\bibfnamefont {A.}~\bibnamefont
      {Divin}}, \bibinfo {author} {\bibfnamefont {G.}~\bibnamefont {Lapenta}},
      \bibinfo {author} {\bibfnamefont {S.}~\bibnamefont {Markidis}}, \bibinfo
      {author} {\bibfnamefont {D.~L.}\ \bibnamefont {Newman}}, \ and\ \bibinfo
      {author} {\bibfnamefont {M.~V.}\ \bibnamefont {Goldman}},\ }\href {\doibase
      10.1063/1.3698621} {\bibfield  {journal} {\bibinfo  {journal} {Phys.
      Plasmas}\ }\textbf {\bibinfo {volume} {19}},\ \bibinfo {pages} {042110}
      (\bibinfo {year} {2012})}\BibitemShut {NoStop}%
    \bibitem [{\citenamefont {Hesse}\ \emph {et~al.}(2018)\citenamefont {Hesse},
      \citenamefont {Norgren}, \citenamefont {Tenfjord}, \citenamefont {Burch},
      \citenamefont {Liu}, \citenamefont {Chen}, \citenamefont {Bessho},
      \citenamefont {Wang}, \citenamefont {Nakamura}, \citenamefont {Eastwood},
      \citenamefont {Hoshino}, \citenamefont {Torbert},\ and\ \citenamefont
      {Ergun}}]{Hesse2018}%
      \BibitemOpen
      \bibfield  {author} {\bibinfo {author} {\bibfnamefont {M.}~\bibnamefont
      {Hesse}}, \bibinfo {author} {\bibfnamefont {C.}~\bibnamefont {Norgren}},
      \bibinfo {author} {\bibfnamefont {P.}~\bibnamefont {Tenfjord}}, \bibinfo
      {author} {\bibfnamefont {J.~L.}\ \bibnamefont {Burch}}, \bibinfo {author}
      {\bibfnamefont {Y.-H.}\ \bibnamefont {Liu}}, \bibinfo {author} {\bibfnamefont
      {L.-J.}\ \bibnamefont {Chen}}, \bibinfo {author} {\bibfnamefont
      {N.}~\bibnamefont {Bessho}}, \bibinfo {author} {\bibfnamefont
      {S.}~\bibnamefont {Wang}}, \bibinfo {author} {\bibfnamefont {R.}~\bibnamefont
      {Nakamura}}, \bibinfo {author} {\bibfnamefont {J.~P.}\ \bibnamefont
      {Eastwood}}, \bibinfo {author} {\bibfnamefont {M.}~\bibnamefont {Hoshino}},
      \bibinfo {author} {\bibfnamefont {R.~B.}\ \bibnamefont {Torbert}}, \ and\
      \bibinfo {author} {\bibfnamefont {R.~E.}\ \bibnamefont {Ergun}},\ }\href
      {\doibase 10.1063/1.5054100} {\bibfield  {journal} {\bibinfo  {journal}
      {Phys. Plasmas}\ }\textbf {\bibinfo {volume} {25}},\ \bibinfo {pages}
      {122902} (\bibinfo {year} {2018})}\BibitemShut {NoStop}%
    \bibitem [{\citenamefont {Chu}(2004)}]{Chu2004a}%
      \BibitemOpen
      \bibfield  {author} {\bibinfo {author} {\bibfnamefont {K.~R.}\ \bibnamefont
      {Chu}},\ }\href {\doibase 10.1103/RevModPhys.76.489} {\bibfield  {journal}
      {\bibinfo  {journal} {Rev. Mod. Phys.}\ }\textbf {\bibinfo {volume} {76}},\
      \bibinfo {pages} {489} (\bibinfo {year} {2004})}\BibitemShut {NoStop}%
    \bibitem [{\citenamefont {Hewitt}, \citenamefont {Melrose},\ and\ \citenamefont
      {R{\"{o}}nnmark}(1982)}]{Hewitt1982}%
      \BibitemOpen
      \bibfield  {author} {\bibinfo {author} {\bibfnamefont {R.}~\bibnamefont
      {Hewitt}}, \bibinfo {author} {\bibfnamefont {D.}~\bibnamefont {Melrose}}, \
      and\ \bibinfo {author} {\bibfnamefont {K.}~\bibnamefont {R{\"{o}}nnmark}},\
      }\href {\doibase 10.1071/PH820447} {\bibfield  {journal} {\bibinfo  {journal}
      {J. Phys.}\ }\textbf {\bibinfo {volume} {35}},\ \bibinfo {pages} {447}
      (\bibinfo {year} {1982})}\BibitemShut {NoStop}%
    \bibitem [{\citenamefont {Winglee}\ and\ \citenamefont
      {Dulk}(1986)}]{Winglee1986a}%
      \BibitemOpen
      \bibfield  {author} {\bibinfo {author} {\bibfnamefont {R.~M.}\ \bibnamefont
      {Winglee}}\ and\ \bibinfo {author} {\bibfnamefont {G.~A.}\ \bibnamefont
      {Dulk}},\ }\href {\doibase 10.1086/164467} {\bibfield  {journal} {\bibinfo
      {journal} {Astrophys. J.}\ }\textbf {\bibinfo {volume} {307}},\ \bibinfo
      {pages} {808} (\bibinfo {year} {1986})}\BibitemShut {NoStop}%
    \bibitem [{\citenamefont {Treumann}\ and\ \citenamefont
      {Baumjohann}(2017)}]{Treumann2017}%
      \BibitemOpen
      \bibfield  {author} {\bibinfo {author} {\bibfnamefont {R.~A.}\ \bibnamefont
      {Treumann}}\ and\ \bibinfo {author} {\bibfnamefont {W.}~\bibnamefont
      {Baumjohann}},\ }\href {\doibase 10.5194/angeo-35-999-2017} {\bibfield
      {journal} {\bibinfo  {journal} {Ann. Geophys.}\ }\textbf {\bibinfo {volume}
      {35}},\ \bibinfo {pages} {999} (\bibinfo {year} {2017})}\BibitemShut
      {NoStop}%
    \bibitem [{\citenamefont {Melrose}, \citenamefont {Hewitt},\ and\ \citenamefont
      {Dulk}(1984)}]{Melrose1984}%
      \BibitemOpen
      \bibfield  {author} {\bibinfo {author} {\bibfnamefont {D.~B.}\ \bibnamefont
      {Melrose}}, \bibinfo {author} {\bibfnamefont {R.~G.}\ \bibnamefont {Hewitt}},
      \ and\ \bibinfo {author} {\bibfnamefont {G.~A.}\ \bibnamefont {Dulk}},\
      }\href {\doibase 10.1029/JA089iA02p00897} {\bibfield  {journal} {\bibinfo
      {journal} {J. Geophys. Res.}\ }\textbf {\bibinfo {volume} {89}},\ \bibinfo
      {pages} {897} (\bibinfo {year} {1984})}\BibitemShut {NoStop}%
    \bibitem [{\citenamefont {Treumann}(2006)}]{Treumann2006}%
      \BibitemOpen
      \bibfield  {author} {\bibinfo {author} {\bibfnamefont {R.~A.}\ \bibnamefont
      {Treumann}},\ }\href {\doibase 10.1007/s00159-006-0001-y} {\bibfield
      {journal} {\bibinfo  {journal} {Astron. Astrophys. Rev.}\ }\textbf {\bibinfo
      {volume} {13}},\ \bibinfo {pages} {229} (\bibinfo {year} {2006})}\BibitemShut
      {NoStop}%
    \bibitem [{\citenamefont {Pritchett}(1984)}]{Pritchett1984}%
      \BibitemOpen
      \bibfield  {author} {\bibinfo {author} {\bibfnamefont {P.~L.}\ \bibnamefont
      {Pritchett}},\ }\href {\doibase 10.1029/JA089iA10p08957} {\bibfield
      {journal} {\bibinfo  {journal} {J. Geophys. Res.}\ }\textbf {\bibinfo
      {volume} {89}},\ \bibinfo {pages} {8957} (\bibinfo {year}
      {1984})}\BibitemShut {NoStop}%
    \bibitem [{\citenamefont {Melrose}(1986)}]{Melrose1986}%
      \BibitemOpen
      \bibfield  {author} {\bibinfo {author} {\bibfnamefont {D.~B.}\ \bibnamefont
      {Melrose}},\ }\href {\doibase 10.1017/CBO9780511564123} {\emph {\bibinfo
      {title} {{Instabilities in Space and Laboratory Plasmas}}}}\ (\bibinfo
      {publisher} {Cambridge University Press},\ \bibinfo {year}
      {1986})\BibitemShut {NoStop}%
    \bibitem [{\citenamefont {Moseev}\ and\ \citenamefont
      {Salewski}(2019)}]{Moseev2019}%
      \BibitemOpen
      \bibfield  {author} {\bibinfo {author} {\bibfnamefont {D.}~\bibnamefont
      {Moseev}}\ and\ \bibinfo {author} {\bibfnamefont {M.}~\bibnamefont
      {Salewski}},\ }\href {\doibase 10.1063/1.5085429} {\bibfield  {journal}
      {\bibinfo  {journal} {Phys. Plasmas}\ }\textbf {\bibinfo {volume} {26}},\
      \bibinfo {pages} {020901} (\bibinfo {year} {2019})}\BibitemShut {NoStop}%
    \bibitem [{\citenamefont {Lee}, \citenamefont {Omura},\ and\ \citenamefont
      {Lee}(2011)}]{Lee2011}%
      \BibitemOpen
      \bibfield  {author} {\bibinfo {author} {\bibfnamefont {K.~H.}\ \bibnamefont
      {Lee}}, \bibinfo {author} {\bibfnamefont {Y.}~\bibnamefont {Omura}}, \ and\
      \bibinfo {author} {\bibfnamefont {L.~C.}\ \bibnamefont {Lee}},\ }\href
      {\doibase 10.1063/1.3626562} {\bibfield  {journal} {\bibinfo  {journal}
      {Phys. Plasmas}\ }\textbf {\bibinfo {volume} {18}},\ \bibinfo {pages}
      {092110} (\bibinfo {year} {2011})}\BibitemShut {NoStop}%
    \bibitem [{\citenamefont {Zhou}\ \emph {et~al.}(2020)\citenamefont {Zhou},
      \citenamefont {Mu{\~{n}}oz}, \citenamefont {B{\"{u}}chner},\ and\
      \citenamefont {Liu}}]{Zhou2020}%
      \BibitemOpen
      \bibfield  {author} {\bibinfo {author} {\bibfnamefont {X.}~\bibnamefont
      {Zhou}}, \bibinfo {author} {\bibfnamefont {P.~A.}\ \bibnamefont
      {Mu{\~{n}}oz}}, \bibinfo {author} {\bibfnamefont {J.}~\bibnamefont
      {B{\"{u}}chner}}, \ and\ \bibinfo {author} {\bibfnamefont {S.}~\bibnamefont
      {Liu}},\ }\href {\doibase 10.3847/1538-4357/ab6a0d} {\bibfield  {journal}
      {\bibinfo  {journal} {Astrophys. J.}\ }\textbf {\bibinfo {volume} {891}},\
      \bibinfo {pages} {92} (\bibinfo {year} {2020})}\BibitemShut {NoStop}%
    \bibitem [{\citenamefont {Shuster}\ \emph {et~al.}(2014)\citenamefont
      {Shuster}, \citenamefont {Chen}, \citenamefont {Daughton}, \citenamefont
      {Lee}, \citenamefont {Lee}, \citenamefont {Bessho}, \citenamefont {Torbert},
      \citenamefont {Li},\ and\ \citenamefont {Argall}}]{Shuster2014}%
      \BibitemOpen
      \bibfield  {author} {\bibinfo {author} {\bibfnamefont {J.~R.}\ \bibnamefont
      {Shuster}}, \bibinfo {author} {\bibfnamefont {L.-J.}\ \bibnamefont {Chen}},
      \bibinfo {author} {\bibfnamefont {W.~S.}\ \bibnamefont {Daughton}}, \bibinfo
      {author} {\bibfnamefont {L.~C.}\ \bibnamefont {Lee}}, \bibinfo {author}
      {\bibfnamefont {K.~H.}\ \bibnamefont {Lee}}, \bibinfo {author} {\bibfnamefont
      {N.}~\bibnamefont {Bessho}}, \bibinfo {author} {\bibfnamefont {R.~B.}\
      \bibnamefont {Torbert}}, \bibinfo {author} {\bibfnamefont {G.}~\bibnamefont
      {Li}}, \ and\ \bibinfo {author} {\bibfnamefont {M.~R.}\ \bibnamefont
      {Argall}},\ }\href {\doibase 10.1002/2014GL060608} {\bibfield  {journal}
      {\bibinfo  {journal} {Geophys. Res. Lett.}\ }\textbf {\bibinfo {volume}
      {41}},\ \bibinfo {pages} {5389} (\bibinfo {year} {2014})}\BibitemShut
      {NoStop}%
    \bibitem [{\citenamefont {Egedal}\ \emph
      {et~al.}(2016{\natexlab{b}})\citenamefont {Egedal}, \citenamefont
      {Wetherton}, \citenamefont {Daughton},\ and\ \citenamefont
      {Le}}]{Egedal2016a}%
      \BibitemOpen
      \bibfield  {author} {\bibinfo {author} {\bibfnamefont {J.}~\bibnamefont
      {Egedal}}, \bibinfo {author} {\bibfnamefont {B.}~\bibnamefont {Wetherton}},
      \bibinfo {author} {\bibfnamefont {W.}~\bibnamefont {Daughton}}, \ and\
      \bibinfo {author} {\bibfnamefont {A.}~\bibnamefont {Le}},\ }\href {\doibase
      10.1063/1.4972135} {\bibfield  {journal} {\bibinfo  {journal} {Phys.
      Plasmas}\ }\textbf {\bibinfo {volume} {23}},\ \bibinfo {pages} {122904}
      (\bibinfo {year} {2016}{\natexlab{b}})}\BibitemShut {NoStop}%
    \bibitem [{\citenamefont {Phan}\ \emph {et~al.}(2016)\citenamefont {Phan},
      \citenamefont {Eastwood}, \citenamefont {Cassak}, \citenamefont
      {{\O}ieroset}, \citenamefont {Gosling}, \citenamefont {Gershman},
      \citenamefont {Mozer}, \citenamefont {Shay}, \citenamefont {Fujimoto},
      \citenamefont {Daughton}, \citenamefont {Drake}, \citenamefont {Burch},
      \citenamefont {Torbert}, \citenamefont {Ergun}, \citenamefont {Chen},
      \citenamefont {Wang}, \citenamefont {Pollock}, \citenamefont {Dorelli},
      \citenamefont {Lavraud}, \citenamefont {Giles}, \citenamefont {Moore},
      \citenamefont {Saito}, \citenamefont {Avanov}, \citenamefont {Paterson},
      \citenamefont {Strangeway}, \citenamefont {Russell}, \citenamefont
      {Khotyaintsev}, \citenamefont {Lindqvist}, \citenamefont {Oka},\ and\
      \citenamefont {Wilder}}]{Phan2016}%
      \BibitemOpen
      \bibfield  {author} {\bibinfo {author} {\bibfnamefont {T.~D.}\ \bibnamefont
      {Phan}}, \bibinfo {author} {\bibfnamefont {J.~P.}\ \bibnamefont {Eastwood}},
      \bibinfo {author} {\bibfnamefont {P.~A.}\ \bibnamefont {Cassak}}, \bibinfo
      {author} {\bibfnamefont {M.}~\bibnamefont {{\O}ieroset}}, \bibinfo {author}
      {\bibfnamefont {J.~T.}\ \bibnamefont {Gosling}}, \bibinfo {author}
      {\bibfnamefont {D.~J.}\ \bibnamefont {Gershman}}, \bibinfo {author}
      {\bibfnamefont {F.~S.}\ \bibnamefont {Mozer}}, \bibinfo {author}
      {\bibfnamefont {M.~A.}\ \bibnamefont {Shay}}, \bibinfo {author}
      {\bibfnamefont {M.}~\bibnamefont {Fujimoto}}, \bibinfo {author}
      {\bibfnamefont {W.}~\bibnamefont {Daughton}}, \bibinfo {author}
      {\bibfnamefont {J.~F.}\ \bibnamefont {Drake}}, \bibinfo {author}
      {\bibfnamefont {J.~L.}\ \bibnamefont {Burch}}, \bibinfo {author}
      {\bibfnamefont {R.~B.}\ \bibnamefont {Torbert}}, \bibinfo {author}
      {\bibfnamefont {R.~E.}\ \bibnamefont {Ergun}}, \bibinfo {author}
      {\bibfnamefont {L.~J.}\ \bibnamefont {Chen}}, \bibinfo {author}
      {\bibfnamefont {S.}~\bibnamefont {Wang}}, \bibinfo {author} {\bibfnamefont
      {C.}~\bibnamefont {Pollock}}, \bibinfo {author} {\bibfnamefont {J.~C.}\
      \bibnamefont {Dorelli}}, \bibinfo {author} {\bibfnamefont {B.}~\bibnamefont
      {Lavraud}}, \bibinfo {author} {\bibfnamefont {B.~L.}\ \bibnamefont {Giles}},
      \bibinfo {author} {\bibfnamefont {T.~E.}\ \bibnamefont {Moore}}, \bibinfo
      {author} {\bibfnamefont {Y.}~\bibnamefont {Saito}}, \bibinfo {author}
      {\bibfnamefont {L.~A.}\ \bibnamefont {Avanov}}, \bibinfo {author}
      {\bibfnamefont {W.}~\bibnamefont {Paterson}}, \bibinfo {author}
      {\bibfnamefont {R.~J.}\ \bibnamefont {Strangeway}}, \bibinfo {author}
      {\bibfnamefont {C.~T.}\ \bibnamefont {Russell}}, \bibinfo {author}
      {\bibfnamefont {Y.}~\bibnamefont {Khotyaintsev}}, \bibinfo {author}
      {\bibfnamefont {P.~A.}\ \bibnamefont {Lindqvist}}, \bibinfo {author}
      {\bibfnamefont {M.}~\bibnamefont {Oka}}, \ and\ \bibinfo {author}
      {\bibfnamefont {F.~D.}\ \bibnamefont {Wilder}},\ }\href {\doibase
      10.1002/2016GL069212} {\bibfield  {journal} {\bibinfo  {journal} {Geophys.
      Res. Lett.}\ }\textbf {\bibinfo {volume} {43}},\ \bibinfo {pages} {6060}
      (\bibinfo {year} {2016})}\BibitemShut {NoStop}%
    \bibitem [{\citenamefont {Chen}\ \emph
      {et~al.}(2016{\natexlab{a}})\citenamefont {Chen}, \citenamefont {Hesse},
      \citenamefont {Wang}, \citenamefont {Gershman}, \citenamefont {Ergun},
      \citenamefont {Pollock}, \citenamefont {Torbert}, \citenamefont {Bessho},
      \citenamefont {Daughton}, \citenamefont {Dorelli}, \citenamefont {Giles},
      \citenamefont {Strangeway}, \citenamefont {Russell}, \citenamefont
      {Khotyaintsev}, \citenamefont {Burch}, \citenamefont {Moore}, \citenamefont
      {Lavraud}, \citenamefont {Phan},\ and\ \citenamefont {Avanov}}]{Chen2016c}%
      \BibitemOpen
      \bibfield  {author} {\bibinfo {author} {\bibfnamefont {L.-J.}\ \bibnamefont
      {Chen}}, \bibinfo {author} {\bibfnamefont {M.}~\bibnamefont {Hesse}},
      \bibinfo {author} {\bibfnamefont {S.}~\bibnamefont {Wang}}, \bibinfo {author}
      {\bibfnamefont {D.}~\bibnamefont {Gershman}}, \bibinfo {author}
      {\bibfnamefont {R.}~\bibnamefont {Ergun}}, \bibinfo {author} {\bibfnamefont
      {C.}~\bibnamefont {Pollock}}, \bibinfo {author} {\bibfnamefont
      {R.}~\bibnamefont {Torbert}}, \bibinfo {author} {\bibfnamefont
      {N.}~\bibnamefont {Bessho}}, \bibinfo {author} {\bibfnamefont
      {W.}~\bibnamefont {Daughton}}, \bibinfo {author} {\bibfnamefont
      {J.}~\bibnamefont {Dorelli}}, \bibinfo {author} {\bibfnamefont
      {B.}~\bibnamefont {Giles}}, \bibinfo {author} {\bibfnamefont
      {R.}~\bibnamefont {Strangeway}}, \bibinfo {author} {\bibfnamefont
      {C.}~\bibnamefont {Russell}}, \bibinfo {author} {\bibfnamefont
      {Y.}~\bibnamefont {Khotyaintsev}}, \bibinfo {author} {\bibfnamefont
      {J.}~\bibnamefont {Burch}}, \bibinfo {author} {\bibfnamefont
      {T.}~\bibnamefont {Moore}}, \bibinfo {author} {\bibfnamefont
      {B.}~\bibnamefont {Lavraud}}, \bibinfo {author} {\bibfnamefont
      {T.}~\bibnamefont {Phan}}, \ and\ \bibinfo {author} {\bibfnamefont
      {L.}~\bibnamefont {Avanov}},\ }\href {\doibase 10.1002/2016GL069215}
      {\bibfield  {journal} {\bibinfo  {journal} {Geophys. Res. Lett.}\ }\textbf
      {\bibinfo {volume} {43}},\ \bibinfo {pages} {6036} (\bibinfo {year}
      {2016}{\natexlab{a}})}\BibitemShut {NoStop}%
    \bibitem [{\citenamefont {Genestreti}\ \emph {et~al.}(2018)\citenamefont
      {Genestreti}, \citenamefont {Varsani}, \citenamefont {Burch}, \citenamefont
      {Cassak}, \citenamefont {Torbert}, \citenamefont {Nakamura}, \citenamefont
      {Ergun}, \citenamefont {Phan}, \citenamefont {Toledo-Redondo}, \citenamefont
      {Hesse}, \citenamefont {Wang}, \citenamefont {Giles}, \citenamefont
      {Russell}, \citenamefont {V{\"{o}}r{\"{o}}s}, \citenamefont {Hwang},
      \citenamefont {Eastwood}, \citenamefont {Lavraud}, \citenamefont {Escoubet},
      \citenamefont {Fear}, \citenamefont {Khotyaintsev}, \citenamefont {Nakamura},
      \citenamefont {Webster},\ and\ \citenamefont {Baumjohann}}]{Genestreti2018}%
      \BibitemOpen
      \bibfield  {author} {\bibinfo {author} {\bibfnamefont {K.~J.}\ \bibnamefont
      {Genestreti}}, \bibinfo {author} {\bibfnamefont {A.}~\bibnamefont {Varsani}},
      \bibinfo {author} {\bibfnamefont {J.~L.}\ \bibnamefont {Burch}}, \bibinfo
      {author} {\bibfnamefont {P.~A.}\ \bibnamefont {Cassak}}, \bibinfo {author}
      {\bibfnamefont {R.~B.}\ \bibnamefont {Torbert}}, \bibinfo {author}
      {\bibfnamefont {R.}~\bibnamefont {Nakamura}}, \bibinfo {author}
      {\bibfnamefont {R.~E.}\ \bibnamefont {Ergun}}, \bibinfo {author}
      {\bibfnamefont {T.-D.}\ \bibnamefont {Phan}}, \bibinfo {author}
      {\bibfnamefont {S.}~\bibnamefont {Toledo-Redondo}}, \bibinfo {author}
      {\bibfnamefont {M.}~\bibnamefont {Hesse}}, \bibinfo {author} {\bibfnamefont
      {S.}~\bibnamefont {Wang}}, \bibinfo {author} {\bibfnamefont {B.~L.}\
      \bibnamefont {Giles}}, \bibinfo {author} {\bibfnamefont {C.~T.}\ \bibnamefont
      {Russell}}, \bibinfo {author} {\bibfnamefont {Z.}~\bibnamefont
      {V{\"{o}}r{\"{o}}s}}, \bibinfo {author} {\bibfnamefont {K.-J.}\ \bibnamefont
      {Hwang}}, \bibinfo {author} {\bibfnamefont {J.~P.}\ \bibnamefont {Eastwood}},
      \bibinfo {author} {\bibfnamefont {B.}~\bibnamefont {Lavraud}}, \bibinfo
      {author} {\bibfnamefont {C.~P.}\ \bibnamefont {Escoubet}}, \bibinfo {author}
      {\bibfnamefont {R.~C.}\ \bibnamefont {Fear}}, \bibinfo {author}
      {\bibfnamefont {Y.}~\bibnamefont {Khotyaintsev}}, \bibinfo {author}
      {\bibfnamefont {T.~K.~M.}\ \bibnamefont {Nakamura}}, \bibinfo {author}
      {\bibfnamefont {J.~M.}\ \bibnamefont {Webster}}, \ and\ \bibinfo {author}
      {\bibfnamefont {W.}~\bibnamefont {Baumjohann}},\ }\href {\doibase
      10.1002/2017JA025019} {\bibfield  {journal} {\bibinfo  {journal} {J. Geophys.
      Res.}\ }\textbf {\bibinfo {volume} {123}},\ \bibinfo {pages} {1806} (\bibinfo
      {year} {2018})}\BibitemShut {NoStop}%
    \bibitem [{\citenamefont {Norgren}\ \emph {et~al.}(2016)\citenamefont
      {Norgren}, \citenamefont {Graham}, \citenamefont {Khotyaintsev},
      \citenamefont {Andr{\'{e}}}, \citenamefont {Vaivads}, \citenamefont {Chen},
      \citenamefont {Lindqvist}, \citenamefont {Marklund}, \citenamefont {Ergun},
      \citenamefont {Magnes}, \citenamefont {Strangeway}, \citenamefont {Russell},
      \citenamefont {Torbert}, \citenamefont {Paterson}, \citenamefont {Gershman},
      \citenamefont {Dorelli}, \citenamefont {Avanov}, \citenamefont {Lavraud},
      \citenamefont {Saito}, \citenamefont {Giles}, \citenamefont {Pollock},\ and\
      \citenamefont {Burch}}]{Norgren2016}%
      \BibitemOpen
      \bibfield  {author} {\bibinfo {author} {\bibfnamefont {C.}~\bibnamefont
      {Norgren}}, \bibinfo {author} {\bibfnamefont {D.~B.}\ \bibnamefont {Graham}},
      \bibinfo {author} {\bibfnamefont {Y.~V.}\ \bibnamefont {Khotyaintsev}},
      \bibinfo {author} {\bibfnamefont {M.}~\bibnamefont {Andr{\'{e}}}}, \bibinfo
      {author} {\bibfnamefont {A.}~\bibnamefont {Vaivads}}, \bibinfo {author}
      {\bibfnamefont {L.-J.}\ \bibnamefont {Chen}}, \bibinfo {author}
      {\bibfnamefont {P.-A.}\ \bibnamefont {Lindqvist}}, \bibinfo {author}
      {\bibfnamefont {G.~T.}\ \bibnamefont {Marklund}}, \bibinfo {author}
      {\bibfnamefont {R.~E.}\ \bibnamefont {Ergun}}, \bibinfo {author}
      {\bibfnamefont {W.}~\bibnamefont {Magnes}}, \bibinfo {author} {\bibfnamefont
      {R.~J.}\ \bibnamefont {Strangeway}}, \bibinfo {author} {\bibfnamefont
      {C.~T.}\ \bibnamefont {Russell}}, \bibinfo {author} {\bibfnamefont {R.~B.}\
      \bibnamefont {Torbert}}, \bibinfo {author} {\bibfnamefont {W.~R.}\
      \bibnamefont {Paterson}}, \bibinfo {author} {\bibfnamefont {D.~J.}\
      \bibnamefont {Gershman}}, \bibinfo {author} {\bibfnamefont {J.~C.}\
      \bibnamefont {Dorelli}}, \bibinfo {author} {\bibfnamefont {L.~A.}\
      \bibnamefont {Avanov}}, \bibinfo {author} {\bibfnamefont {B.}~\bibnamefont
      {Lavraud}}, \bibinfo {author} {\bibfnamefont {Y.}~\bibnamefont {Saito}},
      \bibinfo {author} {\bibfnamefont {B.~L.}\ \bibnamefont {Giles}}, \bibinfo
      {author} {\bibfnamefont {C.~J.}\ \bibnamefont {Pollock}}, \ and\ \bibinfo
      {author} {\bibfnamefont {J.~L.}\ \bibnamefont {Burch}},\ }\href {\doibase
      10.1002/2016GL069205} {\bibfield  {journal} {\bibinfo  {journal} {Geophys.
      Res. Lett.}\ }\textbf {\bibinfo {volume} {43}},\ \bibinfo {pages} {6724}
      (\bibinfo {year} {2016})}\BibitemShut {NoStop}%
    \bibitem [{\citenamefont {Yu}\ \emph {et~al.}(2019)\citenamefont {Yu},
      \citenamefont {Wang}, \citenamefont {Lu}, \citenamefont {Russell},\ and\
      \citenamefont {Wang}}]{Yu2019}%
      \BibitemOpen
      \bibfield  {author} {\bibinfo {author} {\bibfnamefont {X.}~\bibnamefont
      {Yu}}, \bibinfo {author} {\bibfnamefont {R.}~\bibnamefont {Wang}}, \bibinfo
      {author} {\bibfnamefont {Q.}~\bibnamefont {Lu}}, \bibinfo {author}
      {\bibfnamefont {C.~T.}\ \bibnamefont {Russell}}, \ and\ \bibinfo {author}
      {\bibfnamefont {S.}~\bibnamefont {Wang}},\ }\href {\doibase
      10.1029/2019GL082538} {\bibfield  {journal} {\bibinfo  {journal} {Geophys.
      Res. Lett.}\ }\textbf {\bibinfo {volume} {46}},\ \bibinfo {pages} {10744}
      (\bibinfo {year} {2019})}\BibitemShut {NoStop}%
    \bibitem [{\citenamefont {Hesse}\ \emph {et~al.}(2014)\citenamefont {Hesse},
      \citenamefont {Aunai}, \citenamefont {Sibeck},\ and\ \citenamefont
      {Birn}}]{Hesse2014}%
      \BibitemOpen
      \bibfield  {author} {\bibinfo {author} {\bibfnamefont {M.}~\bibnamefont
      {Hesse}}, \bibinfo {author} {\bibfnamefont {N.}~\bibnamefont {Aunai}},
      \bibinfo {author} {\bibfnamefont {D.}~\bibnamefont {Sibeck}}, \ and\ \bibinfo
      {author} {\bibfnamefont {J.}~\bibnamefont {Birn}},\ }\href {\doibase
      10.1002/2014GL061586} {\bibfield  {journal} {\bibinfo  {journal} {Geophys.
      Res. Lett.}\ }\textbf {\bibinfo {volume} {41}},\ \bibinfo {pages} {8673}
      (\bibinfo {year} {2014})}\BibitemShut {NoStop}%
    \bibitem [{\citenamefont {Bessho}\ \emph {et~al.}(2019)\citenamefont {Bessho},
      \citenamefont {Chen}, \citenamefont {Wang},\ and\ \citenamefont
      {Hesse}}]{Bessho2019}%
      \BibitemOpen
      \bibfield  {author} {\bibinfo {author} {\bibfnamefont {N.}~\bibnamefont
      {Bessho}}, \bibinfo {author} {\bibfnamefont {L.-J.}\ \bibnamefont {Chen}},
      \bibinfo {author} {\bibfnamefont {S.}~\bibnamefont {Wang}}, \ and\ \bibinfo
      {author} {\bibfnamefont {M.}~\bibnamefont {Hesse}},\ }\href {\doibase
      10.1063/1.5092809} {\bibfield  {journal} {\bibinfo  {journal} {Phys.
      Plasmas}\ }\textbf {\bibinfo {volume} {26}},\ \bibinfo {pages} {082310}
      (\bibinfo {year} {2019})}\BibitemShut {NoStop}%
    \bibitem [{\citenamefont {Shay}\ \emph {et~al.}(2016)\citenamefont {Shay},
      \citenamefont {Phan}, \citenamefont {Haggerty}, \citenamefont {Fujimoto},
      \citenamefont {Drake}, \citenamefont {Malakit}, \citenamefont {Cassak},\ and\
      \citenamefont {Swisdak}}]{Shay2016}%
      \BibitemOpen
      \bibfield  {author} {\bibinfo {author} {\bibfnamefont {M.~A.}\ \bibnamefont
      {Shay}}, \bibinfo {author} {\bibfnamefont {T.~D.}\ \bibnamefont {Phan}},
      \bibinfo {author} {\bibfnamefont {C.~C.}\ \bibnamefont {Haggerty}}, \bibinfo
      {author} {\bibfnamefont {M.}~\bibnamefont {Fujimoto}}, \bibinfo {author}
      {\bibfnamefont {J.~F.}\ \bibnamefont {Drake}}, \bibinfo {author}
      {\bibfnamefont {K.}~\bibnamefont {Malakit}}, \bibinfo {author} {\bibfnamefont
      {P.~A.}\ \bibnamefont {Cassak}}, \ and\ \bibinfo {author} {\bibfnamefont
      {M.}~\bibnamefont {Swisdak}},\ }\href {\doibase 10.1002/2016GL069034}
      {\bibfield  {journal} {\bibinfo  {journal} {Geophys. Res. Lett.}\ }\textbf
      {\bibinfo {volume} {43}},\ \bibinfo {pages} {4145} (\bibinfo {year}
      {2016})}\BibitemShut {NoStop}%
    \bibitem [{\citenamefont {Chen}\ \emph
      {et~al.}(2016{\natexlab{b}})\citenamefont {Chen}, \citenamefont {Hesse},
      \citenamefont {Wang}, \citenamefont {Bessho},\ and\ \citenamefont
      {Daughton}}]{Chen2016}%
      \BibitemOpen
      \bibfield  {author} {\bibinfo {author} {\bibfnamefont {L.}~\bibnamefont
      {Chen}}, \bibinfo {author} {\bibfnamefont {M.}~\bibnamefont {Hesse}},
      \bibinfo {author} {\bibfnamefont {S.}~\bibnamefont {Wang}}, \bibinfo {author}
      {\bibfnamefont {N.}~\bibnamefont {Bessho}}, \ and\ \bibinfo {author}
      {\bibfnamefont {W.}~\bibnamefont {Daughton}},\ }\href {\doibase
      10.1002/2016GL068243} {\bibfield  {journal} {\bibinfo  {journal} {Geophys.
      Res. Lett.}\ }\textbf {\bibinfo {volume} {43}},\ \bibinfo {pages} {2405}
      (\bibinfo {year} {2016}{\natexlab{b}})}\BibitemShut {NoStop}%
    \bibitem [{\citenamefont {Zenitani}, \citenamefont {Hasegawa},\ and\
      \citenamefont {Nagai}(2017)}]{Zenitani2017}%
      \BibitemOpen
      \bibfield  {author} {\bibinfo {author} {\bibfnamefont {S.}~\bibnamefont
      {Zenitani}}, \bibinfo {author} {\bibfnamefont {H.}~\bibnamefont {Hasegawa}},
      \ and\ \bibinfo {author} {\bibfnamefont {T.}~\bibnamefont {Nagai}},\ }\href
      {\doibase 10.1002/2017JA023969} {\bibfield  {journal} {\bibinfo  {journal}
      {J. Geophys. Res.}\ }\textbf {\bibinfo {volume} {122}},\ \bibinfo {pages}
      {7396} (\bibinfo {year} {2017})}\BibitemShut {NoStop}%
    \bibitem [{\citenamefont {B{\"{u}}chner}\ and\ \citenamefont
      {Kuska}(1996)}]{Buchner1996}%
      \BibitemOpen
      \bibfield  {author} {\bibinfo {author} {\bibfnamefont {J.}~\bibnamefont
      {B{\"{u}}chner}}\ and\ \bibinfo {author} {\bibfnamefont {J.-P.}\ \bibnamefont
      {Kuska}},\ }\href {\doibase 10.5636/jgg.48.781} {\bibfield  {journal}
      {\bibinfo  {journal} {J. Geomagn. Geoelectr.}\ }\textbf {\bibinfo {volume}
      {48}},\ \bibinfo {pages} {781} (\bibinfo {year} {1996})}\BibitemShut
      {NoStop}%
    \bibitem [{\citenamefont {Lee}\ \emph {et~al.}(2004)\citenamefont {Lee},
      \citenamefont {Wilber}, \citenamefont {Parks}, \citenamefont {Min},\ and\
      \citenamefont {Lee}}]{Lee2004}%
      \BibitemOpen
      \bibfield  {author} {\bibinfo {author} {\bibfnamefont {E.}~\bibnamefont
      {Lee}}, \bibinfo {author} {\bibfnamefont {M.}~\bibnamefont {Wilber}},
      \bibinfo {author} {\bibfnamefont {G.~K.}\ \bibnamefont {Parks}}, \bibinfo
      {author} {\bibfnamefont {K.~W.}\ \bibnamefont {Min}}, \ and\ \bibinfo
      {author} {\bibfnamefont {D.-Y.}\ \bibnamefont {Lee}},\ }\href {\doibase
      10.1029/2004GL020331} {\bibfield  {journal} {\bibinfo  {journal} {Geophys.
      Res. Lett.}\ }\textbf {\bibinfo {volume} {31}},\ \bibinfo {pages} {L21806}
      (\bibinfo {year} {2004})}\BibitemShut {NoStop}%
    \bibitem [{\citenamefont {Usami}, \citenamefont {Horiuchi},\ and\ \citenamefont
      {Ohtani}(2017)}]{Usami2017}%
      \BibitemOpen
      \bibfield  {author} {\bibinfo {author} {\bibfnamefont {S.}~\bibnamefont
      {Usami}}, \bibinfo {author} {\bibfnamefont {R.}~\bibnamefont {Horiuchi}}, \
      and\ \bibinfo {author} {\bibfnamefont {H.}~\bibnamefont {Ohtani}},\ }\href
      {\doibase 10.1063/1.4997453} {\bibfield  {journal} {\bibinfo  {journal}
      {Phys. Plasmas}\ }\textbf {\bibinfo {volume} {24}},\ \bibinfo {pages}
      {092101} (\bibinfo {year} {2017})}\BibitemShut {NoStop}%
    \bibitem [{\citenamefont {Lapenta}\ \emph {et~al.}(2017)\citenamefont
      {Lapenta}, \citenamefont {Berchem}, \citenamefont {Zhou}, \citenamefont
      {Walker}, \citenamefont {El-Alaoui}, \citenamefont {Goldstein}, \citenamefont
      {Paterson}, \citenamefont {Giles}, \citenamefont {Pollock}, \citenamefont
      {Russell}, \citenamefont {Strangeway}, \citenamefont {Ergun}, \citenamefont
      {Khotyaintsev}, \citenamefont {Torbert},\ and\ \citenamefont
      {Burch}}]{Lapenta2017d}%
      \BibitemOpen
      \bibfield  {author} {\bibinfo {author} {\bibfnamefont {G.}~\bibnamefont
      {Lapenta}}, \bibinfo {author} {\bibfnamefont {J.}~\bibnamefont {Berchem}},
      \bibinfo {author} {\bibfnamefont {M.}~\bibnamefont {Zhou}}, \bibinfo {author}
      {\bibfnamefont {R.~J.}\ \bibnamefont {Walker}}, \bibinfo {author}
      {\bibfnamefont {M.}~\bibnamefont {El-Alaoui}}, \bibinfo {author}
      {\bibfnamefont {M.~L.}\ \bibnamefont {Goldstein}}, \bibinfo {author}
      {\bibfnamefont {W.~R.}\ \bibnamefont {Paterson}}, \bibinfo {author}
      {\bibfnamefont {B.~L.}\ \bibnamefont {Giles}}, \bibinfo {author}
      {\bibfnamefont {C.~J.}\ \bibnamefont {Pollock}}, \bibinfo {author}
      {\bibfnamefont {C.~T.}\ \bibnamefont {Russell}}, \bibinfo {author}
      {\bibfnamefont {R.~J.}\ \bibnamefont {Strangeway}}, \bibinfo {author}
      {\bibfnamefont {R.~E.}\ \bibnamefont {Ergun}}, \bibinfo {author}
      {\bibfnamefont {Y.~V.}\ \bibnamefont {Khotyaintsev}}, \bibinfo {author}
      {\bibfnamefont {R.~B.}\ \bibnamefont {Torbert}}, \ and\ \bibinfo {author}
      {\bibfnamefont {J.~L.}\ \bibnamefont {Burch}},\ }\href {\doibase
      10.1002/2016JA023290} {\bibfield  {journal} {\bibinfo  {journal} {J. Geophys.
      Res.}\ }\textbf {\bibinfo {volume} {122}},\ \bibinfo {pages} {2024} (\bibinfo
      {year} {2017})}\BibitemShut {NoStop}%
    \bibitem [{\citenamefont {Voitcu}\ and\ \citenamefont
      {Echim}(2018)}]{Voitcu2018a}%
      \BibitemOpen
      \bibfield  {author} {\bibinfo {author} {\bibfnamefont {G.}~\bibnamefont
      {Voitcu}}\ and\ \bibinfo {author} {\bibfnamefont {M.}~\bibnamefont {Echim}},\
      }\href {\doibase 10.5194/angeo-36-1521-2018} {\bibfield  {journal} {\bibinfo
      {journal} {Ann. Geophys.}\ }\textbf {\bibinfo {volume} {36}},\ \bibinfo
      {pages} {1521} (\bibinfo {year} {2018})}\BibitemShut {NoStop}%
    \bibitem [{\citenamefont {Egedal}\ \emph {et~al.}(2008)\citenamefont {Egedal},
      \citenamefont {Fox}, \citenamefont {Katz}, \citenamefont {Porkolab},
      \citenamefont {{\O}Ieroset}, \citenamefont {Lin}, \citenamefont {Daughton},\
      and\ \citenamefont {Drake}}]{Egedal2008}%
      \BibitemOpen
      \bibfield  {author} {\bibinfo {author} {\bibfnamefont {J.}~\bibnamefont
      {Egedal}}, \bibinfo {author} {\bibfnamefont {W.}~\bibnamefont {Fox}},
      \bibinfo {author} {\bibfnamefont {N.}~\bibnamefont {Katz}}, \bibinfo {author}
      {\bibfnamefont {M.}~\bibnamefont {Porkolab}}, \bibinfo {author}
      {\bibfnamefont {M.}~\bibnamefont {{\O}Ieroset}}, \bibinfo {author}
      {\bibfnamefont {R.}~\bibnamefont {Lin}}, \bibinfo {author} {\bibfnamefont
      {W.}~\bibnamefont {Daughton}}, \ and\ \bibinfo {author} {\bibfnamefont
      {J.}~\bibnamefont {Drake}},\ }\href {\doibase 10.1029/2008JA013520}
      {\bibfield  {journal} {\bibinfo  {journal} {J. Geophys. Res.}\ }\textbf
      {\bibinfo {volume} {113}},\ \bibinfo {pages} {A12207} (\bibinfo {year}
      {2008})}\BibitemShut {NoStop}%
    \bibitem [{\citenamefont {Che}, \citenamefont {Drake},\ and\ \citenamefont
      {Swisdak}(2011)}]{Che2011}%
      \BibitemOpen
      \bibfield  {author} {\bibinfo {author} {\bibfnamefont {H.}~\bibnamefont
      {Che}}, \bibinfo {author} {\bibfnamefont {J.~F.}\ \bibnamefont {Drake}}, \
      and\ \bibinfo {author} {\bibfnamefont {M.}~\bibnamefont {Swisdak}},\ }\href
      {\doibase 10.1038/nature10091} {\bibfield  {journal} {\bibinfo  {journal}
      {Nature}\ }\textbf {\bibinfo {volume} {474}},\ \bibinfo {pages} {184}
      (\bibinfo {year} {2011})}\BibitemShut {NoStop}%
    \bibitem [{\citenamefont {Mu{\~{n}}oz}\ and\ \citenamefont
      {B{\"{u}}chner}(2018)}]{Munoz2018}%
      \BibitemOpen
      \bibfield  {author} {\bibinfo {author} {\bibfnamefont {P.~A.}\ \bibnamefont
      {Mu{\~{n}}oz}}\ and\ \bibinfo {author} {\bibfnamefont {J.}~\bibnamefont
      {B{\"{u}}chner}},\ }\href {\doibase 10.3847/1538-4357/aad5e9} {\bibfield
      {journal} {\bibinfo  {journal} {Astrophys. J.}\ }\textbf {\bibinfo {volume}
      {864}},\ \bibinfo {pages} {92} (\bibinfo {year} {2018})}\BibitemShut
      {NoStop}%
    \bibitem [{\citenamefont {Price}\ \emph {et~al.}(2016)\citenamefont {Price},
      \citenamefont {Swisdak}, \citenamefont {Drake}, \citenamefont {Cassak},
      \citenamefont {Dahlin},\ and\ \citenamefont {Ergun}}]{Price2016}%
      \BibitemOpen
      \bibfield  {author} {\bibinfo {author} {\bibfnamefont {L.}~\bibnamefont
      {Price}}, \bibinfo {author} {\bibfnamefont {M.}~\bibnamefont {Swisdak}},
      \bibinfo {author} {\bibfnamefont {J.~F.}\ \bibnamefont {Drake}}, \bibinfo
      {author} {\bibfnamefont {P.~A.}\ \bibnamefont {Cassak}}, \bibinfo {author}
      {\bibfnamefont {J.~T.}\ \bibnamefont {Dahlin}}, \ and\ \bibinfo {author}
      {\bibfnamefont {R.~E.}\ \bibnamefont {Ergun}},\ }\href {\doibase
      10.1002/2016GL069578} {\bibfield  {journal} {\bibinfo  {journal} {Geophys.
      Res. Lett.}\ }\textbf {\bibinfo {volume} {43}},\ \bibinfo {pages} {6020}
      (\bibinfo {year} {2016})}\BibitemShut {NoStop}%
    \bibitem [{\citenamefont {Le}\ \emph {et~al.}(2017)\citenamefont {Le},
      \citenamefont {Daughton}, \citenamefont {Chen},\ and\ \citenamefont
      {Egedal}}]{Le2017}%
      \BibitemOpen
      \bibfield  {author} {\bibinfo {author} {\bibfnamefont {A.}~\bibnamefont
      {Le}}, \bibinfo {author} {\bibfnamefont {W.}~\bibnamefont {Daughton}},
      \bibinfo {author} {\bibfnamefont {L.}~\bibnamefont {Chen}}, \ and\ \bibinfo
      {author} {\bibfnamefont {J.}~\bibnamefont {Egedal}},\ }\href {\doibase
      10.1002/2017GL072522} {\bibfield  {journal} {\bibinfo  {journal} {Geophys.
      Res. Lett.}\ }\textbf {\bibinfo {volume} {44}},\ \bibinfo {pages} {2096}
      (\bibinfo {year} {2017})}\BibitemShut {NoStop}%
    \bibitem [{\citenamefont {Dupuis}\ \emph {et~al.}(2020)\citenamefont {Dupuis},
      \citenamefont {Goldman}, \citenamefont {Newman}, \citenamefont {Amaya},\ and\
      \citenamefont {Lapenta}}]{Dupuis2020}%
      \BibitemOpen
      \bibfield  {author} {\bibinfo {author} {\bibfnamefont {R.}~\bibnamefont
      {Dupuis}}, \bibinfo {author} {\bibfnamefont {M.~V.}\ \bibnamefont {Goldman}},
      \bibinfo {author} {\bibfnamefont {D.~L.}\ \bibnamefont {Newman}}, \bibinfo
      {author} {\bibfnamefont {J.}~\bibnamefont {Amaya}}, \ and\ \bibinfo {author}
      {\bibfnamefont {G.}~\bibnamefont {Lapenta}},\ }\href {\doibase
      10.3847/1538-4357/ab5524} {\bibfield  {journal} {\bibinfo  {journal}
      {Astrophys. J.}\ }\textbf {\bibinfo {volume} {889}},\ \bibinfo {pages} {22}
      (\bibinfo {year} {2020})}\BibitemShut {NoStop}%
    \bibitem [{\citenamefont {Aunai}, \citenamefont {Hesse},\ and\ \citenamefont
      {Kuznetsova}(2013)}]{Aunai2013}%
      \BibitemOpen
      \bibfield  {author} {\bibinfo {author} {\bibfnamefont {N.}~\bibnamefont
      {Aunai}}, \bibinfo {author} {\bibfnamefont {M.}~\bibnamefont {Hesse}}, \ and\
      \bibinfo {author} {\bibfnamefont {M.}~\bibnamefont {Kuznetsova}},\ }\href
      {\doibase 10.1063/1.4820953} {\bibfield  {journal} {\bibinfo  {journal}
      {Phys. Plasmas}\ }\textbf {\bibinfo {volume} {20}},\ \bibinfo {pages}
      {092903} (\bibinfo {year} {2013})}\BibitemShut {NoStop}%
    \bibitem [{\citenamefont {Zenitani}\ \emph {et~al.}(2011)\citenamefont
      {Zenitani}, \citenamefont {Hesse}, \citenamefont {Klimas},\ and\
      \citenamefont {Kuznetsova}}]{Zenitani2011}%
      \BibitemOpen
      \bibfield  {author} {\bibinfo {author} {\bibfnamefont {S.}~\bibnamefont
      {Zenitani}}, \bibinfo {author} {\bibfnamefont {M.}~\bibnamefont {Hesse}},
      \bibinfo {author} {\bibfnamefont {A.}~\bibnamefont {Klimas}}, \ and\ \bibinfo
      {author} {\bibfnamefont {M.}~\bibnamefont {Kuznetsova}},\ }\href {\doibase
      10.1103/PhysRevLett.106.195003} {\bibfield  {journal} {\bibinfo  {journal}
      {Phys. Rev. Lett.}\ }\textbf {\bibinfo {volume} {106}},\ \bibinfo {pages}
      {195003} (\bibinfo {year} {2011})}\BibitemShut {NoStop}%
    \bibitem [{\citenamefont {Kilian}, \citenamefont {Burkart},\ and\ \citenamefont
      {Spanier}(2012)}]{Kilian2012}%
      \BibitemOpen
      \bibfield  {author} {\bibinfo {author} {\bibfnamefont {P.}~\bibnamefont
      {Kilian}}, \bibinfo {author} {\bibfnamefont {T.}~\bibnamefont {Burkart}}, \
      and\ \bibinfo {author} {\bibfnamefont {F.}~\bibnamefont {Spanier}},\ }in\
      \href {\doibase 10.1007/978-3-642-23869-7_1} {\emph {\bibinfo {booktitle}
      {High Perform. Comput. Sci. Eng. '11}}},\ \bibinfo {editor} {edited by\
      \bibinfo {editor} {\bibfnamefont {W.~E.}\ \bibnamefont {Nagel}}, \bibinfo
      {editor} {\bibfnamefont {D.~B.}\ \bibnamefont {Kr{\"{o}}ner}}, \ and\
      \bibinfo {editor} {\bibfnamefont {M.~M.}\ \bibnamefont {Resch}}}\ (\bibinfo
      {publisher} {Springer Berlin Heidelberg},\ \bibinfo {address} {Berlin,
      Heidelberg},\ \bibinfo {year} {2012})\ pp.\ \bibinfo {pages}
      {5--13}\BibitemShut {NoStop}%
    \bibitem [{\citenamefont {Harris}(1962)}]{Harris1962}%
      \BibitemOpen
      \bibfield  {author} {\bibinfo {author} {\bibfnamefont {E.~G.}\ \bibnamefont
      {Harris}},\ }\href {\doibase 10.1007/BF02733547} {\bibfield  {journal}
      {\bibinfo  {journal} {Nuovo Cim.}\ }\textbf {\bibinfo {volume} {23}},\
      \bibinfo {pages} {115} (\bibinfo {year} {1962})}\BibitemShut {NoStop}%
    \bibitem [{\citenamefont {Daughton}(2005)}]{Daughton2005}%
      \BibitemOpen
      \bibfield  {author} {\bibinfo {author} {\bibfnamefont {W.}~\bibnamefont
      {Daughton}},\ }\href {\doibase 10.1029/2004JA010751} {\bibfield  {journal}
      {\bibinfo  {journal} {J. Geophys. Res.}\ }\textbf {\bibinfo {volume} {110}},\
      \bibinfo {pages} {A03217} (\bibinfo {year} {2005})}\BibitemShut {NoStop}%
    \bibitem [{\citenamefont {Goldman}, \citenamefont {Newman},\ and\ \citenamefont
      {Lapenta}(2016)}]{Goldman2016}%
      \BibitemOpen
      \bibfield  {author} {\bibinfo {author} {\bibfnamefont {M.~V.}\ \bibnamefont
      {Goldman}}, \bibinfo {author} {\bibfnamefont {D.~L.}\ \bibnamefont {Newman}},
      \ and\ \bibinfo {author} {\bibfnamefont {G.}~\bibnamefont {Lapenta}},\ }\href
      {\doibase 10.1007/s11214-015-0154-y} {\bibfield  {journal} {\bibinfo
      {journal} {Space Sci. Rev.}\ }\textbf {\bibinfo {volume} {199}},\ \bibinfo
      {pages} {651} (\bibinfo {year} {2016})}\BibitemShut {NoStop}%
    \bibitem [{\citenamefont {Bishop}(2006)}]{Bishop2006}%
      \BibitemOpen
      \bibfield  {author} {\bibinfo {author} {\bibfnamefont {C.~M.}\ \bibnamefont
      {Bishop}},\ }\href@noop {} {\emph {\bibinfo {title} {{Pattern Recognition and
      Machine Learning}}}}\ (\bibinfo  {publisher} {Springer},\ \bibinfo {year}
      {2006})\BibitemShut {NoStop}%
    \bibitem [{\citenamefont {Livadiotis}(2017)}]{Livadiotis2017}%
      \BibitemOpen
      \bibfield  {author} {\bibinfo {author} {\bibfnamefont {G.}~\bibnamefont
      {Livadiotis}},\ }\href@noop {} {\bibfield  {journal} {\bibinfo  {journal}
      {Kappa Distrib. Theory Appl. Plasmas}\ }\textbf {\bibinfo {volume} {10003}},\
      \bibinfo {pages} {1} (\bibinfo {year} {2017})}\BibitemShut {NoStop}%
    \bibitem [{\citenamefont {Livadiotis}(2018)}]{Livadiotis2018}%
      \BibitemOpen
      \bibfield  {author} {\bibinfo {author} {\bibfnamefont {G.}~\bibnamefont
      {Livadiotis}},\ }\href {\doibase 10.3390/universe4120144} {\bibfield
      {journal} {\bibinfo  {journal} {Universe}\ }\textbf {\bibinfo {volume} {4}},\
      \bibinfo {pages} {144} (\bibinfo {year} {2018})}\BibitemShut {NoStop}%
    \bibitem [{\citenamefont {B{\"{u}}chner}\ and\ \citenamefont
      {Zelenyi}(1989)}]{Buchner1989}%
      \BibitemOpen
      \bibfield  {author} {\bibinfo {author} {\bibfnamefont {J.}~\bibnamefont
      {B{\"{u}}chner}}\ and\ \bibinfo {author} {\bibfnamefont {L.~M.}\ \bibnamefont
      {Zelenyi}},\ }\href {\doibase 10.1029/JA094iA09p11821} {\bibfield  {journal}
      {\bibinfo  {journal} {J. Geophys. Res.}\ }\textbf {\bibinfo {volume} {94}},\
      \bibinfo {pages} {11821} (\bibinfo {year} {1989})}\BibitemShut {NoStop}%
    \bibitem [{\citenamefont {B{\"{u}}chner}\ and\ \citenamefont
      {Zelenyi}(1991)}]{Buchner1991a}%
      \BibitemOpen
      \bibfield  {author} {\bibinfo {author} {\bibfnamefont {J.}~\bibnamefont
      {B{\"{u}}chner}}\ and\ \bibinfo {author} {\bibfnamefont {L.~M.}\ \bibnamefont
      {Zelenyi}},\ }\href {\doibase 10.1016/0273-1177(91)90030-N} {\bibfield
      {journal} {\bibinfo  {journal} {Adv. Space Res.}\ }\textbf {\bibinfo {volume}
      {11}},\ \bibinfo {pages} {177} (\bibinfo {year} {1991})}\BibitemShut
      {NoStop}%
    \bibitem [{\citenamefont {Horiuchi}\ and\ \citenamefont
      {Sato}(1997)}]{Horiuchi1997}%
      \BibitemOpen
      \bibfield  {author} {\bibinfo {author} {\bibfnamefont {R.}~\bibnamefont
      {Horiuchi}}\ and\ \bibinfo {author} {\bibfnamefont {T.}~\bibnamefont
      {Sato}},\ }\href {\doibase 10.1063/1.872088} {\bibfield  {journal} {\bibinfo
      {journal} {Phys. Plasmas}\ }\textbf {\bibinfo {volume} {4}},\ \bibinfo
      {pages} {277} (\bibinfo {year} {1997})}\BibitemShut {NoStop}%
    \bibitem [{\citenamefont {Ricci}\ \emph {et~al.}(2004)\citenamefont {Ricci},
      \citenamefont {Brackbill}, \citenamefont {Daughton},\ and\ \citenamefont
      {Lapenta}}]{Ricci2004}%
      \BibitemOpen
      \bibfield  {author} {\bibinfo {author} {\bibfnamefont {P.}~\bibnamefont
      {Ricci}}, \bibinfo {author} {\bibfnamefont {J.~U.}\ \bibnamefont
      {Brackbill}}, \bibinfo {author} {\bibfnamefont {W.}~\bibnamefont {Daughton}},
      \ and\ \bibinfo {author} {\bibfnamefont {G.}~\bibnamefont {Lapenta}},\ }\href
      {\doibase 10.1063/1.1768552} {\bibfield  {journal} {\bibinfo  {journal}
      {Phys. Plasmas}\ }\textbf {\bibinfo {volume} {11}},\ \bibinfo {pages} {4102}
      (\bibinfo {year} {2004})}\BibitemShut {NoStop}%
    \bibitem [{\citenamefont {Cassak}, \citenamefont {Liu},\ and\ \citenamefont
      {Shay}(2017)}]{Cassak2017}%
      \BibitemOpen
      \bibfield  {author} {\bibinfo {author} {\bibfnamefont {P.~A.}\ \bibnamefont
      {Cassak}}, \bibinfo {author} {\bibfnamefont {Y.-H.}\ \bibnamefont {Liu}}, \
      and\ \bibinfo {author} {\bibfnamefont {M.~A.}\ \bibnamefont {Shay}},\ }\href
      {\doibase 10.1017/S0022377817000666} {\bibfield  {journal} {\bibinfo
      {journal} {J. Plasma Phys.}\ }\textbf {\bibinfo {volume} {83}},\ \bibinfo
      {pages} {715830501} (\bibinfo {year} {2017})}\BibitemShut {NoStop}%
    \bibitem [{\citenamefont {Liu}\ \emph {et~al.}(2017)\citenamefont {Liu},
      \citenamefont {Hesse}, \citenamefont {Guo}, \citenamefont {Daughton},
      \citenamefont {Li}, \citenamefont {Cassak},\ and\ \citenamefont
      {Shay}}]{Liu2017}%
      \BibitemOpen
      \bibfield  {author} {\bibinfo {author} {\bibfnamefont {Y.-H.}\ \bibnamefont
      {Liu}}, \bibinfo {author} {\bibfnamefont {M.}~\bibnamefont {Hesse}}, \bibinfo
      {author} {\bibfnamefont {F.}~\bibnamefont {Guo}}, \bibinfo {author}
      {\bibfnamefont {W.}~\bibnamefont {Daughton}}, \bibinfo {author}
      {\bibfnamefont {H.}~\bibnamefont {Li}}, \bibinfo {author} {\bibfnamefont
      {P.~A.}\ \bibnamefont {Cassak}}, \ and\ \bibinfo {author} {\bibfnamefont
      {M.~A.}\ \bibnamefont {Shay}},\ }\href {\doibase
      10.1103/PhysRevLett.118.085101} {\bibfield  {journal} {\bibinfo  {journal}
      {Phys. Rev. Lett.}\ }\textbf {\bibinfo {volume} {118}},\ \bibinfo {pages}
      {085101} (\bibinfo {year} {2017})}\BibitemShut {NoStop}%
    \bibitem [{\citenamefont {Dempster}, \citenamefont {Laird},\ and\ \citenamefont
      {Rubin}(1977)}]{Dempster1977}%
      \BibitemOpen
      \bibfield  {author} {\bibinfo {author} {\bibfnamefont {A.~P.}\ \bibnamefont
      {Dempster}}, \bibinfo {author} {\bibfnamefont {N.~M.}\ \bibnamefont {Laird}},
      \ and\ \bibinfo {author} {\bibfnamefont {D.~B.}\ \bibnamefont {Rubin}},\
      }\href {\doibase 10.1111/j.2517-6161.1977.tb01600.x} {\bibfield  {journal}
      {\bibinfo  {journal} {J. R. Stat. Soc. Ser. B}\ }\textbf {\bibinfo {volume}
      {39}},\ \bibinfo {pages} {1} (\bibinfo {year} {1977})}\BibitemShut {NoStop}%
    \bibitem [{\citenamefont {McLachlan}\ and\ \citenamefont
      {Krishnan}(2008)}]{McLachlan2008}%
      \BibitemOpen
      \bibfield  {author} {\bibinfo {author} {\bibfnamefont {G.~J.}\ \bibnamefont
      {McLachlan}}\ and\ \bibinfo {author} {\bibfnamefont {T.}~\bibnamefont
      {Krishnan}},\ }\href {\doibase 10.1002/9780470191613} {\emph {\bibinfo
      {title} {{The EM Algorithm and Extensions}}}},\ Wiley Series in Probability
      and Statistics\ (\bibinfo  {publisher} {John Wiley {\&} Sons, Inc.},\
      \bibinfo {address} {Hoboken, NJ, USA},\ \bibinfo {year} {2008})\BibitemShut
      {NoStop}%
    \bibitem [{\citenamefont {Burnham}\ and\ \citenamefont
      {Anderson}(2004)}]{Burnham2004}%
      \BibitemOpen
      \bibfield  {author} {\bibinfo {author} {\bibfnamefont {K.~P.}\ \bibnamefont
      {Burnham}}\ and\ \bibinfo {author} {\bibfnamefont {D.~R.}\ \bibnamefont
      {Anderson}},\ }\href {\doibase 10.1007/b97636} {\emph {\bibinfo {title}
      {{Model Selection and Multimodel Inference}}}},\ edited by\ \bibinfo {editor}
      {\bibfnamefont {K.~P.}\ \bibnamefont {Burnham}}\ and\ \bibinfo {editor}
      {\bibfnamefont {D.~R.}\ \bibnamefont {Anderson}}\ (\bibinfo  {publisher}
      {Springer New York},\ \bibinfo {address} {New York, NY},\ \bibinfo {year}
      {2004})\ pp.\ \bibinfo {pages} {352--436}\BibitemShut {NoStop}%
    \end{thebibliography}

%

\end{document}